\documentclass[graybox]{svmult}

\usepackage{type1cm}        
\usepackage{makeidx}         
\usepackage{graphicx}        
\usepackage{multicol}        
\usepackage[bottom]{footmisc}

\usepackage{newtxtext}       %
\usepackage[varvw]{newtxmath}       

\usepackage{cite}
\usepackage{float}
\usepackage{scrextend} 
\usepackage{subcaption}
\usepackage{url}
\usepackage{verbatim}           
\usepackage{xcolor}
\usepackage{xspace}

\usepackage{lineno}

\usepackage[pagebackref=true,colorlinks=true]{hyperref}
\usepackage[colorlinks=true]{hyperref}
\hypersetup{
     unicode=true,          
     pdfnewwindow=true,      
     bookmarksnumbered=true,
     colorlinks=true,       
     linkcolor=blue,          
     citecolor=blue,        
     filecolor=magenta,      
     urlcolor=cyan,           
     menucolor=blue,
     linktocpage=true
}

\def\TeV{\ifmmode {\mathrm{\ Te\kern -0.1em V}}\else
                   \textrm{Te\kern -0.1em V}\fi}%
\def\GeV{\ifmmode {\mathrm{\ Ge\kern -0.1em V}}\else
                   \textrm{Ge\kern -0.1em V}\fi}%

\makeindex             

\usepackage{epigraph}
\begin{document}

\let\Xspace\xspace

\DeclareRobustCommand{\Pepem}{\HepParticle{\Pe}{}{+}\HepParticle{\Pe}{}{-}\Xspace} 
\DeclareRobustCommand{\PGmpGmm}{\HepParticle{\PGm}{}{+}\HepParticle{\PGm}{}{-}\Xspace} 
\newcommand{\pT}{\ensuremath{p\sb{\scriptstyle\mathrm{T}}}\Xspace}

\newcommand{\sqrts}{\ensuremath{\sqrt{s}}\Xspace}
\newcommand{\sNNraw}{\ensuremath{s_{\mbox{\tiny NN}}}}
\newcommand{\sqrtsNN}{\ensuremath{\sqrt{\sNNraw}}\Xspace}
\newcommand{\sNN}{\sqrtsNN}
\newcommand{\Npart}{\ensuremath{N_{\rm part}}\Xspace}
\newcommand{\Ncoll}{\ensuremath{N_{\rm coll}}\Xspace}
\newcommand{\RpPb}{\ensuremath{R_{\rm pPb}}\Xspace}
\newcommand{\RdAu}{\ensuremath{R_{\rm dAu}}\Xspace}
\newcommand{\RAA}{\ensuremath{R_{\rm AA}}\Xspace}
\newcommand{\RpA}{\ensuremath{R_{\rm pA}}\Xspace}
\newcommand{\TAA}{\ensuremath{T_{\rm AA}}\Xspace}
\newcommand{\RCP}{\ensuremath{R_{\rm CP}}\Xspace}
\newcommand{\vtwo}{\ensuremath{v_{\rm 2}}\Xspace}
\newcommand{\vone}{\ensuremath{v_{\rm 1}}\Xspace}
\newcommand{\vthree}{\ensuremath{v_{\rm 3}}\Xspace}
\newcommand{\vfour}{\ensuremath{v_{\rm 4}}\Xspace}
\newcommand{\vfive}{\ensuremath{v_{\rm 5}}\Xspace}
\newcommand{\vsix}{\ensuremath{v_{\rm 6}}\Xspace}
\newcommand{\vseven}{\ensuremath{v_{\rm 7}}\Xspace}
\newcommand{\veight}{\ensuremath{v_{\rm 8}}\Xspace}
\newcommand{\vnine}{\ensuremath{v_{\rm 9}}\Xspace}
\newcommand{\vn}{\ensuremath{v_{\rm n}}\Xspace}

\newcommand{\nch}         {\ensuremath{N}\Xspace}
\newcommand{\mch}         {\ensuremath{M_{\mathrm {ch}}}\Xspace}
\newcommand{\meannch}     {\ensuremath{\langle N_{\mathrm {ch}} \rangle}\Xspace}
\newcommand{\meanpT}      {\ensuremath{\langle\pT\rangle}}
\newcommand{\dNdeta}      {\mathrm{d}N_\mathrm{ch}/\mathrm{d}\eta}
\newcommand{\dNdy}        {\mathrm{d}N_\mathrm{ch}/\mathrm{d}y}
\newcommand{\kT}          {\ensuremath{k\sb{\scriptstyle\mathrm{T}}}}
\newcommand{\ptt}         {\ensuremath{p_{\mathrm{T, trig}}}}
\newcommand{\pta}         {\ensuremath{p_{\mathrm{T, assoc}}}}

\newcommand{\ee}          {\ensuremath{\mathrm{e^+e^-}}\Xspace}
\newcommand{\pp}          {$\mathrm{pp}$\Xspace}
\newcommand{\pPb}         {$\mathrm{p}$--$\mathrm{Pb}$\Xspace}
\newcommand{\Pbp}         {$\mathrm{Pb}$--$\mathrm{p}$\Xspace}
\newcommand{\pO}          {$\mathrm{p}$--$\mathrm{O}$\Xspace}
\newcommand{\Op}          {$\mathrm{O}$--$\mathrm{p}$\Xspace}
\newcommand{\OO}          {$\mathrm{O}$--$\mathrm{O}$\Xspace}
\newcommand{\pA}          {$\mathrm{p}$--$\mathrm{A}$\Xspace}
\newcommand{\dA}          {$\mathrm{d}$--$\mathrm{A}$\Xspace}
\newcommand{\AOnA}        {$\mathrm{A}$--$\mathrm{A}$\Xspace}
\newcommand{\PbPb}        {$\mathrm{Pb}$--$\mathrm{Pb}$\Xspace}
\newcommand{\ArAr}        {$\mathrm{Ar}$--$\mathrm{Ar}$\Xspace}
\newcommand{\XeXe}        {$\mathrm{Xe}$--$\mathrm{Xe}$\Xspace}
\newcommand{\KrKr}        {$\mathrm{Kr}$--$\mathrm{Kr}$\Xspace}
\newcommand{\AuAu}        {$\mathrm{Au}$--$\mathrm{Au}$\Xspace}
\newcommand{\CuCu}        {$\mathrm{Cu}$--$\mathrm{Cu}$\Xspace}
\newcommand{\pAu}         {$\mathrm{p}$--$\mathrm{Au}$\Xspace}
\newcommand{\dAu}         {$\mathrm{d}$--$\mathrm{Au}$\Xspace}
\newcommand{\HeAu}         {$\mathrm{^{3}He}$--$\mathrm{Au}$\Xspace}

\newcommand{\sigmaBFPP}   {\ensuremath{\sigma_\mathrm{BFPP}}\Xspace}

\newcommand{\lsim}        {\,{\buildrel < \over {_\sim}}\,}
\newcommand{\gsim}        {\,{\buildrel > \over {_\sim}}\,}
\newcommand{\co}[1]       {\relax}
\newcommand{\nl}          {\newline}
\newcommand{\el}          {\\\hline\\[-0.4cm]}

\newcommand{\gmom}{\ensuremath{\mathrm{GeV}\kern-0.05em/\kern-0.02em c}}
\newcommand{\antip}{\ensuremath{\overline{\mathrm{p}}}}
\newcommand{\antid}{\ensuremath{\overline{\mathrm{d}}}}
\newcommand{\tritium}{\ensuremath{{}^{3}\mathrm{H}}}
\newcommand{\antitritium}{\ensuremath{{}^{3}\overline{\mathrm{\mathrm{He}}}}}
\newcommand{\hethree}{\ensuremath{{}^{3}\mathrm{He}}}
\newcommand{\hefour}{\ensuremath{{}^{4}\mathrm{He}}}
\newcommand{\antihethree}{\ensuremath{{}^{3}\overline{\mathrm{He}}}}
\newcommand{\antihefour}{\ensuremath{{}^{4}\overline{\mathrm{He}}}}
\newcommand{\hyp}        {\ensuremath{^{3}_{\Lambda}\mathrm{H}}}
\newcommand{\antihyp}{\ensuremath{^{3}_{\bar{\Lambda}}\overline{\mathrm{H}}}}
\newcommand{\hypfour}    {\ensuremath{^{4}_{\Lambda}\mathrm{H}}}
\newcommand{\antihypfour}{\ensuremath{^{4}_{\bar{\Lambda}}\overline{\mathrm{H}}}}
\newcommand{\hyphefour}    {\ensuremath{^{4}_{\Lambda}\mathrm{He}}}
\newcommand{\antihehypfour}{\ensuremath{^{4}_{\bar{\Lambda}}\overline{\mathrm{He}}}}
\newcommand{\sigmahyp}     {\ensuremath{^{3}_{\Sigma^{0}}\mathrm{H}}}
\newcommand{\antisigmahyp} {\ensuremath{^{3}_{\bar{\Sigma}^{0}}\overline{\mathrm{H}}}}

\newcommand{\sla}{\slash \hspace{-0.2cm}}
\newcommand{\slam}{\slash \hspace{-0.25cm}}
\newcommand{\no}{\nonumber}
\def\lsim{\mathrel{\rlap{\lower4pt\hbox{\hskip1pt$\sim$}}
    \raise1pt\hbox{$<$}}}         
\def\gsim{\mathrel{\rlap{\lower4pt\hbox{\hskip1pt$\sim$}}
    \raise1pt\hbox{$>$}}}         

\newcommand{\Anucl}{$\mathrm{A}$\Xspace}
\newcommand{\nbInv}{$\mathrm{nb}^{-1}$\Xspace}
\newcommand{\pbInv}{$\mathrm{pb}^{-1}$\Xspace}
\newcommand{\isospin}{$I$\Xspace}
\newcommand{\spinJ}{$J$\Xspace}
\newcommand{\BA}{$B_{\mathrm{A}}$\Xspace}
\newcommand{\significance}{$\frac{\mathrm{S}}{\sqrt{\mathrm{S}+\mathrm{B}}}$\Xspace}
\newcommand{\Tchem}{\ensuremath{T_{\mathrm{chem}}}\Xspace}

\def\Bs{{\overline{B}}_s}
\def\R{\mathcal{R}}
\newcommand{\gev}{\mathrm{GeV}}
\newcommand{\tev}{\mathrm{TeV}}
\newcommand{\mev}{\mathrm{MeV}}
\newcommand{\e}{\epsilon}
\newcommand{\tce}{\frac{t_{\rm cool}(\e)}{t_{\rm esc}(\e)}}
\newcommand{\tcer}{\frac{t_{\rm c}(\R)}{t_{\rm esc}(\R)}}
\def\Xe{X_{\rm esc}}
\def\X{X_{\rm esc}}
\def\te{t_{\rm esc}}
\def\tc{t_{\rm cool}}
\def\nb{n_{\rm B}}
\def\nc{n_{\rm C}}
\def\ni{n_{i}}
\def\rism{\rho_{\rm ISM}}
\def\nism{n_{\rm ISM}}
\def\x{(\R,\vec r,t)}
\def\xo{(\R,\vec r_\odot,t_\odot)}
\def\ap{\overline{\rm p}}
\def\ad{\overline{\rm d}}
\def\ep{e^+}
\def\Qep{Q_{e^+}}
\def\epm{$e^\pm$\ }
\def\ah{\overline{\rm ^3He}}
\def\at{\overline{\rm t}}
\def\s{$(*)$}
\newcommand{\dd}{\text{d}}
\newcommand{\gaga}{\gamma\gamma}
\newcommand{\Rp}{\mathcal{R}^\prime}
\newcommand{\Lp}{L^{\prime}}

\newcommand{\Dsc}{\ensuremath{D_{\rm s}}\Xspace}
\newcommand{\twopiTDsc}{\ensuremath{2 \pi T D_{\rm s}}\Xspace}
\newcommand{\ToverTc}{\ensuremath{T/T_{\rm c}}\Xspace}
\newcommand{\Tc}{\ensuremath{T_{\rm c}}\Xspace}
\newcommand{\chisquared}{\ensuremath{\chi^{\rm 2}}\Xspace}

\newcommand{\RunsThreeFour}{Runs~3 \& 4\Xspace}

\newcommand{\ttbar}{\ensuremath{t\overline{t}}\Xspace}



\newcommand{\qty}[2]{\ensuremath{#1\,\mathrm{#2}}}  
\newcommand{\enum}[2]{\ensuremath{#1\times10^{#2}}} 
\newcommand{\NQTY}[2]{\mbox{$[#1/{\rm #2}]$}}     
\newcommand{\UQTY}[2]{\ensuremath{#1/\mathrm{#2}}}  
\newcommand{\eqty}[3]{\qty{\enum{#1}{#2}}{#3}}  
\newcommand{\invnb}{\mathrm{nb}^{-1}}
\newcommand{\invpb}{\mathrm{pb}^{-1}}

\newcommand{\elumi}[2]{\qty{\enum{#1}{#2}}{cm^{-2}s^{-1}}}
\newcommand{\murad}[1]{\qty{#1}{\mu rad}}
\newcommand{\intlumimub}[1]{\qty{#1}{\mu b^{-1}}}

\newcommand{\yNN}{\ensuremath{y_{\mbox{\tiny NN}}}}
\newcommand{\bstar}{\ensuremath{\beta^{*}}}
\newcommand{\emittn}{\ensuremath{\varepsilon_n}}
\newcommand{\LAA}{\ensuremath{L_\text{AA}}}
\newcommand{\LpA}{\ensuremath{L_{pA}}}
\newcommand{\Lpp}{\ensuremath{L_{pp}}}
\newcommand{\Lpeak}{\ensuremath{\hat{L}}}
\newcommand{\LNN}{\ensuremath{ L_{\text{NN}}}}

\newcommand{\isotope}[3]{\ensuremath{^{#1}\mathrm{#2}^{#3}}}

\newcommand{\speciesheader}{ &
\isotope{16}{O}{8+}&
\isotope{40}{Ar}{18+}&
\isotope{40}{Ca}{20+}&
\isotope{78}{Kr}{36+}&
\isotope{129}{Xe}{54+}&
\isotope{208}{Pb}{82+}
}

\newcommand{\bfunc}{$\beta$-function}
\newcommand{\bstarval}[1]{$\bstar = #1\,\mbox{m}$}
\newcommand{\betarel}{\ensuremath{\beta_\text{rel}}}
\newcommand{\emittnx}{\ensuremath{\epsilon_{n,x}}}
\newcommand{\emittny}{\ensuremath{\epsilon_{n,y}}}
\newcommand{\emittnxy}{\ensuremath{\epsilon_{n,xy}}}
\newcommand{\emitts}{\ensuremath{\epsilon_s}}
\newcommand{\sigs}{\ensuremath{\sigma_s}}
\newcommand{\sigp}{\ensuremath{\sigma_p}}
\newcommand{\kb}{\ensuremath{k_b}}
\newcommand{\frev}{\ensuremath{f_0}}
\newcommand{\Nb}{\ensuremath{N_b}}
\newcommand{\Eb}{\ensuremath{E_b}}
\newcommand{\emittval}[1]{\ensuremath{\emittn=\qty{#1}{\mu m\,rad}}}
\newcommand{\Nbval}[2]{\ensuremath{\Nb=\enum{#1}{#2}}}
\newcommand{\taul}{\ensuremath{\tau_l}}
\newcommand{\taulval}[1]{\ensuremath{\taul=\qty{#1}{ns}}}
\newcommand{\sigzval}[1]{\ensuremath{\sigz=\qty{#1}{cm}}}
\newcommand{\etev}[1]{\ensuremath{\Eb=\qty{#1}{TeV}}}
\newcommand{\VRF}{\ensuremath{V_{\mathrm{RF}}}}
\newcommand{\lumival}[2]{\ensuremath{L=\qty{#1\times 10^{#2}}{cm^{-2} s^{-1}}}}

\newcommand{\aibsx}{\ensuremath{\alpha_{\mathrm{IBS},x}}}
\newcommand{\aibsy}{\ensuremath{\alpha_{\mathrm{IBS},y}}}
\newcommand{\aibsxy}{\ensuremath{\alpha_{\mathrm{IBS},x,y}}}
\newcommand{\aradd}{\ensuremath{\alpha_{\mathrm{rad}}}}
\newcommand{\aradds}{\ensuremath{\alpha_{\mathrm{rad},s}}}
\newcommand{\araddx}{\ensuremath{\alpha_{\mathrm{rad},x}}}
\newcommand{\araddy}{\ensuremath{\alpha_{\mathrm{rad},y}}}
\newcommand{\araddxy}{\ensuremath{\alpha_{\mathrm{rad},x,y}}}
\newcommand{\Z}{\ensuremath{Z_\text{ion}}}
\newcommand{\A}{\ensuremath{A_\text{ion}}}
\newcommand{\Circ}{\ensuremath{C_\text{ring}}}
\newcommand{\lumi}{\ensuremath{\mathcal{L}}}
\newcommand{\Lb}{\ensuremath{\mathcal{L}_b}}
\newcommand{\Lint}{\ensuremath{L_{\text{int}}}}
\newcommand{\Lbint}{\ensuremath{L_{b,\text{int}}}}
\newcommand{\Lbpeak}{\ensuremath{\mathcal{L}_{b,\text{peak}}}}


\newcommand{\mlna}{\langle \ln\!A \rangle}
\newcommand{\nmu}{N_\mu}
\newcommand{\lnnmu}{\ln\!\nmu}
\newcommand{\xmax}{X_\text{max}}
\newcommand{\nmult}{N_\text{mult}}
\newcommand{\tocite}{{\bf REF}}
\newcommand{\si}[1]{\ensuremath{\text{#1}}}
\newcommand{\SI}[2]{\ensuremath{#1\,\si{#2}}}

\newcommand{\Ntrig}        {\ensuremath{N_{\mathrm{trig}}}}
\newcommand{\Nassoc}       {\ensuremath{N_{\mathrm{assoc}}}}
\newcommand{\dNassoc}      {\ensuremath{\frac{\dd^2N_{\mathrm{assoc}}}{\dd\Deta\dd\Dphi}}}
\newcommand{\Dphi}         {\ensuremath{\Delta\varphi}}
\newcommand{\Deta}         {\ensuremath{\Delta\eta}}

\newcommand{\pbar}{\ensuremath{\overline{p}}\xspace}
\newcommand{\p}{\ensuremath{p}\xspace}
\newcommand{\nbar}{\ensuremath{\overline{n}}}
\newcommand{\dbar}{\ensuremath{\overline{d}}}
\newcommand{\pim}{\ensuremath{\pi^-}\xspace}
\newcommand{\pip}{\ensuremath{\pi^+}\xspace}
\newcommand{\km}{\ensuremath{K^-}\xspace}
\newcommand{\kp}{\ensuremath{K^+}\xspace}
\newcommand{\hm}{\ensuremath{h^-}\xspace}
\newcommand{\pt}{\ensuremath{p_{\rm T}}\xspace}
\newcommand{\pl}{\ensuremath{p_{\rm L}}\xspace}

\newcommand{\GeVc}{\ensuremath{\mbox{~Ge\kern-0.1em V}\!/\!c}\xspace}
\newcommand{\AGeVc}{\ensuremath{A\,\mbox{Ge\kern-0.1em V}\!/\!c}\xspace}
\newcommand{\AGeV}{\ensuremath{A\,\mbox{Ge\kern-0.1em V}}\xspace}

\title*{QGP@50: More than Four Decades of Jet Quenching}

\author{Xin-Nian Wang, Urs Achim Wiedemann}

\institute
{
Xin-Nian Wang \at Central China Center for Nuclear Theory and Institute of Particle Physics, Central China Normal University, Wuhan, China 430079 and 
Institut f\"{u}r Theoretische Physik, Johann Wolfgang Goethe–Universit\"{a}t, 
Max-von-Laue-Str.~1, D-60438 Frankfurt am Main, Germany;\\
\email{xnwang@ccnu.edu.cn} \and
Urs Achim Wiedemann \at CERN, 1211 Geneva, Switzerland;
\email{urs.wiedemann@cern.ch}
}

\maketitle

\abstract{
How are high-momentum transfer processes modified when embedded in the Quark-Gluon Plasma (QGP) instead of the vacuum? How can fundamental properties of the QGP be inferred from their medium-modifications? And what can be learnt about QCD? These questions have motivated theoretical and experimental studies for more than four decades almost since the beginning of QGP research. Here we review with a historical perspective the main theoretical developments and the resulting interplay of theory and experiment at RHIC and at the LHC. }






\epigraph{
It is not our purpose here to discuss the evolution of the plasma per se, but to study only one consequence of this picture. This has to do with the influence of this phenomenon on hard-collision processes. From the space-time picture of the collision, we may infer that the two initial- state partons are, because of causality, unaffected by the plasma which is subsequently produced. However, the secondary high-$p_T$ quarks or gluons are affected.}{J.D. Bjorken, 1982}

\section{Introduction}

Jet quenching refers to the suppression and modification of energetic jets as they traverse the quark-gluon plasma (QGP), resulting from multiple interactions between the jet partons and the strongly interacting matter. These interactions lead to energy loss and transverse momentum broadening of the jet partons, and modification of the observed jet and hadron spectra in the final state. Since the magnitude of the energy loss is closely related to the strength of jet-medium interaction and the density of the medium along the propagation path inside the QGP, jet quenching has become a powerful tool to study properties of QGP in high-energy heavy-ion collisions.

The production of hard jets in hadronic collisions can be calculated using perturbative QCD (pQCD) due to the large momentum transfer involved in the initial scattering processes. In proton-proton (p+p) collisions, pQCD calculations based on collinear factorization can describe the experimental measurements 
well  and with controlled accuracy. The calculations can be extended to proton-nucleus (p+A) collisions by including nuclear modifications to parton distribution functions inside nuclei. These calculations serve as an essential baseline for initial jet production, against which modifications due to jet quenching can be identified.

The theory of jet quenching has been reviewed multiple times at different stages of its development ~\cite{Baier:2000mf,Gyulassy:2003mc,Kovner:2003zj,Casalderrey-Solana:2007knd,Wiedemann:2009sh,Blaizot:2015lma,Qin:2015srf,Cao:2020wlm,Cao:2024pxc}. Measurements of jet quenching phenomena are summarized in the RHIC white papers~\cite{PHENIX:2004vcz,STAR:2005gfr,PHOBOS:2004zne,BRAHMS:2004adc}, in recent overviews of the ALICE~\cite{ALICE:2022wpn} and CMS~\cite{CMS:2024krd} collaborations, and in other comprehensive reviews~\cite{dEnterria:2009xfs,Majumder:2010qh,Connors:2017ptx,Cunqueiro:2021wls}. In addition, there is a number of working group reports aimed at clarifying and unifying the community-wide understanding of important aspects of the jet quenching formalisms and their phenomenological applicability~\cite{Accardi:2004gp,Armesto:2011ht,Andrews:2018jcm}. 

The present article does not aim at providing yet another update to this long list of excellent summaries nor having a comprehensive discussion of an exhaustive list of topics in jet quenching. While many of existing reviews aimed primarily at documenting the state of the art at the time of their appearance, here we put more weight on the historical and major developments of jet quenching throughout the decades.

\section{Bjorken’s phenomenological prelude}
Already in 1974, almost immediately after the discovery of the asymptotic freedom of Quantum Chromodynamics (QCD)~\cite{Gross:1973id,Politzer:1973fx},  T.~D.~Lee had suggested~\cite{Lee:1974kn} that phase transitions of fundamental quantum fields could be studied by colliding heavy nuclei. Around the same time, it had been concluded \cite{Collins:1974ky,Cabibbo:1975ig} that the high-temperature phase of QCD is the deconfined, chirally symmetric Quark Gluon Plasma (QGP), and one can produce such matter in high-energy collisions\cite{Shuryak:1978ij}. And yet, it still took almost a decade to move from these profound concepts to formulating the phenomenological strategies of how QGP physics can be studied via heavy-ion collisions. In modern heavy-ion phenomenology, there are essentially two strategies for studying QGP properties.
Either one studies how the {\it bulk} evolution of the plasma depends on QGP properties, or one asks how specific particle production processes embedded in the QGP can be used to {\it probe} its properties. The advancement of both of these modern phenomenological strategies can be attributed to two preprints of Bjorken in the summer of 1982~\footnote{ A few years earlier, Shuryak had formulated an encompassing phenomenological strategy for QGP studies, including electro-magnetic probes but not mentioning hard probes such as jets~\cite{Shuryak:1978ij}.}.
    \begin{figure}[t]
    \centering
    \includegraphics[width=0.3\textwidth]{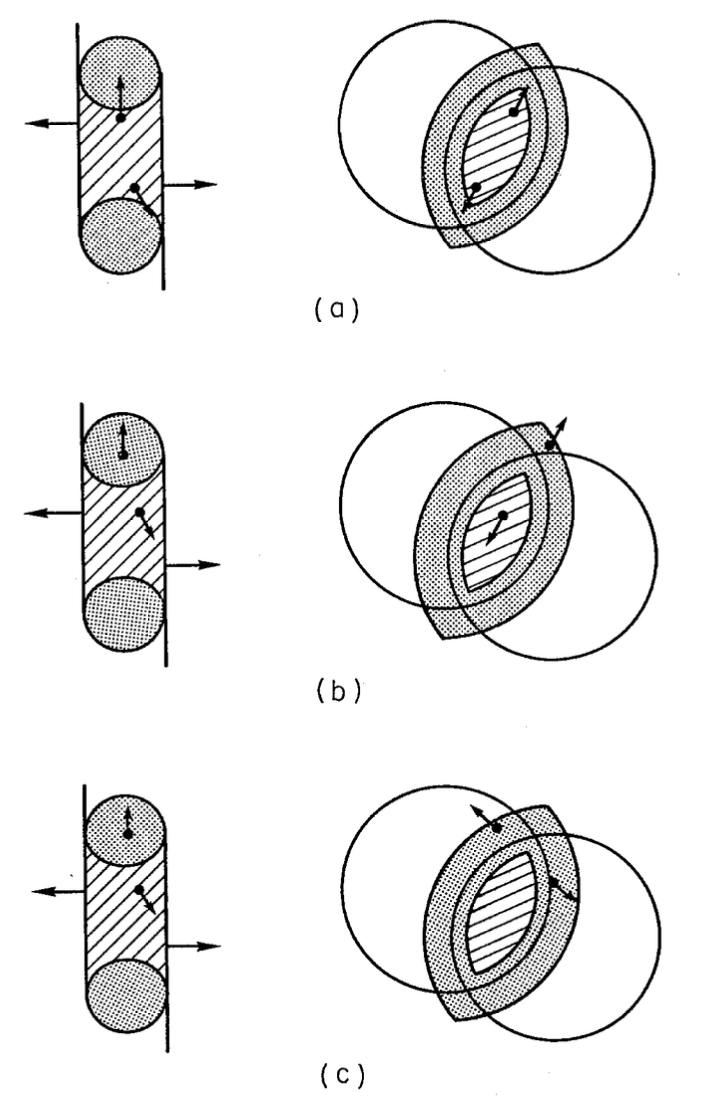}
    \hskip1.5cm
     \includegraphics[width=0.55\textwidth]{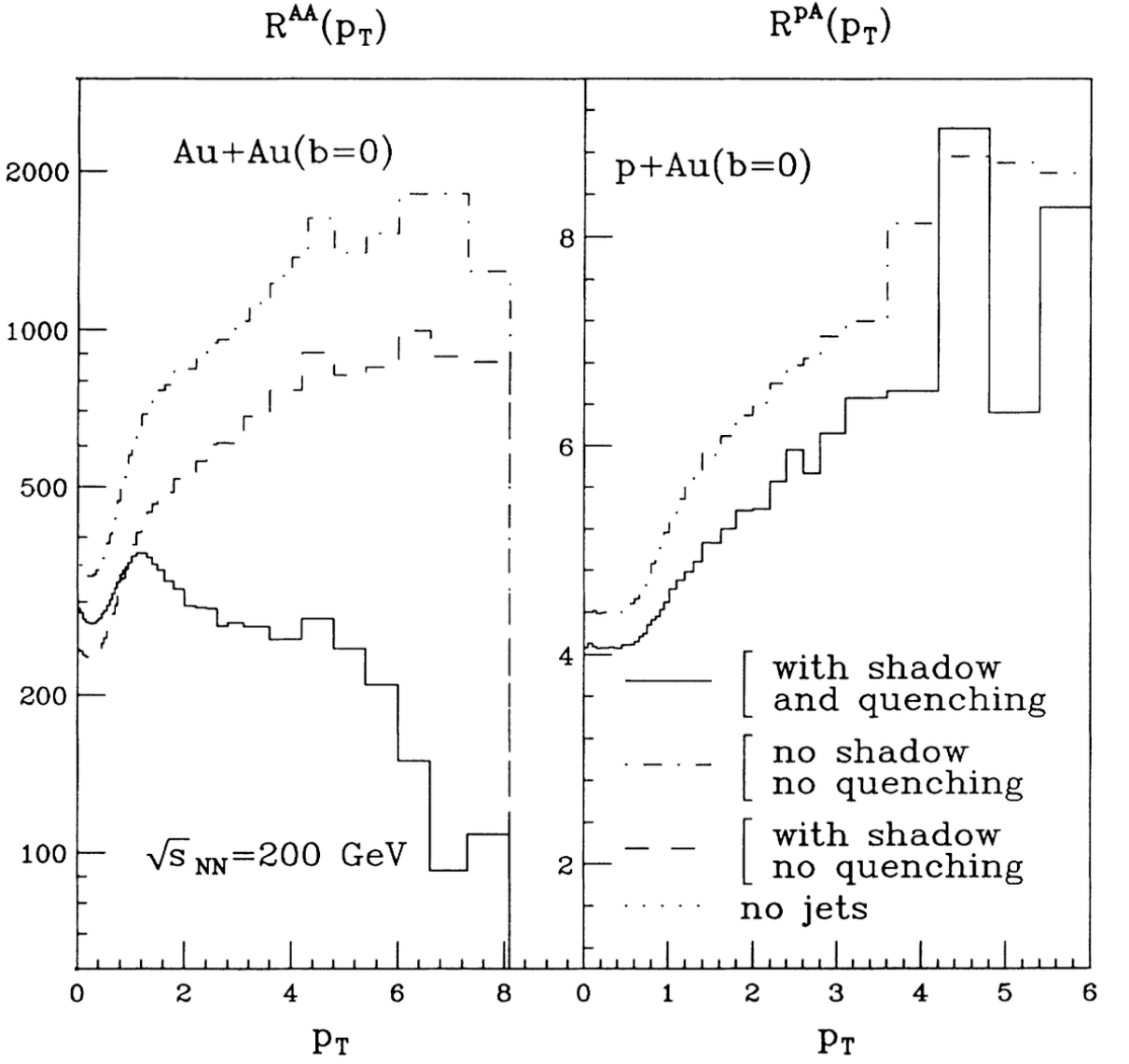}
    \caption{(\textit{Left}): Bjorken's original sketch of the ``fate of secondary high-$p_T$ jets". Depending on how the hard process is embedded in a non-central heavy-ion collision, both (a), one (b) or none (c) of the secondary jets propagates through the medium over a significant distance and is degraded in energy. (\textit{Right}): The first prediction by Gyulassy and Wang of jet quenching  measured via the nuclear modification factor in A+A and p+A collisions. [Different from today's practice, these nuclear modification factors were not divided by the nuclear overlap function, as first proposed by Wang \cite{Wang:1998bha} in 1998, and are thus not normalized to unity.] Right Figure adopted from Ref.~\cite{Wang:1992qdg}. }
    \label{fig1}
    \end{figure}    

In his July 1982 preprint, Bjorken reformulated Landau's hydrodynamical model in terms of boost-invariant initial conditions~\cite{Bjorken:1982qr},  introducing a central idea of the phenomenological description of the bulk evolution of the plasma in high-energy heavy-ion collisions. His estimate of the initial energy density provided an early indication that QGP physics was indeed within reach of relativistic heavy-ion collisions. The manuscript also mentioned in passing  electromagnetic photon and dilepton radiation, as well as an enhanced $K/\pi$ ratio as possible QGP signatures. Just one month later, Bjorken's August 1982 preprint~\cite{Bjorken:1982tu} considered for the very first time the physics of high-momentum transfer processes embedded in the QGP.
Due to the Lorentz-contraction in ultra-relativistic collisions, most partons in the incoming nuclei do not have time to interact before a high-momentum-transfer process occurs. Since the hard process happens before plasma formation, it is -- by causality -- unaffected by the plasma. Bjorken concluded that only the outgoing secondary high-$p_T$ partons can sense the QGP. This marks the birth of jet quenching phenomenology. In particular, if the secondary partons were to propagate through vacuum, their transverse momenta would balance\footnote{This is true only in leading order pQCD. At the next-to-leading order, radiative corrections lead to acoplanarity of two secondary partons.}. However, in the QGP, differing in-medium path lengths lead to unequal parton energy loss, as illustrated by Bjorken in Fig.~\ref{fig1}(left panel). In this context, Bjorken predicted a monojet phenomenon ``in which the hard collision occurs near the edge of the overlap region, with one jet escaping without absorption and the other fully absorbed." This is indeed one of the most striking of the many jet quenching signatures observed at RHIC~\cite{STAR:2002svs} and the LHC~\cite{ATLAS:2010isq,CMS:2011iwn}. 

Today's usual formulation of high-momentum transfer processes in hadronic collisions uses collinear factorized QCD, which expresses the process as a convolution of parton distributions, the hard partonic cross section and fragmentation functions for the outgoing secondary partons. Within this framework,  Bjorken's conclusion may be summarized schematically as\footnote{To be specific, we write this equation for single inclusive hadron production in nucleus-nucleus (A+A) collisions. Similar expressions could be written for other high-$p_T$ production processes.}
\begin{equation}
    d\sigma_{\rm med}^{AA\to h+X} = \sum_{i,j,k,f} \underbrace{
        f_{i/A}\otimes f_{j/A} \otimes \hat{\sigma}_{ij\to f+k} }_{\hbox{indep. of QGP-effects}}
        \otimes \underbrace{D_{f\to h}}_{\hbox{QGP-dependent}}\, ,
        \label{eq1}
\end{equation}
where $f_{i/A}$ is the parton distribution inside a nucleus, $\hat{\sigma}_{ij\to f+k}$ is the pQCD cross section of $i+j\to f+k$ processes and $D_{f\to h}$ is the fragmentation function.
Around the time of Bjorken's preprint, particle physicists had already begun to formalize the scale separation between short- and long-distance physics in QCD factorization theorems which form the basis of expressions like Eq.~\eqref{eq1}. The scale evolution of the parton distributions and fragmentation functions
were established a few years earlier in 1977 by Dokshitzer, Gribov, Lipatov, Altarelli and Parisi (DGLAP) \cite{Dokshitzer:1977sg,Gribov:1972ri,Altarelli:1977zs}. One might wonder how the field would have evolved differently, if the full implications of Eq.~\eqref{eq1} and the DGLAP evolution equations had been understood at the time.  Namely, it was already known that the DGLAP evolution of fragmentation functions like $D_{f\to h}$ is dictated by perturbative $1\to 2$ parton splittings and one might have therefore conjectured already then that the QGP affects predominantly these final state $1\to 2$ processes, i.e., that the dominant or equally important effect could be inelastic, resulting in medium-modified splitting functions for the evolution of parton fragmentation functions in the QGP medium. But that's not how the field evolved historically. Instead, Bjorken at that time considered only the possibility that elastic small-angle scattering degrades the energy of secondary partons. Initially, he estimated collisional energy losses of up to 10 GeV -- an overestimate due to a numerical error that prompted him to sketch the general characteristic experimental signatures of jet quenching. When it was realized that these estimates were numerically flawed, the idea went dormant and he never published his seminal preprint. After that prelude, it took another decade and the work of many researchers to establish QCD-based calculations of
the dominant medium-modification of high-$p_T$ partons propagating through the QGP.

\section{Parton energy loss and jet quenching - theory }

Gyulassy, Pl\"umer and Wang were the first to emphasize the possibly dominant role of radiative processes in jet quenching\cite{Gyulassy:1990ye,Wang:1990bka}. Exploiting the new opportunities arising with the dedicated Monte Carlo event generator HIJING~\cite{Wang:1991hta} under development at that time for jet-medium interaction in relativistic heavy-ion collisions, Gyulassy and Wang had demonstrated in model studies as early as 1990 the suppression of single inclusive high-$p_T$ hadron spectra~\cite{Wang:1990bka}. Emphasizing the importance of establishing jet quenching on top of a well-controlled baseline, they had presented the first nuclear modification factors~\cite{Wang:1992qdg} for nucleus-nucleus and proton-nucleus collisions, see Fig.\ref{fig1} (right panel).  Around the same time, Braaten and Thoma had put the calculation of collisional parton energy loss on a more rigorous footing within QCD finite temperature perturbation theory~\cite{Thoma:1990fm,Braaten:1991jj,Braaten:1991we}. This led to a community-wide consensus that radiative contributions to parton energy loss may dominate in the limit of very high parton energy~\cite{Gyulassy:1991xb}. We use the word ``may" in the last sentence, since at the time of these early conclusions, today's framework of radiative energy loss was not yet developed and major parametric dependencies of parton energy loss were not yet understood. 

In this section, we discuss the main elements of the formulation of radiative energy loss that was advanced throughout the 1990s. These early calculations are limited to a close-to-eikonal approximation in which the energy $E$ of the fragmenting parent parton is much larger than the energy $\omega$ of the emitted gluon which is also much larger than the transverse momentum ${\bf k}$ of that gluon and the transverse momentum ${\bf q}$ transferred from the medium,
\begin{equation}
    E \gg \omega \gg {\bf k}, {\bf q}\, \quad \hbox{(close-to-eikonal approximation)}.
    \label{closetoeikonal}
\end{equation}
We note that this condition is more stringent than the collinear limit $E,\omega \gg {\bf k}, {\bf q}$, and that it therefore can give access only to splitting functions in which one of the daughters is soft. Our presentation will expose both the insights gained from this approach, its limitations and efforts to go beyond them.

\subsection{Jet-medium interactions: What is it on which the partons in a quenched jet scatter? }
\label{weakjet}

Any answer to the question asked in the above title will deliver on the main promise of jet quenching as a {\it hard probe}, namely, how the embedding of a {\it hard} process into a QCD medium turns that process into a {\it probe} of the material properties of that medium. However, the sensitivity to material properties depends on the kinematic limits within which a probe is studied. So what is it that a highly energetic parton in close-to-eikonal approximation is sensitive to? Multiple explicit calculations in different theoretical set-ups support the answer: such probes are described in a recoilless approximation and their medium-modification is sensitive to the average transverse color field strength that the medium exerts on the partonic projectile per unit path length. Here, we list several lines of arguments that support this statement and that bring to the forefront key elements of the underlying theory:
\begin{enumerate}
    \item {\it Jet-medium interactions in the Gyulassy-Wang model}\\
With the solid calculations of elastic energy loss within the finite temperature field theory \cite{Thoma:1990fm, Braaten:1991jj,Braaten:1991we} in the early 1990's, it became clear that radiative parton energy loss may be the dominant mechanism for jet quenching. In the first study of radiative energy loss induced by multiple parton-medium interactions \cite{Gyulassy:1993hr,Wang:1994fx},  Gyulassy and Wang (GW) assumed a model of the medium as a collection of scattering centers of static Debye screened potential 
 \begin{equation}
     V_i^a({\bf q})=gT^a_i\frac{e^{-i{\bf q}\cdot{\bf x}_i}}{{\bf q}^2+\mu_D^2},
     \label{Debye-potential}
 \end{equation}
with the elastic scattering cross section,
\begin{equation}
    \frac{d\sigma^R_{\rm el}(T)}{d{\bf q}^2}=\frac{C_2(R)C_2(T)}{N_c^2-1}\frac{4\pi\alpha_s^2}{({\bf q}^2+\mu_D^2)^2},
    \label{elscat}
\end{equation}
where $\mu_D$ is the Debye screening mass,  $C_2(R)$ and $C_2(T)$ are the quadratic Casimirs of the jet and medium parton that generates the potential, respectively.  The ensemble average of the static potential gives the transverse density of the scattering centers $N/S_\perp$ which provides the opacity parameter $n=\sigma_{\rm el}N/S_\perp$ in the framework of opacity expansion  by Gyulassy, Levai and Vitev (GLV)\cite{Gyulassy:2000fs,Gyulassy:2000er} and Wiedemann \cite{Wiedemann:2000za}. 

Though the original GW model used Coulomb gauge, one can also use other gauges such as the covariant gauge in which only the $A^+$ component of the potential is important. In this case, the differential elastic cross section can be formulated as \cite{Zhang:2019toi},
\begin{equation}
    \frac{d\sigma^R_{\rm el}(T)}{d{\bf q}^2}=\frac{4\pi\alpha_sC_2(R)}{N_c^2-1}\frac{\phi_T(0,{\bf q})}{4\pi{\bf q}^2}\, ,
    \label{eq:tmdpdf}
\end{equation}
where $\phi_T(x,{\bf q})$ is the transverse momentum dependent (TMD) gluon distribution of the medium parton that generates the potential,
\begin{equation}
    \phi_T(x,{\bf q})\equiv \int \frac{dz^-}{2\pi p^+}d^2{\bf x}
    e^{-ixp^+z^- + {\bf q}\cdot{\bf x}} \langle\langle F_i^+(z^-,{\bf x}) F^{+i}(0,{\bf 0})\rangle\rangle_T\, .
    \label{eq:tmd2}
\end{equation}
 One should note that the issue of gauge-invariance is subtle: radiating a gluon from a partonic $2\rightarrow 2$ elastic scattering amplitude involves five emission amplitudes whose sum is gauge-invariant, of course. But when one replaces the ``target" parton in these radiation amplitudes by an idealized static Debye-screened potential in Eq.~\eqref{Debye-potential}, gauge-invariance is lost. The radiation off the target legs needs to be kept for a gauge-invariant formulation. As clarified in Ref.~\cite{Wang:1994fx}, if gluon radiation is formulated in the target rest frame (in which the energy $E$ of the projectile parton is much larger than that of the target parton) and if one works in the close-to-eikonal approximation [Eq.~\eqref{closetoeikonal}], then the radiative contribution from emission off the target legs is suppressed by $O(1/E^2)$. In short, potential scattering in QCD cannot be formulated in a gauge-invariant way but gauge-dependence arises only at the subleading power in projectile energy, and the close-to-eikonal high energy limit can still be formulated by idealizing the colored constituents of the QGP in terms of Eq.~\eqref{Debye-potential}. 

Any collisional energy loss (i.e., energy loss via elastic scattering) enters at the subleading order in projectile energy $E$. The insight~\cite{Wang:1994fx} of how QCD potential scattering allows for a gauge-invariant formulation of radiative phenomena is thus a double-sided sword: On the one hand, it allows for an arguably simple and explicit formulation of radiative parton energy loss that continues to form the basis of jet quenching phenomenology three decades after its initial formulation. On the other hand, it complicates formulating radiative and collisional mechanisms in the same computational set-up. The notion of recoilless-ness is hard-wired in the formulation and difficult to overcome. 

    \item {\it Dirac equation in a spatially extended color field: eikonal limit.}\\
    There are several formulations of the generic physics idea that a highly energetic parton exchanges color and momentum with the QGP while transversing it. Upon closer inspection, they all reveal limitations very similar to those of the GW model. For instance, one may formulate the in-medium propagation of a parton by solving the Dirac equation for the parton wave function $\psi_\alpha$ in the spatially extended color field $A_\mu$ provided by the medium 
    \begin{equation}
        \gamma^\mu \left(\partial_\mu + iA_\mu \right) \psi_\alpha = 0\, .
        \label{eq3}
    \end{equation}
    In the eikonal limit ($E\to \infty$), 
    the solution is given in terms of an eikonal Wilson line that in the light-cone gauge $A^-=0$ reads
    \begin{equation}
        W\left[{\bf x}; z_i, z_f \right] = {\cal P} \exp\left[i \int_{z_i}^{z_f} dz^-\, A^+({\bf x},z^-) \right]\, .
        \label{eq4}
    \end{equation}
    Here, the propagation path of the parton is parametrized by the light-cone time $z^-$, and $A_\mu = T^a\, A_\mu^a$ is a matrix in color space specified by the generators $T^a$ of the $SU(3)_{\rm color}$-representation of the projectile. The parton thus propagates through the medium by rotating in color space from initial color $\alpha$ to final color $\beta$ without changing its position ${\bf x}$  transverse to the initial direction of motion, $\psi_\beta({\bf x}) = W_{\beta\alpha} ({\bf x}) \psi_\alpha({\bf x})$. For the projectile, the target appears Lorentz-contracted to $\delta(z^-)$ along the light-cone time.

    Since radiative phenomena take finite time and can resolve distance-scales within the medium, one should include in parton energy loss formulations the order $O(1/E)$ corrections to the phase of Eq.~\eqref{eq4}. To this desired accuracy, 
    the Green's function
    \begin{eqnarray}
        &&G\left[{\bf x}_f,z_f; {\bf x}_i,z_i \| E \right] 
        \nonumber \\
        && = \int_{{\bf r}(z_i)={\bf x}_i}^{{\bf r}(z_f)={\bf x}_f} {\cal D}{\bf r} \exp\left[\frac{i\, E}{2} i \int_{z_i}^{z_f} dz^-\, \dot{\bf r}^2(z^-)  \right] W\left[{\bf r}; z_i, z_f \right]
        \label{eq7}
    \end{eqnarray}
    gives the solution $\psi_\beta({\bf x}_f, z_f) = \int d{\bf x}_i\, 
    G\left[{\bf x}_f,z_f; {\bf x}_i,z_i \| E \right]\, \psi_{\alpha}({\bf x}_i,z_i)$ to the Dirac equation in Eq.~\eqref{eq3}.
    Physically, Eq.~\eqref{eq7} describes the propagation of a parton that rotates in color space while exhibiting Brownian motion in the plane transverse to the propagation direction. This formula is the basis of the path-integral formulation of parton energy loss of Zakharov~\cite{Zakharov:1996fv,Zakharov:1997uu} and Wiedemann~\cite{Wiedemann:2000za}.  
    
    If we parametrize the spatially extended color field in the Dirac equation [Eq.~\eqref{eq3}] in terms of an ensemble of Debye-screened static scattering centers in Eq.~\eqref{Debye-potential}, then multiple scattering as formulated in the GW-model and multiple scattering formulated as a solution to Eq.~\eqref{eq3} are equivalent. In particular, the Green's function in Eq.~\eqref{eq7} formulates a recoilless propagation in which 
    transverse momentum is exchanged with the medium but no fraction of the initial energy $E$ of the parton is transferred to the medium.

    Numerous reviews including Refs~\cite{Baier:2000mf,Kovner:2003zj,Casalderrey-Solana:2007knd,Wiedemann:2009sh} have documented the technical steps and tricks with which the in-medium propagation [Eq.~\eqref{eq7}] of a single parton can be turned into a compact expression for medium-induced gluon radiation. Here, we only recall that in the calculation of such radiative cross sections, medium properties enter as target averages. 
Of particular importance is the two-point correlation function  $\langle\langle A_\mu^a({\bf x})\,  A_\nu^b({\bf y}) \rangle\rangle$ which is typically assumed to be
    \begin{itemize}
        \item local in light-cone time, $\langle\langle A_\mu^a({\bf x})\,  A_\nu^b({\bf y}) \rangle\rangle \propto \delta(x^- - y^-)$, 
        \item translation invariant in the plane transverse to the direction of propagation  $\langle\langle A_\mu^a({\bf x})\,  A_\nu^b({\bf y}) \rangle\rangle \propto f({\bf x}- {\bf y})$. 
    \end{itemize} 
    One can show that the GW model, formulated for a translationally invariant distribution of scattering centers, provides just one particularly simple and explicit ansatz for a correlation function $\langle\langle A_\mu^a({\bf x})\,  A_\nu^b({\bf y}) \rangle\rangle$ that satisfies the above-listed properties and that parametrizes the color field strength of the medium.
    Target averages of pairs of Wilson lines take the form
    \begin{equation}
        \langle\langle {\rm Tr}\left[ W({\bf x})\, W({\bf y})\right]\rangle\rangle_{\rm target}
            = {\cal N} \exp\left[- \frac{1}{2} \int dz^-\, n(z^-)\, \sigma_R({\bf x}-{\bf y}) \right]\, ,
            \label{eq5}
    \end{equation}
   where ${\cal N}$ is the normalization. 
   In the path-integral formulation of parton energy loss, the transverse separation ${\bf r}= {\bf x} - {\bf y}$ depends typically on $z^-$ and the expression becomes a path integral in ${\bf r}(z^-)$. Here, the locality of $\langle\langle A_\mu^a({\bf x})\,  A_\nu^b({\bf y}) \rangle\rangle$ in light-cone time translates in Eq.~\eqref{eq5} into an exponent that factorizes into a longitudinal density of scattering centers $n(z^-)$ and a purely transverse information $\sigma_R({\bf r})$. The transverse translation invariance of $\langle\langle A_\mu^a({\bf x})\,  A_\nu^b({\bf y}) \rangle\rangle$ is reflected in $\sigma_R({\bf r})$
   depending only on the difference ${\bf r}={\bf x}-{\bf y} $. For the GW model, this dipole cross section takes the explicit form of a Fourier transform of a single elastic scattering center,
   \begin{equation}
       \sigma_R({\bf r}) = \int \frac{d{\bf q}}{(2\pi)^2} \frac{d\sigma^R_{\rm el}(T)}{d{\bf q}^2} \left(1 - \exp\left[ i {\bf q}.{\bf r} \right] \right)\, .
       \label{dipolecross}
   \end{equation}
   Early calculations in the field were performed by either expanding the integrand of the cross section in powers of $n\, \sigma$ (which is one way of realizing the opacity expansion), or by making a saddle point approximation of expectation values like Eq.~\eqref{eq5} with the help of 
   \begin{equation}
    n(z^-) \sigma_R({\bf r}) \approx \frac{1}{2} \hat{q}_R(z^-)\, {\bf r}^2\, . 
    \label{dipole}
   \end{equation}
    Here, $\hat{q}_R$\footnote{Throughout this article, we use the subscript $R$ in $\hat q_R$ to denote the color representation of the propagating parton. When no subscript is used, $\hat q$ is by default for a quark in the fundamental representation.}, known as the jet quenching parameter or jet  transport coefficient,  provides a particularly simple one-parameter description of the transverse color field strength seen by a high-energy colored (in representation $R$) parton propagating through the QGP. This jet quenching parameter $\hat{q}_R$ measures the average squared transverse momentum transferred from the QGP to the partonic projectile per unit path-length. The take-home message is that jet quenching in the close-to-eikonal approximation is sensitive to the transverse color field strength of the QGP, and that there are numerous ways of parameterizing this field strength, for instance via the GW model, or via the quenching parameter $\hat{q}_R$, or via the transverse momentum dependent gluon distribution in Eq.~\eqref{eq:tmdpdf}. For a more detailed discussion of such parameterizations, see Ref.\cite{Armesto:2011ht}.
    
    \item {\it Relation to other formulations of eikonal scattering}\\
    Formulations of eikonal scattering have been used repeatedly in the formulation of DIS structure functions and diffractive phenomena at small $x$~\cite{McLerran:1994vd,Buchmuller:1995mr,Kovner:2003zj}. In particular, there are strong technical and physical commonalities between jet quenching and saturation physics, as developed in the 1990s~\cite{McLerran:1993ni,Iancu:2003xm,Kovner:2003zj}. In both cases, energetic partons propagate through a spatially extended color field. In saturation physics, they typically propagate from time $z^- = - \infty$ along the beam direction through the color field of a cold and highly Lorentz-contracted nucleus whose gluon distribution is saturated up to a momentum scale $Q_s$. As in jet quenching, the scattering on this cold gluon field is parametrized by target averages of Wilson lines such as 
    \begin{equation}
        \langle\langle {\rm Tr}\left[ W({\bf x})\, W({\bf y})\right]\rangle\rangle_{\rm target}
            = {\cal N} \exp\left[- \frac{({\bf x}-{\bf y})^2}{4} Q_s^2 \right]\, . 
            \label{eq6}
    \end{equation}  
    Identifying $Q_s^2 \leftrightarrow \int dx^- \hat{q}(z^-)$, one sees that this target average is equivalent to the one given by Eqs.~\eqref{eq5} and \eqref{dipole}. DIS scattering processes in saturation physics are sensitive to essentially the same transverse color field strength as jet quenching in the cold nucleus\cite{Liang:2008vz}. Nevertheless, jet quenching in QGP and cold nucleus or saturation are sensitive to very different physics: in saturation physics, it is the color field strength of a cold saturated gluon distribution, in jet quenching it is the field strength exhibited by a heated collision region that evolves rapidly to equilibrium.
    
    We finally note that there are also remarkable commonalities with some formulations of high-energy electron multiple scattering in QED~\cite{Zakharov:1996fv}. For instance, in electron scattering on amorphous collections of atoms, the target material has been described in terms of a Debye-screened potential akin of the GW model. The physics picture is that dependent on the momentum-transfer from the atoms to the electron projectile, the electron sees the central charge of the nucleus partially screened via the electron shells of the atom. 

\end{enumerate} 

So far, we gave a rather sobering answer to the central question: ``which QGP properties are probed by jet quenching?". We answered that it is the transverse color field strength seen by a highly energetic parton, parametrized in terms of $\hat{q}_R$ or equivalent formulations. We hasten to remark, however, that the answer to this question is much richer if we do not restrict ourselves to probes that are described in the close-to-eikonal approximation. For instance, jet quenching has the sensitivity of resolving physical degrees of freedom in Rutherford-type large-angle scattering experiments~\cite{DEramo:2012uzl,Kurkela:2014tla}. This sensitivity can already be exhibited in the discussion above: in the GW-model, scattering centers in Eq.~ \eqref{elscat} have a perturbative $1/({\bf q}^2)^2$ power-law tail that would show up in hard scattering on a {\it single} Debye-screened potential. But any measure of the squared momentum transfer per unit path-length is sensitive to the second moment $\langle {\bf q}^2\rangle$ of the elastic scattering cross section of the GW model, and the power-law tails would manifest themselves only in a weak logarithmic correction of Eq.~\eqref{dipolecross} on $ {\bf r}$ which is normally not kept, see Eq.~\eqref{dipole}. The physics problem here is that of Molière-scattering, that is the question in which kinematical region in a multiple scattering scenario the effects of multiple soft or single-hard scattering dominate. In yet another very substantial and phenomenologically rich way, jet quenching gives access to QGP transport properties once one understands how much energy a quenched jet injects into the QGP and one becomes sensitive to measuring how that energy is transported in the medium. For instance, if that radiated energy propagates in Mach-like cones~\cite{Stoecker:2004qu,Ruppert:2005uz,Casalderrey-Solana:2004fdk}, the probe is automatically sensitive to properties of the equation of state of the QGP such as sound velocity, that dictate the cone-like emission pattern. In short, there is a rich phenomenology that goes substantially beyond the techniques developed in the 1990s. 

\subsection{The jet quenching parameter $\hat{q}$}
\begin{figure}[t]
 \centering
    \includegraphics[width=0.50\textwidth]{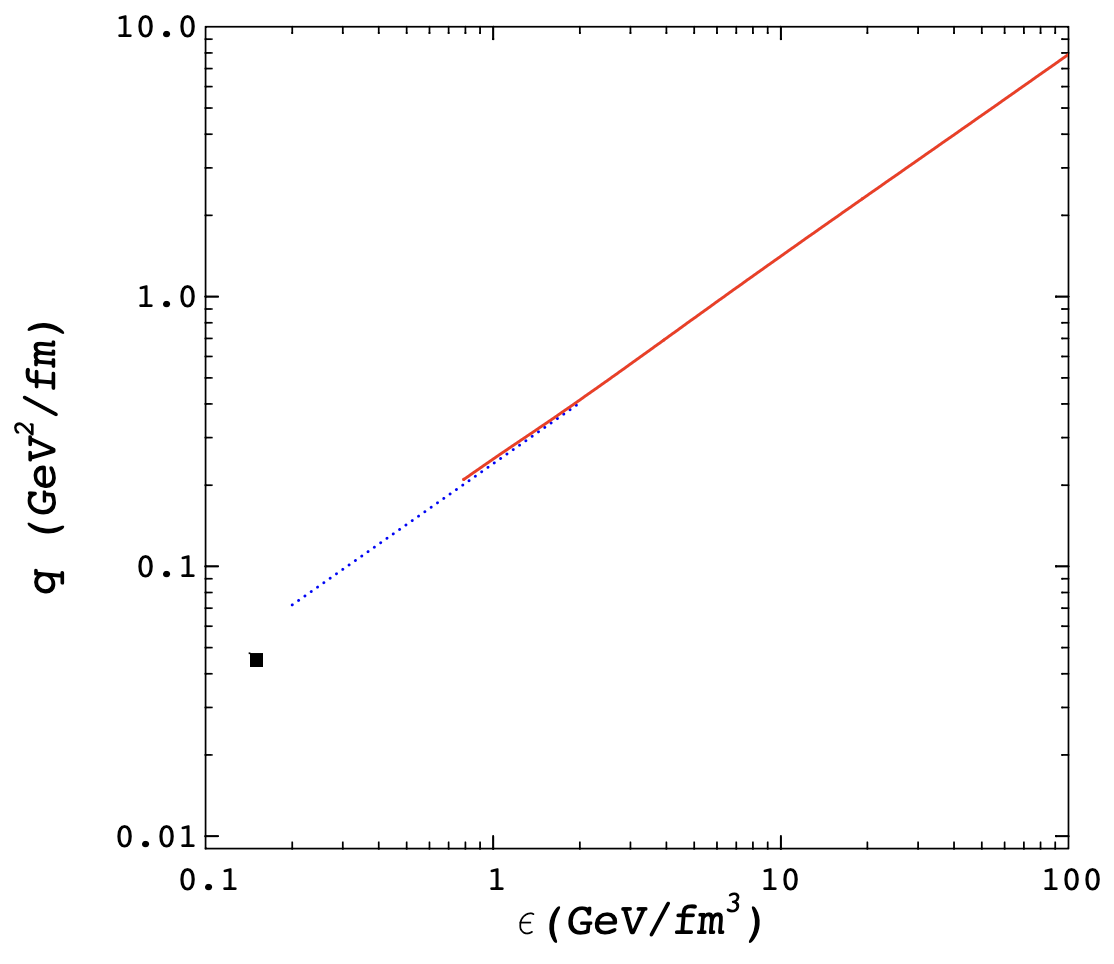}
    \caption{The first estimate of the jet-quenching parameter $\hat{q}$ as a function of energy density $\epsilon$, given by Baier in his plenary talk at Quark Matter 2002~\cite{Baier:2002tc}. Here, blue-dashed and red curves are estimates for a hadron gas and a QGP, respectively, the black point is an estimate for normal nuclear matter density. Numerical values have been revised since then. }
    \label{fig:qhat1}
    \end{figure}

In jet quenching calculations, the production of the initial parton in a hard process factorizes from the properties of the medium the parton interacts with. The jet-medium interaction can be characterized by the jet transport coefficient $\hat q_R$ which describes the squared transverse momentum broadening per unit path length,
\begin{equation}
     \hat q_R=\frac{4\pi \alpha_sC_2(R)}{N_c^2-1}\rho\int \frac{d^2{\bf q}}{(2\pi)^2}\phi_T(0,{\bf q}),
\end{equation}
according to Eq.~(\ref{eq:tmdpdf}) as originally defined by BDMPS \cite{Baier:1996kr}, where $\phi_T(0,{\bf q})$ is the transverse-momentum dependent gluon distribution of the scattering-centers defined in Eq.~(\ref{eq:tmd2}). Shown in Fig.~\ref{fig:qhat1} is the first numerical estimate of $\hat q_F$ by Baier~\cite{Baier:2002tc} as a function of the energy density $\epsilon$. In QGP, $\hat q \propto T^3 \sim \epsilon^{3/4}$ according to a dimensional counting. This jet transport coefficient can also be estimated in the leading order perturbative thermal QCD\cite{Wang:2000uj},
\begin{equation}
    \hat q_R=8\pi C_2(R)\alpha_s^2 T^31.42\ln\frac{3ET}{2\mu_D^2},
\end{equation}
with resummed hot-thermal-loop gluon propagator that includes both chromo-electric and magnetic interaction. In a normal cold nucleus, one can determine $\hat q$ experimentally through the transverse momentum broadening of a quark in deep inelastic scattering or Drell-Yan dilepton in p+A collisions\cite{Deng:2009ncl,Ru:2019qvz}, $\hat q_F\approx 0.02$ GeV$^2$/fm.

Beyond such estimates, we have by now numerical calculations of $\hat{q}$ on lattice. These start from the observation that the jet quenching parameter can be determined from the short-distance behavior of a certain light-like Wilson loop. As proposed by Caron-Huot~\cite{Caron-Huot:2008zna}, such null Wilson loops can be determined in terms of a particular modified Wilson loop of the dimensionally reduced long-distance effective theory for QCD. The earliest calculation following this proposal was presented in Ref.~\cite{Panero:2013pla} and found -- for temperatures close to the QCD transition temperature -- values of $\hat{q}$ that were numerically comparable to those found in holographic calculations of strongly coupled non-Abelian plasmas [see our discussion of  Eq.~\eqref{LRWqhat} below]. More recent work includes Refs.~\cite{DOnofrio:2014mld, Moore:2019lua}; the current state of the art of lattice calculations is set by Ref.~\cite{Moore:2021jwe}.

In heavy-ion collisions, the initially produced matter will go through a rapid expansion and cooling. The quenching parameter will also experience a rapid time evolution which is thought to be initialized in an early “glasma” phase~\cite{Lappi:2006fp}. QCD effective kinetic theory [see our discussion of Eq.~\eqref{ekt} below] is used to evolve the over-occupied and highly anisotropic out-of-equilibrium initial conditions from that glasma phase to a close-to-equilibrium system described by hydrodynamics. In such dynamical scenarios, the above-mentioned lattice calculations describe $\hat{q}$ in the hydrodynamic phase. The question arises how the value of $\hat{q}$ evolves in the earlier non-equilibrium phases. Within these early phases, the rates of longitudinal and transverse momentum broadening are different, the quenching parameter becomes a tensor $\hat{q}^{ij}$. We have now first calculations of how this asymmetry of $\hat{q}^{ij}$ decreases during the effective kinetic evolution~\cite{Boguslavski:2024ezg, Boguslavski:2024jwr}. Recent calculations also indicate that $\hat{q}$ in the early glasma phase may be anomalously large~\cite{Ipp:2020nfu,Avramescu:2023qvv}. Future reviews of jet quenching may present these works as the beginning of an era in which jet quenching also becomes a tool for the study of the early non-equilibrium stage of heavy-ion collisions. A community-wide effort to explore such possibilities is under way~\cite{CERNTHInstitut}.

\subsection{Jet-medium interactions: early calculations of $1\to 2$ parton splittings}

Within the GW model of interaction between a colored parton and the quark-gluon plasma, Gyulassy and Wang made the first attempt to calculate radiative parton energy loss induced by multiple interaction, taking into account the Landau-Pomeranchuck-Migdal (LPM) interference effect \cite{Landau:1953um,Migdal:1956tc}. Even though they considered gluon emissions from the gluon propagator through the three-gluon vertex in the case of single scattering, they neglected the rescattering of the emitted gluon from such three-gluon vertex with the QGP. They argued that such additional gluon-medium rescattering should be suppressed by $k_\perp/\mu_D$ for small angle emissions in response to the referee report of their work.  This was shown not be true shortly afterwards by  Baier, Dokshitzer, Mueller, Peigne and Schiff (BDMPS)\cite{Baier:1996kr} and also independently by Zakharov \cite{Zakharov:1996fv}. The BDMPS-Z studies showed that these are the dominant processes contributing to radiative energy loss after one takes into account the dependence of the phases in the radiative amplitudes on the accumulated transverse momentum transfer from the QGP medium. 

Because the accumulated transverse momentum in the phase of the radiative amplitudes is neglected in Gyulassy-Wang's original attempt, the LPM interference is QED-like, not capturing the non-Abelian feature of gluon bremsstrahlung in multiple-parton interaction. Nevertheless, the GW model remains a simple one for interaction between a colored parton and the QGP that is still used by many today with its limitations. Gyulassy, Levai and Vitev (GLV) \cite{Gyulassy:2000fs,Gyulassy:2000er} and Wiedemann \cite{Wiedemann:2000za} employed the model in their opacity expansion approach fully taking into account the non-Abelian LPM inference for a few scatterings.  Because of its static nature, there is no energy and longitudinal momentum exchange between the parton and the medium, therefore, no elastic energy loss. Opposite time ordering which is allowed in quantum mechanics becomes negligible in this model if scattering centers are at sufficient spatial distances. Going beyond eikonal approximation in this potential model is difficult, though many recent works have calculated sub-eikonal corrections \cite{He:2020iow,Sadofyev:2021ohn,Barata:2022krd,Andres:2022ndd,Barata:2022utc,Barata:2023qds,Kuzmin:2023hko} to consider the effect of gradient in density and flow on parton energy loss and momentum broadening. 

The simplest case of going beyond the eikonal approximation in the static potential model is to replace the scattering centers with thermal partons. One can then calculate the elastic energy loss \cite{Bjorken:1982tu,Thoma:1990fm},
\begin{equation}
\frac{dE^R_{\rm el}}{dx}=\int\frac{d^3k}{(2\pi)^3}\sum_{T} \gamma_Tf_T(k) \int{d{\bf q}^2 \, \frac{d\sigma^R_{\rm el}(T)}{d{\bf q}^2}} \, \nu,
\end{equation}
where $f_T(k)$ is the phase-space distribution (Fermi-Dirac for quarks and Bose-Einstein for gluons) for thermal partons with the degeneracy $\gamma_T$, and $\nu$ is the energy transfer from the jet parton to the thermal medium.
With a small angle scattering approximation, $\nu\approx {\bf q}^2/2|\bf k|$, the elastic parton energy loss is
\begin{equation}
\frac{dE^R_{\rm el}}{dx}\approx C_2(R)\frac{3\pi}{2}\alpha_{\rm s}^2 T^2\ln\left(\frac{2.6ET}{4\mu_{\rm D}^2}\right),
\end{equation}
in a QGP with 3 quark flavors and the Debye screening mass $\mu_{\rm D}^2=6\pi\alpha_{\rm s} T^2$.

\subsubsection{BDMPS-Z as the first QCD-based close-to-eikonal formulation of parton energy loss }
The formalism of BDMPS-Z has been reviewed so often~\cite{Baier:2000mf,Kovner:2003zj,Casalderrey-Solana:2007knd,Wiedemann:2009sh}
that any detailed exposure in this review would unnecessarily duplicate existing material. We therefore restrict our discussion to giving a parametric back-of-the-envelope argument that exhibits some of the key conclusions of the BDMPS-Z formalism without addressing the many interesting technical details of the calculation. To start, let us recall that any energetic parton (say a quark) reveals at higher resolution gluonic degrees of freedom in its wavefunction. Such a gluon becomes an independent, emitted gluon once it has accumulated sufficient transverse momentum due to multiple scattering so that it can decohere from the partonic projectile. This decoherence is measured by the phase $\varphi$ of such a gluon. In general, phases are given by $e^{i\varphi} = e^{i p.x}$ where the four-momentum $p$ is updated due to multiple scattering. In the close-to-eikonal approximation, this phase reduces to the form of a non-relativistic transverse energy times a longitudinal distance $\Delta z$,
\begin{equation}
    \varphi = \Big\langle \frac{{\bf k}_\perp^2}{2\omega} \Delta z \Big\rangle  \sim \frac{\hat{q} L}{2\omega} L = \frac{\omega_c}{\omega}\, .
    \label{phase}
\end{equation}
Here, we have assumed that transverse momentum is accumulated by Brownian motion, ${\bf k}_\perp^2 \sim \hat{q} L$. We see that the phase is then characterized by a characteristic gluon frequency $\omega_c = \tfrac{1}{2} \hat{q} L^2$. Gluons with energy $\omega < \omega_c$ have phases larger than unity and can be emitted. Soft gluons are thus emitted first. To estimate the shape of the energy distribution of emitted gluons, consider the number $N_{\rm coh}$ of scattering centers which add coherently to the gluon phase, ${\bf k}_\perp^2 \sim N_{\rm coh} \langle q_\perp^2\rangle_{\rm med}$ where we parametrize more explicitly $\hat{q} = \langle q_\perp^2\rangle_{\rm med}/\lambda $ as squared momentum transfer per unit path-length $\lambda$. The coherence time that a gluon needs to accumulate a phase of order one is then $t_{\rm coh} \simeq \omega/2{\bf k}_\perp^2 \simeq \sqrt{\omega/\hat{q}}$ and the number of scattering centers that act coherently is
$N_{\rm coh} = t_{\rm coh}/\lambda = \sqrt{\omega/\langle q_\perp^2\rangle_{\rm med}\lambda}$.
This leads to  a gluon energy spectrum per unit path-length $dz$ that has the form
\begin{equation}
    \omega \frac{dI}{d\omega\, dz} \simeq \frac{1}{N_{\rm coh}}\, \omega \frac{dI^{\rm 1\, scatt}}{d\omega\, dz}
    \simeq \frac{\alpha_s}{t_{\rm coh}} \simeq \alpha_s \sqrt{\frac{\hat{q}}{\omega}}.
    \label{LPMest}
\end{equation}
This is the characteristic $1/\sqrt{\omega}$-dependence of an LPM-suppressed gluon radiation spectrum. The total average energy $\Delta E$ ``lost" by the energetic parton due to such emission can then be estimated by integrating Eq.~\eqref{LPMest} over the in-medium path-length up to $L$, and integrating over gluon energy up to the hightest gluon energy that can be emitted according to Eq.~\eqref{phase}. One thus obtains
\begin{equation}
    \Delta E = \int_0^L dz \int_0^{\omega_c} d\omega\, \omega \frac{dI}{d\omega\, dz}  \sim \alpha_s\omega_c
    \sim \alpha_s \hat{q}\, L^2\, .
    \label{BDMPSeloss}
\end{equation}
These back-of-the-envelope estimates illustrate both the key physics picture and the main physics conclusions of the BDMPS-Z formalism which provides a QCD-based treatment of the medium-modification of $1\to 2$ parton splitting in the close-to-eikonal approximation. 

At the time of its derivation, this characteristic BDMPS $L^2$-dependence was an eye-opener. At face value, it implies that if the average in-medium path length changes by a factor $O(10)$ in going from proton-proton to nucleus-nucleus collisions, then the expected parton energy loss changes by $O(100)$. Not observing any signal of energy loss above background in small p+p collisions can thus be compatible with observing numerically large effects in nucleus-nucleus collisions. We caution, however, that this is only a parametric pocket estimate derived for an idealized situation of scattering on a non-expanding medium and for fixed in-medium path length $L$. Also, the relation between $\Delta E$ and experimentally observable signals is more complicated due to trigger bias effects that we discuss in the following. The take-home message here is that we have a QCD-based formulation in which the characteristic scales of transverse momentum broadening $\sim \hat{q}L$ and of medium-induced radiation $\sim \omega_c = \hat{q} L^2$ are governed by the same medium properties and where the expected jet quenching phenomena can increase strongly with increasing in-medium path length, i.e., with increasing system size. 

\subsubsection{High-twist approach to multiple parton interaction}
Power or higher-twist (HT) corrections to hard processes had been pursued by many people since the establishment of the collinear factorized pQCD model of hadronic collisions. Qiu and Sterman examined these higher-twist corrections systematically in deeply inelastic scattering (DIS) and Drell-Yan dileption production processes \cite{Qiu:1990xy,Qiu:1990xxa}. They proved the generalized factorization of these higher-twist corrections which can be expressed as the convolution of multiple parton scattering cross sections and the correlation matrix elements of higher-twist multi-parton operators. These higher-twist corrections are power-suppressed $\sim 1/Q^2$ which can be neglected in the limit of $Q^2\gg \Lambda_{\rm QCD}^2$ in hadronic collisions.

However, Luo, Qiu and Sterman (LQS) later pointed out \cite{Luo:1992fz,Luo:1993ui,Luo:1994np} that these higher-twist corrections in collisions involving a nuclear target can be enhanced by the nuclear size $L_A/Q^2$ if multiple partons in the high-twist matrix-elements come from different nucleons in the nuclear target. These enhanced higher-twist corrections can then become significant for a large nucleus even at a reasonably large scale $Q^2$. 

Around the same time when GLV-W opacity expansion was developed in 2000, Guo and Wang used the high-twist technique to calculate parton energy loss and medium modification of the parton fragmentation functions \cite{Guo:2000nz,Wang:2001ifa}.  In the LQS calculations, the LPM interference between hard gluon or photon emissions and soft ones induced by secondary scatterings is neglected since formation time is much shorter than the distance between scattering centers. The calculation of parton energy loss by Guo and Wang considers soft gluon emissions and therefore the LPM interference which suppresses gluon emissions whose formation time is larger than the medium size. This requires the emitted gluon to have a minimum transverse momentum ${\bf k}^2 \sim Q^2r_0/L_A$. Consequently, the parton energy loss and the medium correction to the fragmentation function has a quadratic dependence on the medium size. 

Going beyond the collinear high-twist approach, a generalized high-twist (GHT) approach (without collinear expansion) \cite{Zhang:2019toi,Zhang:2021tcc} has been adopted recently in which the induced gluon spectra from a quark propagating in a medium can be written as
\begin{eqnarray}
\frac{dN_g}{d{\bf k}^2dz}\approx C_A\frac{\pi\alpha_s^2}{N_c}P^0_{qg}(z)
&&\hspace{-0.5cm}\int dy\rho(y)\frac{d^2{\bf q}}{(2\pi)^2} \frac{\phi_T(0,{\bf q})}{\bf q^2}
\nonumber \\
&\times&\frac{2{\bf q}\cdot{\bf k}}{{\bf k}^2({\bf k}-{\bf q})^2} 
\left( 1-\cos[\frac{({\bf k} -{\bf q})^2}{2Ez(1-z)}y]\right),
\end{eqnarray}
where $\rho(y)$ is the density of partonic scattering centers, $z$ is the momentum fractions of emitted gluons and $P^0_{qg}(z)$ is the parton splitting function in vacuum. This result is proved to be equivalent to the opacity expansion \cite{Zhang:2019toi}. Here the factorization of the initial parton production and the gluon distribution density of the medium is assumed. One can also show that the final result on parton energy loss in GHT is similar to the collinear HT \cite{Cao:2020wlm} with the effective jet transport coefficient differing by a factor $2\log(Q^2/\mu_D^2)$. In principle, one can also take into account the energy transfer to the medium during the scattering, i.e., elastic energy loss in HT approach. One, however, would need to know the generalized or off-forward high-twist matrix elements \cite{Osborne:2002st}.

\subsection{In-medium parton propagation and parton splitting: perturbative developments in the RHIC and LHC era}

With the start of the RHIC program, the main thrust of research moved from  the theoretical development of the QCD-based formalism of parton energy loss to its phenomenological study in theory-data comparisons. However, this review would be incomplete without listing at least shortly several important further developments of the QCD-based formalism:

\subsubsection{Effective Kinetic Theory of Arnold, Moore and Yaffe (AMY)}
Within the framework of perturbative QCD finite temperature field theory, Arnold Moore and Yaffe (AMY) managed to formulate an effective kinetic theory (EKT)~\cite{Arnold:2002zm} that faithfully describes (to leading order in $\lambda f$ and for not too high occupancy $f\ll 1/\lambda$, where $\lambda = 4\pi N_c \alpha_s$) the evolution of gluon occupancies $f$ that have momenta significantly larger than the in-medium screening mass. This EKT is an effective Boltzmann equation for the color and spin-averaged gluon distribution $f$, written in terms of a collision kernel that involves effective $2\to 2$ scattering and $1\to 2$ splitting,
\begin{equation}
    - \frac{df_{\bf p} }{d\tau} = {\cal C}_{1\to 2}[f_{\bf p}]
    + {\cal C}_{2\to 2}[f_{\bf p}]\, .
    \label{ekt}
\end{equation}
This AMY theory is at the basis of modern perturbative calculations of QCD transport properties~\cite{Arnold:2000dr} and it has been shown\cite{Kurkela:2014tea,Kurkela:2015qoa} to provide a dynamical framework that exhibits the bottom-up thermalization dynamics first formulated by Baier, Mueller, Schiff and Son~\cite{Baier:2000sb}. In this sense, the range of applicability of AMY extends far beyond that of jet quenching calculations. For highly energetic momenta, however, the $1\to 2$ splitting term dominates in Eq.~\eqref{ekt} and that term exhibits the same non-Abelian LPM effect as mechanisms of radiative parton energy loss~\cite{Caron-Huot:2010qjx}. On the one hand, the acronym AMY stands in jet quenching phenomenology for one of several closely related derivations (BDMPS-Z, GLV, HT, ASW, AMY)\cite{Arnold:2008iy, Caron-Huot:2010qjx,Mehtar-Tani:2019tvy} of 
radiative parton energy loss. On the other hand, AMY is more. As the long-time limit of a Boltzmann equation is the thermal equilibrium distribution, the AMY EKT in Eq.~\eqref{ekt} suggests to view jet quenching as the high-energy limit of a more widely applicable kinetic theory that explains how highly energetic quanta evolve towards equilibrium. 

\subsubsection{Beyond the soft gluon radiation}
It was realized early on that a quantitative jet quenching phenomenology
comes with large uncertainties if based on
results in the close-to-eikonal approximation in Eq.~\eqref{closetoeikonal}. In a community-wide effort, the Theory-Experiment Collaboration for Hot QCD Matter (TEC-HQM) realized \cite{Armesto:2011ht} that these uncertainties resulted from the need to account for medium-induced gluon radiation beyond the kinematic range in which these gluons are soft ($E\gg \omega$) and collinear ($\omega \gg {\bf k}$). Several developments allow by now to go beyond the soft limit. 

Using the soft collinear eﬀective theory (SCET) supplemented with the
Glauber modes of soft-gluon field for interaction between a fast parton and static scattering centers, Vitev and collaborators 
have developed the SCET$_{\rm G}$ formalism \cite{Ovanesyan:2011xy,Ovanesyan:2011kn} that goes beyond soft-gluon radiation approximation. The BDMPS-Z path-integral approach to parton energy loss has been extended to determine double differential medium-induced gluon radiation beyond the soft approximation~\cite{Blaizot:2012fh,Apolinario:2014csa}. This problem is closely related to the problem of dealing with in-medium averages of more complicated products of Wilson lines. To this end, numerical techniques for evaluating medium-induced gluon emission have been developed in several improved 
approximation schemes~\cite{Mehtar-Tani:2019tvy,Feal:2019xfl,Andres:2020vxs,Andres:2020kfg,Barata:2021wuf,Schlichting:2021idr,Isaksen:2022pkj}, and for the more 
complicated medium-averages that arise in some double-differential distributions~\cite{Isaksen:2020npj}.

In general, analytical expressions beyond the close-to-eikonal limit become significantly more complicated. A noteworthy exception is the $g\to Q\bar{Q} $ splitting function whose vacuum term is free of a collinear singularity and is dominated by democratic splittings. In this case, a compact analytical formula based on the standard BDMPS-Z path integral can be given without invoking a soft limit~\cite{Attems:2022ubu,Attems:2022otp}. 

\subsubsection{Beyond $1\to 2$}
Studies of radiative corrections to the medium-modified $1\to 2$ splitting have been pursued mainly for two reasons. First, radiative corrections lead to logarithmic enhancements with in-medium path length that result in a renormalization of the quenching parameter~\cite{Liou:2013qya,Blaizot:2014bha,Wu:2014nca}. Second, the study of final states with three partons allows one to understand destructive interference effects in more involved set-ups that occur in parton showers. An important first step in this direction was the study of medium-induced gluon radiation from a QCD antenna~\cite{Armesto:2011ir,Casalderrey-Solana:2011ule,Mehtar-Tani:2012mfa,Barata:2021byj}. More generally, the problem of overlapping formation times in sequential bremsstrahlung gluons has motivated a complete analysis of medium-induced QCD bremsstrahlung beyond leading order in $\alpha_s$~\cite{Arnold:2015qya,Arnold:2016kek,Arnold:2020uzm,Arnold:2022mby}.

\subsection{Jet quenching in strongly coupled non-Abelian plasmas}
\label{strongjet}
In the late 1990s, with Maldacena’s conjecture of the AdS/CFT correspondence~\cite{Maldacena:1997re}, a completely new theoretical tool set became available for studying strongly coupled non-Abelian plasmas.  An early result of Gubser, Klebanov and Peet~\cite{Gubser:1996de} indicated that, for a large class of non-Abelian plasmas, the only difference between the zero-coupling and infinite-coupling entropies $s$ is an overall numerical factor 3/4 that was suspiciously close to the $\approx 20 \%$  difference between the high-temperature lattice QCD equation of state and its Stefan-Boltzmann limit. Then, in  2001, the work of Policastro, Son and Starinets~\cite{Policastro:2001yc} employed the gauge-gravity conjecture to calculate the shear viscosity for the plasma of  ${\cal N}=4$ supersymmetric Yang-Mills fields, yielding the famous result $\eta/s = 1/4\pi$ in the ’t Hooft strong coupling limit. This set off an avalanche of theoretical studies of strongly coupled non-Abelian plasmas (for an early review, see ~\cite{Casalderrey-Solana:2011dxg}).

One cannot overemphasize the transformative power of theoretical ideas  that may not be directly applicable but that make completely new physics phenomena thinkable. Strictly speaking, the AdS/CFT correspondence does not apply directly to the QCD plasma since the gravity dual of QCD is not known and may not exist. But results established for a wide class of non-Abelian plasmas make it conceivable that these results 
are relevant for the strongly coupled QCD plasma as well. Perturbative calculations of $\eta/s$ yield numerically large value $\eta/s \propto 1/(\alpha_s^2 \log(1/\alpha_s))$~\cite{Arnold:2000dr} in QCD and they imply similarly $\eta/s \propto 1/(\lambda^2 \log(1/\lambda))$ for perturbatively small ’t Hooft coupling $\lambda = g^2 N_C$ in ${\cal N}=4$ SYM. Perturbative NLO corrections~\cite{Ghiglieri:2018dib} are known almost completely. The difference with the non-perturbative results $1/4\pi$ is parametrically large and the result is universal for a large class of non-Abelian plasmas with gravity duals~\cite{Kovtun:2004de}. Teaney was the first to realize the dramatic phenomenological implications~\cite{Teaney:2003kp}: a minimal $\eta/s$ implies minimal dissipation which implies perfect fluidity and that the collective dynamics translates pressure gradients most efficiently into collective flow. The measured flow momentum anisotropies required $\eta/s\ll 1$~\cite{Teaney:2003kp}. These theory developments, combined with a wealth of RHIC data and phenomenological analyses, triggered the perfect fluid paradigm. As of today, the AdS/CFT correspondence provides for many problems of real-time dynamics the only known setting, in which physical consequences of this perfect fluid paradigm can be explored in quantum field theory. 

So, how do energetic partons interact with and propagate through a strongly coupled, perfectly fluid plasma?  In a perfect fluid, there is no hierarchy between the size of a particle wave packet and its mean scattering time, the notion of mean-free path is meaningless, one should not think of jet-medium interactions in terms of microscopic scatterings. The energy lost by a projectile shot into such a plasma can only be carried away by fluid dynamic excitations since these are the only long-lived excitations in the plasma. To go beyond such generic statements requires to employ the AdS/CFT tool set, to embed  energetic quarks as external probes in the ${\cal N}=4$ SYM plasma, and to calculate their dynamical evolution. Here, we highlight only a small number of such studies:

\begin{figure}[t]
    \centering
    \includegraphics[width=0.5\textwidth]{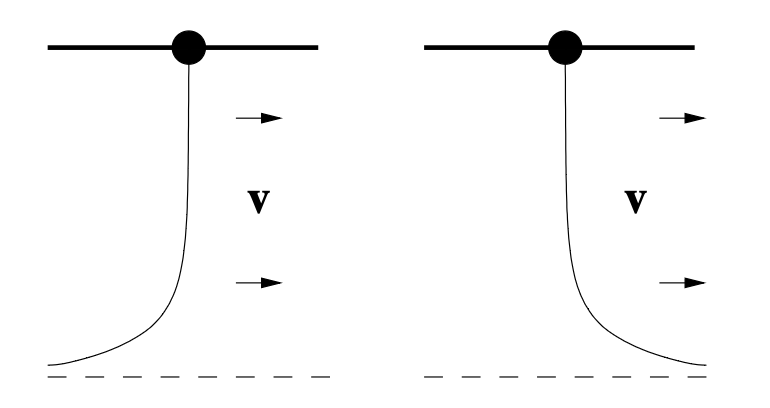}
     \includegraphics[width=0.45\textwidth]{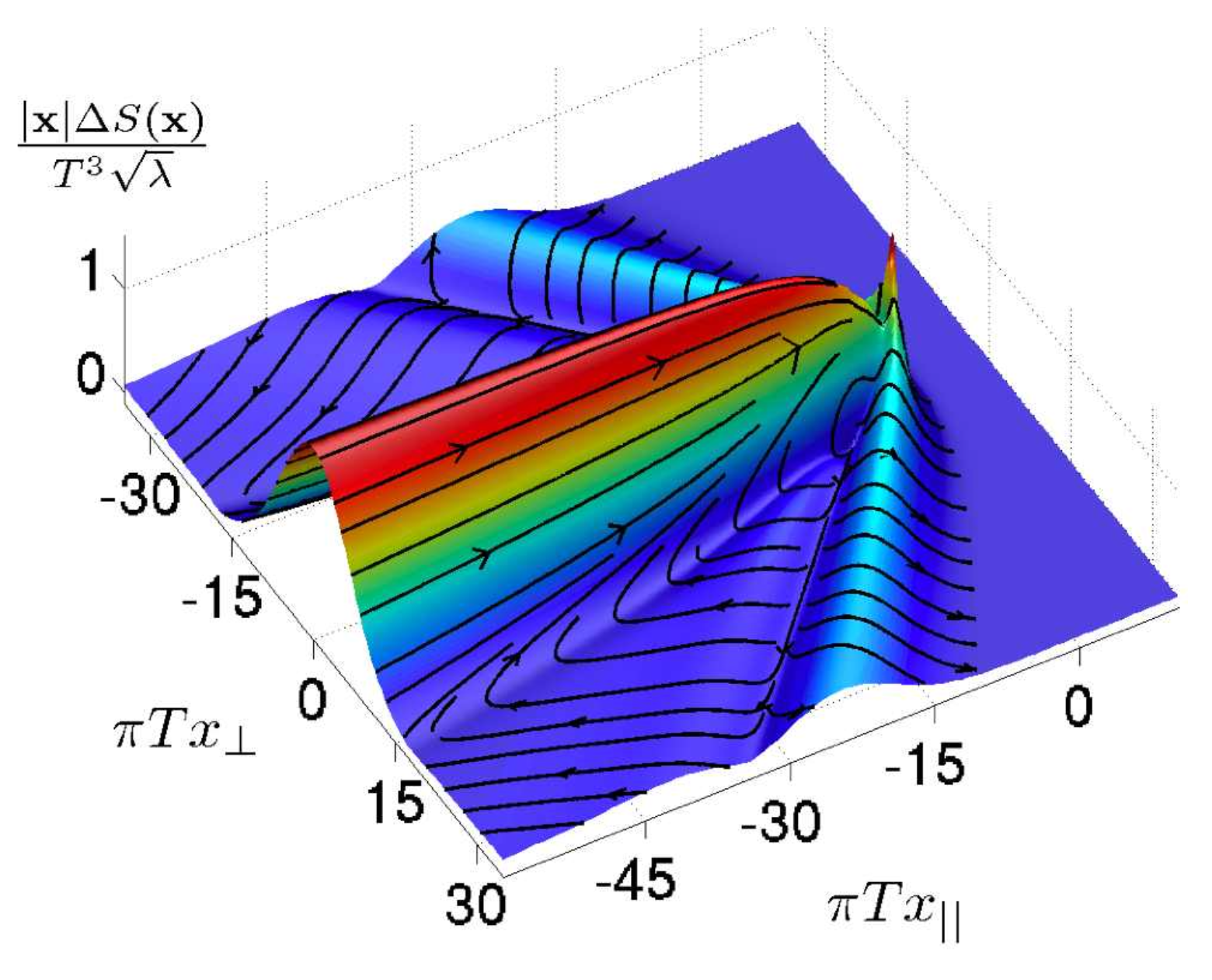}
    \caption{ (\textit{Left}) The first schematic drawing of the trailing string solution(s), Figure from Ref.~\cite{Herzog:2006gh}. In the gravity dual description, the quark is the endpoint of a string. At finite velocity, the string resembles a tail dragged behind along which energy flows to the horizon. An unphysical solution, in which the tail is oriented opposite, can be discarded on physical grounds. 
    (\textit{Right}) The perturbation in energy density induced by the trailing string solution in the ${\cal N}=4$ SYM plasma. Clearly visible is a Mach cone due to supersonic motion, and a wake that is excited by the passing string. Figure from Ref.\cite{Chesler:2007sv}.}
    \label{fig:ads}
    \end{figure}

\subsubsection{In-medium propagation of single partons in AdS/CFT}

The study of jet quenching with string-theory inspired techniques of gauge-gravity duality started with publications from four different collaborations in the same week of May 2006. The four works took maximally different approaches to the above question of how strong coupling techniques can inform jet quenching. Liu, Rajagopal and Wiedemann~\cite{Liu:2006ug} started from the observation that the only non-perturbative input in perturbative jet quenching calculations is the expectation value in Eq.~\eqref{eq5} of a Wilson loop. With the help of the AdS/CFT correspondence, they calculated the quenching weight $\hat{q}$ that characterizes the small transverse distance behavior in Eq.~\eqref{dipole} of that Wilson loop, finding~\cite{Liu:2006ug,DEramo:2010wup}
\begin{equation}
    \hat{q}_{\rm SYM} = \frac{\pi^{3/2} \Gamma(\tfrac{3}{4})}{\Gamma(\tfrac{5}{4})} \sqrt{\lambda} T^3 \approx 26.69\, \sqrt{\alpha_s\, N_c}\, T^3\, .
    \label{LRWqhat}
\end{equation}
Naively applying this formula to QCD-like input ($N_c=3$ and $\alpha_s = 0.5$  i.e. $\lambda = 6 \pi$) yields for the quenching parameter estimates 4.5 (10.6)
${\rm GeV}^2/{\rm fm}$ for temperatures $T=300$ MeV  ($T=400$ MeV)
that lie in the ballpark of phenomenologically favored values, see Section~\ref{sec432}. Remarkably, Eq.~\eqref{LRWqhat} is proportional to $\sqrt{\lambda}$, and not to the number of degrees of freedom $\sim N_c^2$. Ref.~\cite{Liu:2006ug} is not a non-perturbative calculation of parton energy loss, but a calculation of the non-perturbative quantity $\hat{q}$ that a weakly coupled, radiative parton energy loss calculation in QCD requires as input.

In contrast, the works of Casalderrey-Solana and Teaney~\cite{Casalderrey-Solana:2006fio}, of Gubser~\cite{Gubser:2006bz} and of Herzog et al.~\cite{Herzog:2006gh}, considered the problem of a massive quark that interacts on all scales non-perturbatively strongly with the ${\cal N}=4$ SYM plasma. These works start either from the so-called trailing string solution~\cite{Herzog:2006gh,Gubser:2006bz}, or they consider a formulation of heavy-quark drag and momentum diffusion in an AdS/CFT version of the Schwinger-Keldysh formalism~\cite{Casalderrey-Solana:2006fio,Casalderrey-Solana:2007ahi} that turns out to yield equivalent results. In principle, these works describe a Langevin dynamics of heavy quarks of mass $M$ and energy $E = \gamma M$ by calculating the drag coefficient
\begin{equation}
    \eta_D= \frac{\pi}{2} \sqrt{\lambda} \frac{T^2}{M}\, .
    \label{drag}
\end{equation}
The momentum loss of heavy quarks is thus proportional to the momentum of the quark, $dp/dt = -\eta_D\, p$, which is qualitatively different from radiative energy loss mechanisms. One can also calculate the transverse momentum diffusion coefficient~\cite{Casalderrey-Solana:2006fio,Gubser:2006bz}
\begin{equation}
    \kappa_T=\pi\sqrt{\lambda}\gamma^{1/2}T^3\, ,
    \label{ktbroad}
\end{equation}
and the longitudinal diffusion coefficient~\cite{Gubser:2006nz} 
\begin{equation}
    \kappa_L = \pi \sqrt{\lambda} \gamma^{5/2} T^3\, .
    \label{klbroad}
\end{equation}
However, a problem was noted early on. When these parameters are used as inputs for a Langevin description, the two parameters $\kappa_L$ and $\eta_D$ should satisfy the Einstein relation
\begin{equation}
    \kappa_L = 2 T E\, \eta_D \, .
\label{eq:Einstein}
\end{equation}
At zero velocity, this condition follows from the fluctuation-dissipation theorem. At finite velocity, the validity of Eq.~\eqref{eq:Einstein} is not guaranteed by a theorem, but it arises in Langevin formulations as a consistency condition from the requirement that a heavy quark in a plasma with a constant temperature $T$ should eventually equilibrate. However, in strongly coupled ${\cal N}=4$ SYM, the Einstein relation Eq.~\eqref{eq:Einstein} is satisfied only at vanishing velocity. In the words of Steven Gubser~\cite{Gubser:2006nz}:
{\it ``Because of the dramatic failure of the Einstein relation, it doesn’t make sense to plug the trailing string predictions for $\eta_D$, $\kappa_L$ and $\kappa_T$ into a Langevin description of a finite mass quark: it won’t equilibrate to a Maxwell-Boltzmann distribution, due to the largeness of $\kappa_L$ as compared to $\eta_D$ at highly relativistic speeds ...''}.
An analogous problem arises in weakly coupled finite temperature QCD where Moore and Teaney~\cite{Moore:2004tg} demonstrated that Einstein's equation holds only up to leading log accuracy for $v > 0$.

For ${\cal N}=4$ SYM, Gubser's problem could be resolved recently by calculating the entire momentum transfer distribution ${\cal P}({\bf k})$ according to which a massive quark of velocity $v$ changes its momentum by ${\bf k}$ in the next time-step~\cite{Rajagopal:2025ukd}. The first and second moments of ${\cal P}({\bf k})$ are $\eta_D$, $\kappa_T$ and $\kappa_L$ as given above. However, higher than second moments are crucial for equilibration to the Maxwell-Boltzmann distribution. Equilibrium is achieved with a so-called Kolmogorov equation that updates the heavy quark momentum distribution with knowledge of the full ${\cal P}({\bf k})$. Truncating this dynamics to second order leads to a Langevin-formulation that displays Gubser's problem, while the Kolmogorov equation solves Gubser's problem~\cite{Rajagopal:2025ukd}. We do not know how to calculate the full ${\cal P}({\bf k})$ for QCD, but the message that a violation of Einstein's relation at finite $v$ signals the need to go beyond the Langevin formalism is relevant for QCD, too~\cite{Rajagopal:2025rxr}.

\subsubsection{Medium response and jet ``thermalization" in AdS/CFT}

The AdS/CFT toolbox allows one to embed an energetic parton in a strongly coupled plasma without making any ad-hoc assumption about the interaction between that parton and the plasma. Technically, dissipative fluid dynamics arises in the gravity dual description of strongly coupled non-Abelian quantum field theories by studying metric perturbations to lowest order~\cite{Bhattacharyya:2007vjd}. The trailing string solution of the heavy quark is a well-defined source of such metric perturbations and studying its medium-response becomes a numerical problem of solving for small metric perturbations in dual classical supergravity~\cite{Chesler:2007an,Chesler:2007sv,Gubser:2007ga}. Figure~\ref{fig:ads} shows a typical medium-response, calculated in this way for the case of supersonic motion. Two structures are clearly visible: a Mach cone that appears at a characteristic half-opening angle $\cos(\Theta_M) \approx c_s$ and a wake that is excited by the passing partonic projectile. The energy and momentum lost by the quark is lost to these fluid excitations. We note that already prior to these formulations in AdS/CFT, hydrodynamic flow induced by fast particles was studied and the appearance of the Mach cone was identified by solving the fluid dynamic equations $D_\mu T^{\mu\nu} = J^\nu$ with a one-dimensional jet-like line source that deposits a constant amount of energy per unit path length~\cite{Casalderrey-Solana:2004fdk,Casalderrey-Solana:2006lmc}. The calculations in AdS/CFT demonstrate that for a plasma with minimally dissipative properties, this Mach cone also emerges in a fully field-theoretic setting and that it is supplemented by a wake. As of today, Mach cone-like structures are amongst the most sought signals for a collective medium-response from jet-plasma interactions. 

Jet quenching may be viewed as the high-energy approximation of a more complete dynamics that drives highly energetic partons to equilibrium. While the energetic parton that initiates a jet is kinematically well-separated from the degrees of freedom in the plasma, one expects on physical grounds that for sufficiently late times, the jet fragments become indistinguishable from the components of the medium: the jet is thermalized. The AMY  effective kinetic theory Eq.~\eqref{ekt} provides a perturbative set-up in which the entire partonic thermalization process can, in principle, be studied. Also the Kolmogorov equation proposed in Ref.~\cite{Rajagopal:2025ukd} provides a set-up in which the in-medium dynamics of a conserved charge (the heavy quark) can be followed all the way from relativistic velocity to equilibrium. Within the AdS/CFT toolbox, several other thought experiments have been explored to understand the equilibration process of highly energetic quarks. In particular, the penetration depth and energy loss rates of energetic partons in such a non-Abelian plasma have been studied for quarks~\cite{Chesler:2008uy,Chesler:2014jva} and gluons~\cite{Gubser:2008as}, and tentative connections with the phenomenology of jet quenching at RHIC and at the LHC were made~\cite{Ficnar:2013qxa}.

\section{Jet quenching and high-$p_T$ hadron spectra }
Parallel to the formulation of jet-medium interaction and calculations of parton energy loss in the 1990's, there were also coordinated efforts to prepare the field for the forthcoming experimental data on jet quenching and other hard processes. The Hard Probes Collaboration (HPC), coordinated by Satz and Wang began with the first collaboration meeting at CERN in 1994, and aimed to compute the pQCD baseline cross sections of all the hard processes, including jet and high-$p_T$ hadron spectra, in p+p~\cite{Satz:1995cg} and p+A collisions~\cite{hpc2003}. The HPC completed its mission after finishing the p+A baseline calculations in 2003 by organizing the first Hard Probes conference in Ericeira, Portugal in 2004~\cite{Lourenco:2005qp}.

After nearly a decade of groundwork in theoretical studies of parton energy loss, coordinated efforts in baseline calculations and  model investigations of phenomenological consequences \cite{Gyulassy:1990ye,Wang:1992qdg,Wang:1996yh,Wang:1998bha,Gyulassy:1998nc,Wang:1998ww,Wang:2000fq,Gyulassy:2000gk} in high-energy heavy-ion collisions, the observation of jet quenching in Au+Au collisions at $\sqrt{s_{NN}}=130$ GeV at RHIC was reported by both the PHENIX\cite{PHENIX:2001kki} and STAR Collaboration\cite{STAR:2002cmk} at the Quark Matter 2001 Conference when first RHIC data become available. This ushered in the experimental study of jet quenching and its application as a powerful probe of the strong interaction matter in heavy-ion collisions from RHIC and LHC energies. 

\subsection{Discovery of jet quenching at RHIC}

\begin{figure}[t]
    \centering
    \includegraphics[width=0.399\textwidth]{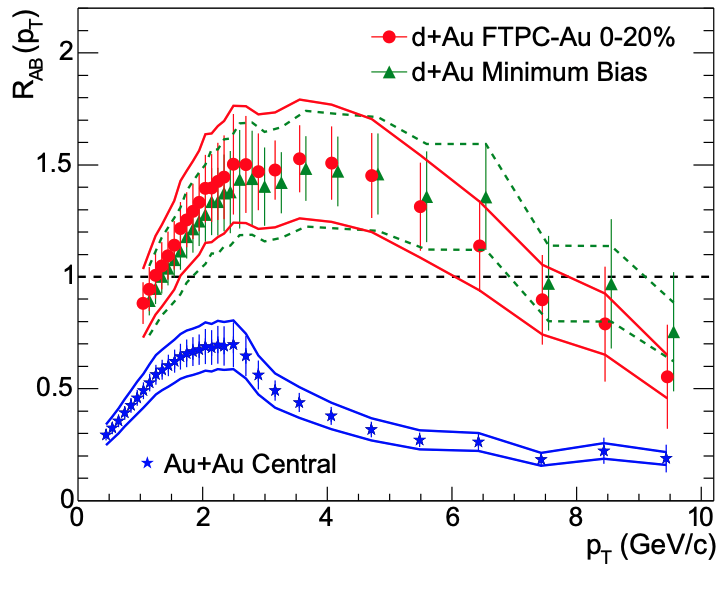}
    \hskip0.0cm
     \includegraphics[width=0.47\textwidth]{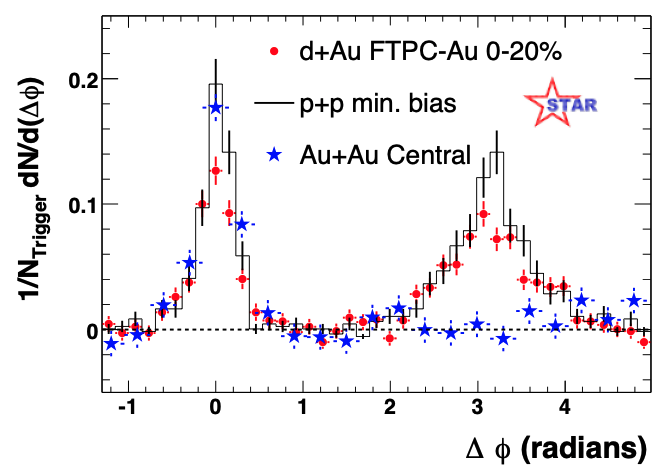}
    \caption{(\textit{Left}) Nuclear modification factors in central Au+Au collisions and d+Au collisions. (\textit{Right}) High-$p_T$ dihadron correlations in azimuthal angle in p+p, d+Au and central Au+Au collisions. Figures from Ref.~\cite{STAR:2003pjh}.}
    \label{fig2}
    \end{figure}

The first observation of jet quenching by the PHENIX \cite{PHENIX:2001hpc} and STAR  Collaboration\cite{STAR:2002ggv} at RHIC was through measurements of the nuclear modification factor
\begin{equation}
    R_{\rm AA}^{\rm h}(p_T,y) = \frac{1}{\langle T_{\rm AA}\rangle}
    \frac{(1/N_{\rm ev}){dN_{\rm AA}^{\rm h}/dp_T\, dy} }
    {d\sigma_{\rm pp}^{\rm h}/dp_T\, dy}
\end{equation}
of neutral and charged hadrons, where $\langle T_{\rm AA}\rangle$ is the averaged thickness function of two colliding nuclei. This modification factor was similarly defined in the first prediction by Wang and Gyulassy \cite{Wang:1992qdg}, later refined by Wang \cite{Wang:1998bha} such that it is normalized to unity by the number of binary nucleon-nucleon collisions in nucleus-nucleus collisions with given centrality. Though the $p_T$ range is not very large, the large suppression of the hadron spectra by up to a factor of 5 was clearly groundbreaking, as it had never been seen before in heavy-ion collisions at AGS and SPS energies \cite{Wang:1998hs}. That the suppression was caused by jet quenching was further confirmed by the observed suppression of high-$p_T$ back-to-back di-hadrons in central Au+Au collisions \cite{STAR:2002svs,STAR:2006vcp} at $\sqrt{s}=200$ GeV exactly as Bjorken speculated almost 20 years earlier in his unpublished work \cite{Bjorken:1982qr}. Even though the produced particles in proton-nucleus collisions might form a small system of interacting matter, its size should be very small and one does not expect to have any significant effect of jet quenching. Proton-nucleus collisions had been proposed as the baseline experiment to measure the effects of the initial-state interaction \cite{Wang:1992qdg,Vitev:2002pf}.  The discovery of jet quenching was finally confirmed by the subsequent observation of absence of suppression of both high-$p_T$ single and di-hadrons in d+Au collisions \cite{PHENIX:2003qdw,STAR:2003pjh,PHOBOS:2003uzz,BRAHMS:2004xry} in 2003 as shown in Fig.~\ref{fig2}, verifying that the observed jet quenching in heavy-ion collisions is indeed caused by final-state interactions and that it cannot originate from initial-state interactions which would have been present in d+Au collisions.  This was also corroborated later by the measurements of direct photon spectra \cite{PHENIX:2005yls} in Au+Au collisions which were not suppressed since photons do not interact (in strong interaction) with the hot QCD matter in the final state. 

The centrality dependence of single inclusive hadron spectra \cite{STAR:2002ggv}  and azimuthal angle anisotropy \cite{STAR:2002pmf} are also consistent with the parton energy loss picture in which the energy loss depends on the path length that varies with the centrality of the nucleus-nucleus collisions and the azimuthal angle of the parton propagation with respect to the event or reaction plane. Because of the trigger bias, the single inclusive hadrons are mostly from jets emitted close to the surface of the formed dense matter\cite{Zhang:2007ja}. This limits the sensitivity of single inclusive hadron suppression to the parton density at the center of the overlap region \cite{Eskola:2004cr}.

Because of the same trigger bias of the leading hadrons in the back-to-back dihadron correlation, the associated hadrons are often from the jet partons that have to traverse the whole volume \cite{Zhang:2007ja}. In $\gamma$-jet events, on the other hand, there is no trigger bias of the direct photons and therefore, the associated hadrons in $\gamma$-hadron correlations reveal an unbiased fragmentation pattern \cite{Zhang:2009rn}. Early experimental data on dihadron \cite{PHENIX:2003qdw,STAR:2003pjh} and $\gamma$-hadron correlations \cite{PHENIX:2009cvn,STAR:2009ojv,STAR:2016jdz} are  consistent with the path-length dependence of the parton energy loss in heavy-ion collisions.

Although the nuclear modification and azimuthal anisotropy of single inclusive light quark hadron spectra at low $p_T$ exhibit a flavor dependence due to collective flow, such a flavor dependence in experimental data continues at intermediate $p_T$ with enhanced baryon to meson ratios in central Au+Au collisions compared to peripheral~\cite{PHENIX:2001vgc} and p+p collisions~\cite{PHENIX:2003tvk,PHENIX:2003iij}. Such a light hadron flavor dependence at low and intermediate $p_T$ led to the parton recombination model \cite{Fries:2003vb,Greco:2003xt} for the hadronization of the quenched jets. Coalescence of jet shower partons and thermal medium partons dominates the hadron spectra at intermediate $p_T$ and is responsible for the enhanced baryon to meson ratio. At high transverse momentum, fragmentation of the quenched jet becomes the dominant mechanism of hadronization. The suppression of single hadrons becomes  independent of the light hadron species \cite{PHENIX:2003tvk,PHENIX:2003djd,PHENIX:2006ujp,ALICE:2019hno}, indicating the partonic nature of the energy loss instead of hadronic interactions in the final state.

 Heavy quark energy loss was predicted to be smaller than light quark because of the so-call ``dead-cone'' effect \cite{Dokshitzer:2001zm} caused by the quark mass that suppresses small angle gluon emission both in vacuum and medium-induced gluon bremsstrahlung \cite{Armesto:2003jh,Zhang:2003wk,Djordjevic:2003zk}. Early measurements of medium modification of heavy quark meson spectra at RHIC were through inclusive non-photonic electrons from semi-leptonic decays of heavy quark mesons at large $p_T$ \cite{PHENIX:2002ecc,PHENIX:2005nhb,STAR:2006btx} and $D^0$ decay products with reconstructed invariant mass using time of flight (TOF) information~\cite{STAR:2014wif}. By this time, the first measurement of $D^0, D^+$ and $D^{*+}$ spectra in Pb+Pb collisions was already made by ALICE \cite{ALICE:2012ab} at LHC. 
 Significant suppression of both the $D^0$ mesons and single non-photonic electrons was observed that are consistent with the energy loss of charmed quarks in Au+Au collisions. Such suppression is, again, not observed in d+Au collisions, thus reconfirming that jet quenching is caused by final state interactions. Measurement of charmed meson suppression through hadronic decay channel in heavy-ion collisions was made later again with the heavy-flavor tracker (HFT) of the STAR experiment~\cite{STAR:2017kkh,STAR:2018zdy} which has a significantly improved signal to background ratio because of the information of the secondary vertex provided by HFT. These measurements of heavy quark suppression are consistent with the picture of parton energy loss as seen in the modification of light quark hadron spectra at large $p_T$. Numerical calculations also show that effects of elastic energy loss becomes non-negligible as compared to the radiative energy loss of a heavy quark, especially at lower energy \cite{Wicks:2005gt}. Though these early data alone cannot definitely prove the mass hierarchy of the parton energy loss as predicted by pQCD calculations, they are nowadays part of a large data set,  including from experiments at LHC, that provides quantitative constraints on jet transport coefficients, see Section~\ref{sec432}.

With the picture of parton energy loss responsible for these early experimental data on jet quenching established, the essential questions arise: What are the properties of the medium that cause the observed jet quenching? Are they different from the normal nuclear matter and how one can characterize these properties? 

\begin{figure}[t]
    \centering
    \includegraphics[width=0.36\textwidth]{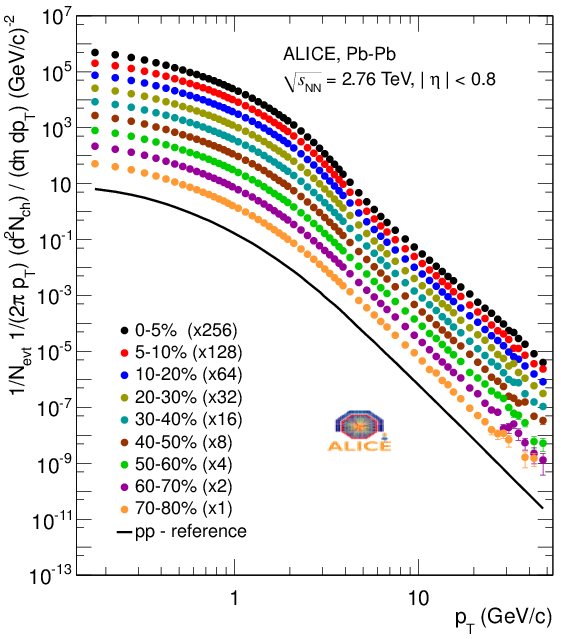}
    \hskip0.0cm
     \includegraphics[width=0.48\textwidth]{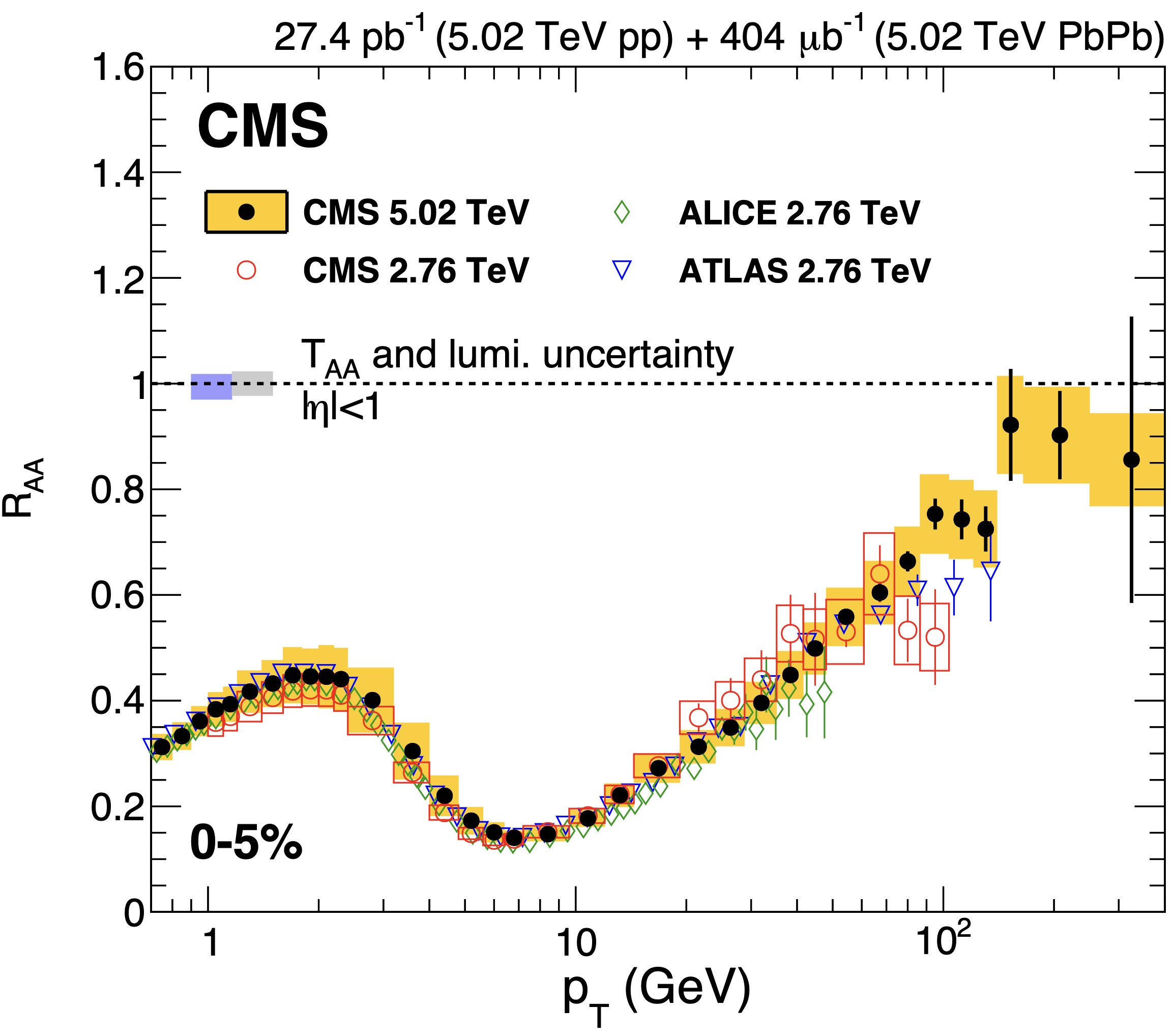}
    \caption{(\textit{Left}) The single inclusive charged hadron spectra in Pb+Pb collisions with different centralities compared to the p+p reference at $\sqrt{s}=2.76$ TeV from ALICE experiment. (\textit{Right}) Nuclear modification factors in 0-5\% central Pb+Pb collisions at $\sqrt{s}=2.67$ and 5.07 TeV from ALICE and CMS experiment. Figures from Refs.~\cite{Otwinowski:2011gq,CMS:2016xef}.}
    \label{fig3}
    \end{figure}    

\subsection{Large $p_T$ hadron spectra at the LHC}

Because of the luminosity and colliding energies, the ranges of transverse momentum reachable in experimental measurements of hadron spectra at RHIC are limited to below 20 GeV/$c$ where effects of medium modification other than parton energy loss such as collective flow and parton recombination only start to diminish. This limits the quantitative study of jet quenching using high-$p_T$ hadron spectra and their power to probe properties of the QGP in heavy-ion collisions. With the start of the heavy-ion program at LHC around 2010, one has access to medium modification of the final hadron spectra in a much larger range of $p_T$ up to a few 100 GeV/$c$ at the highest Pb+Pb colliding energy. The effect of jet quenching is so strong that one can see it even by eye as a characteristic dip in the single inclusive hadron spectra in Pb+Pb \cite{Otwinowski:2011gq,ALICE:2012aqc}, as shown in the left panel of Fig.~\ref{fig3}. The corresponding nuclear modification factors from different experiments \cite{ALICE:2012aqc,ATLAS:2015qmb,CMS:2016xef}  as shown in the right panel are almost independent of the colliding energy in the largest $p_T$ range of 1-200 GeV/$c$. One can clearly see the transition of physics mechanisms behind the medium modification of the single hadron spectra, from collective flow at low $p_T$ ($<2$ GeV/$c$) to non-equilibrium dynamics of parton recombination at intermediate $p_T$ ($2<p_T<10$ GeV/$c$) and domination of parton energy loss at large $p_T$ ($> 10$ GeV/$c$). Only in this high $p_T>10$ GeV/$c$ region, other nonperturbative effects such as parton recombination are negligible and parton energy loss becomes the only mechanism responsible for the modification of the fragmentation functions in the factorized parton model description of inclusive hadron production in heavy-ion collisions [see Eq.~(\ref{eq1})]. This makes it possible to extract medium properties such as the jet transport coefficient through quantitative analyses of the experimental data within a given model of parton energy loss and its implementation in the factorized parton model of high-$p_T$ hadron spectra. 

Dihadron measurements in Pb+Pb collisions at LHC so far have been limited to $p_T<20$ GeV/$c$ \cite{ALICE:2011gpa,ALICE:2016gso} with limited applicability for quantitative phenomenological studies. However, medium modification of the fragmentation functions of $\gamma$-triggered jets are measured at LHC \cite{CMS:2018mqn,ATLAS:2019dsv}, similar to $\gamma$-hadron measurements at RHIC. Because of the clean background and reasonable production rate, $Z$ bosons also become a robust trigger for the study of $Z$-hadron correlations \cite{ATLAS:2020wmg,CMS:2021otx}. Both of these measurements show
significant suppression of leading hadrons inside the away-side jet with a given cone size due to parton energy loss which are compatible to the single inclusive hadron suppression. These measurements also see significant enhancement of soft hadrons due to contributions from induced gluon radiation and medium recoil from the jet medium interaction. Although these enhancements cannot be used to measure parton energy loss directly, they will be essential for the study of medium response to jet propagation inside QGP. 

Measurements of heavy quark meson spectra, both $D$ and $B$ (direct $B$, non-prompt $D^0$, decay $\mu$ and  non-prompt $J/\psi$) at high $p_T$ have shown a mass hierarchy of the suppression factor $R_{AA}^{D,B}$ \cite{CMS:2017uoy,CMS:2018bwt,CMS:2017uuv,ATLAS:2021xtw,ALICE:2022tji}. This hierarchy diminishes at very high $p_T$ where the suppression factors of both charm and beauty mesons approach those observed for light hadrons. This is consistent with the mass dependence of quark energy loss. However, to quantitatively describe the experimental data, one also needs to take into account pQCD higher order processes of gluon splitting into heavy quark and anti-quark pairs \cite{Xing:2019xae}.

\subsection{Constraining $\hat{q}$ with quenched hadron spectra}
\subsubsection{Trigger bias effects for single inclusive spectra}
The phenomenological study of high-$p_T$ hadron spectra is characteristically different from that of fully reconstructed jets since the differential single inclusive $p_T$ spectra are {\it leading} hadron spectra within a jet. More precisely, in the collinear factorized formula in Eq.~\eqref{eq1} for high-$p_T$ hadron spectra, the partonic cross section is steeply falling with transverse momentum. Convoluting this steep distribution with a fragmentation function $D_{f\to h}$ amounts to sampling $D_{f\to h}$ around values for which the energetic parton $f$ fragments significantly harder than the average parton in a parton shower. This trigger bias effect exists already for vacuum fragmentation where it is numerically significant: in a 100 GeV light-flavored hadron jet, the highest momentum hadron carries on average one fourth of the jet energy, but the fragmentation function in Eq.~\eqref{eq1} is typically sampled around three fourth of the parent parton energy.  In medium, the same argument implies that leading hadrons result from energetic partons that have suffered the least medium-induced radiation. Baier et al.~\cite{Baier:2001yt} were the first to formalize this trigger bias effect by calculating the probability ${\cal P}(\Delta E)$ that an arbitrary number $n$ of soft emissions carry away an energy $\Delta E$ from a leading parton, 
\begin{eqnarray}
  {\cal P}(\Delta E) = \sum_{n=0}^\infty \frac{1}{n!}
  \left[ \prod_{i=1}^n \int d\omega_i \frac{dI(\omega_i)}{d\omega}
    \right]
    \delta\left(\Delta E - \sum_{i=1}^n \omega_i\right)
    \exp\left[ - \int_0^\infty \hspace{-0.2cm}
      d\omega \frac{dI}{d\omega}\right]\, ,
   \label{eq:weight}
\end{eqnarray}
where $dI(\omega)/d\omega$ is the emission spectra of a single scattering. Here, the average energy loss in Eq.~\eqref{BDMPSeloss} is the first moment $\langle \Delta E\rangle = \int d\Delta E\, \Delta E\,  {\cal P}(\Delta E)$. However, the typical energy loss relevant for single inclusive spectra results from writing the medium-modified fragmentation function in Eq.~\eqref{eq1} as a convolution
\begin{equation}
D_{f\to h}^{\rm (med)} = {\cal P}(\Delta E) \otimes D_{f\to h}^{\rm (vac)} \, .
\label{fragmed}
\end{equation}
In the HT calculation of the medium modified fragmentation functions\cite{Guo:2000nz,Wang:2001ifa}, the above convolution is between the medium-induced splitting function and the vacuum fragmentation function plus the virtual correction that cancels the infrared divergency. For an early discussion of medium-modified fragmentation functions, see Ref.~\cite{Wang:1996yh}.  
The convolutions in Eq.~\eqref{eq1} then imply that in the calculation of hadron spectra, the probability $ {\cal P}(\Delta E)$ is sampled predominantly for values $\Delta E$ that are significantly smaller than the average $\langle \Delta E\rangle$.

Early calculations of induced gluon spectra and parton energy loss have assumed parton propagation in a static and uniform QGP medium. Since the corrections to the existing calculations of parton energy loss due to gradients and flow of the medium are sub-eikonal \cite{He:2020iow,Sadofyev:2021ohn,Barata:2022krd,Andres:2022ndd,Barata:2022utc,Barata:2023qds,Kuzmin:2023hko}, they can be neglected for very energetic partons. In the framework of opacity expansion (GLV-W), high-twist (HT) and other approaches that rely on local emission rates (AMY, SCET$_{\rm G}$), the effect of flow on parton energy loss can be taken into account through trivial Lorentz boost and the variation of the medium density can be incorporated in the path integral of the interaction kernel. Following Ref.~\cite{Baier:1998yf}, quenching weights in Eq.~\eqref{fragmed} within different models of multiple scattering have been compared and extended to the case of any arbitrary expanding medium~\cite{Salgado:2003gb}. If one characterizes the medium expansion in terms of a power-law $\hat q=\hat q_0 (\tau_0/\tau)^\alpha$ of the time dependence of the jet transport parameter, one finds a scaling law \cite{Salgado:2002cd} that one can use to relate the emission spectra $dI(\omega)/d\omega$ in an expanding system to an equivalent static scenario. 

\subsubsection{Extracting $\hat{q}$}
\label{sec432}
In general, if one knows the dependence of the single gluon emission spectrum $dI(\omega)/d\omega$ on the quenching parameter $\hat{q}$, then one knows the $\hat{q}$-dependence of the energy loss probability ${\cal P}$ in Eq.~\eqref{eq:weight} and one can proceed to extract $\hat{q}$ in a comparison of measured spectra to Eqs.~\eqref{eq1} and \eqref{fragmed}. However, several physical effects need to be considered carefully for such a program. In particular, since quenching samples distributions significantly away from their mean, a detailed understanding of the distribution of in-medium path length~\cite{Dainese:2004te} and of the density-dependence along the path length in the medium is needed. The phenomenology of quenching thus requires control over the calculation of $dI(\omega)/d\omega$ and over the bulk evolution of the medium in which the hard processes are embedded. 

Through phenomenological studies of the early RHIC data on suppression of single inclusive hadron \cite{Wang:2002ri,Vitev:2002pf,Salgado:2003gb,Eskola:2004cr,Qin:2007rn}  and correlated hadron spectra \cite{Wang:2003mm,Renk:2006nd,Zhang:2007ja,Zhang:2009rn,Qin:2009bk,Armesto:2009zi} one can already obtain a glimpse of the properties of the QGP such as the initial parton density or the jet transport parameter in heavy-ion collisions at the RHIC energy. The LHC data in a much larger range of transverse momentum make such phenomenological studies more quantitative, especially with the implementation of more realistic relativistic hydrodynamic description of the space-time profile of the evolving QGP medium in the parton energy loss calculation \cite{Bass:2008rv}. 

\begin{figure}[t]
    \centering
    \includegraphics[width=0.50\textwidth]{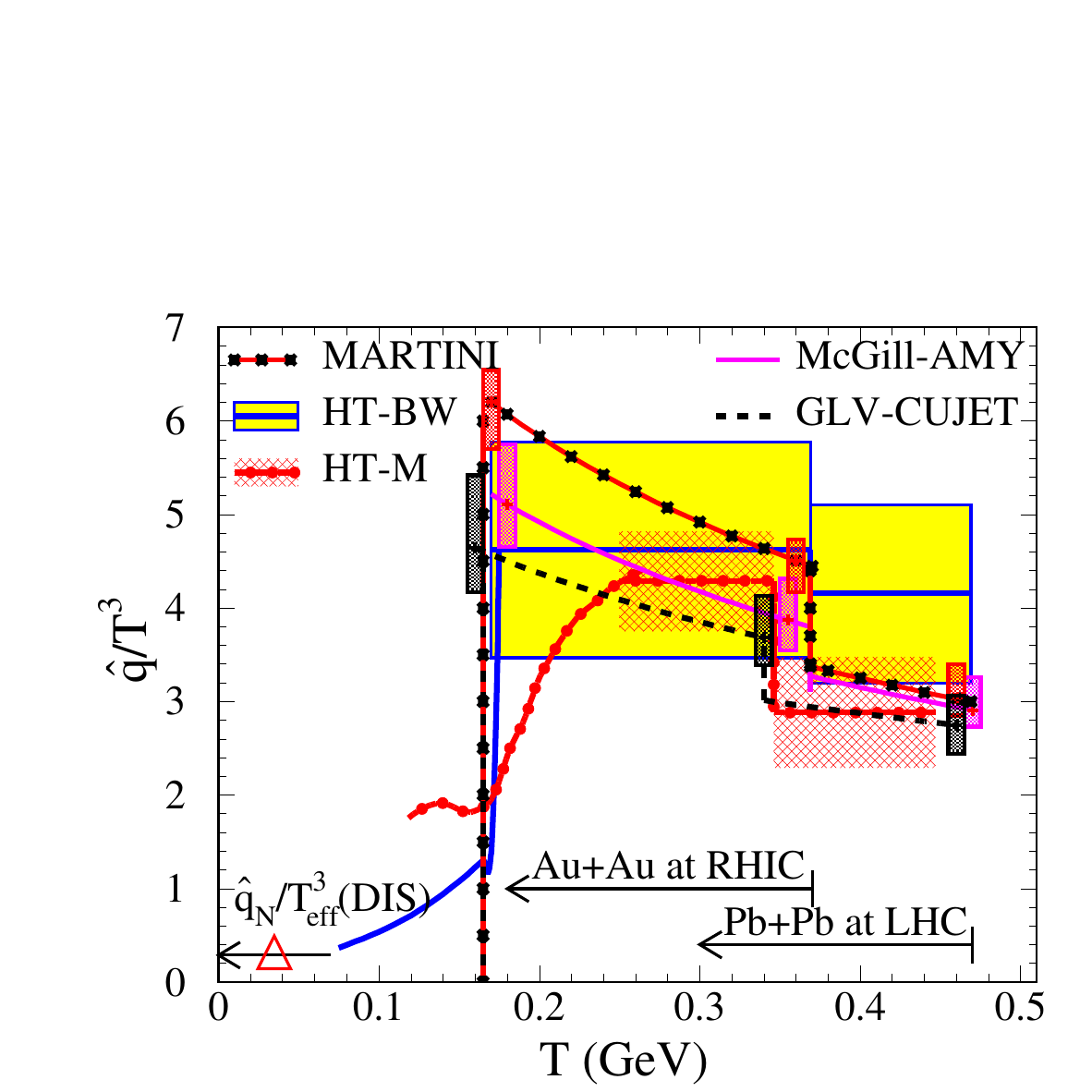}
     \includegraphics[width=0.45\textwidth]{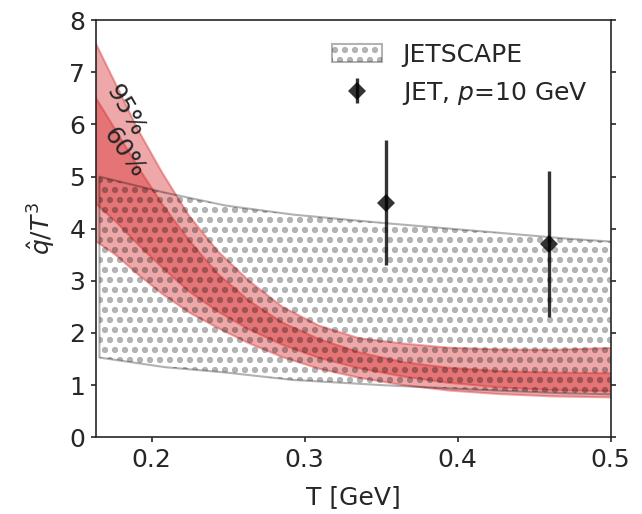}
    \caption{(\textit{Left}) The scaled jet transport parameter $\hat q/T^3$ extracted by the JET Collaboration within different approaches to the parton energy loss. (\textit{Right}) Strong temperature dependence of $\hat q/T^3$ extracted using the information-field approach to Bayesian inference as compared to the results from JETSCAPE \cite{JETSCAPE:2021ehl} and JET Collaboration. Figures from Refs. \cite{JET:2013cls,Xie:2022fak}. }
    \label{fig:qhat}
    \end{figure}

A systematic study of the single hadron spectra at both RHIC and LHC within different frameworks of parton energy loss was carried out by the US DOE JET Collaboration \cite{JET:2013cls} using realistic relativistic hydrodynamic models for the bulk QGP evolution that have been constrained by the experimental data on bulk hadron spectra. The initial values of the jet transport parameter at the center of the most central Au+Au collisions at RHIC and Pb+Pb collisions at LHC energy were extracted through $\chi^2$ fits as shown in 
Fig.~\ref{fig:qhat}~(\textit{Left}).
This corresponds to 
\begin{equation*}
\frac{\hat q_0}{T^3_0}\approx \left\{ 
\begin{array}{l}
4.6\pm 1.2 \qquad \text{T$_0$=370 MeV at RHIC},\\
3.7 \pm 1.4 \qquad \text{T$_0$=470 MeV at LHC},
\end{array}
\right.
\end{equation*}
for a quark jet at the highest initial temperatures estimated from the measured total transverse energy production at the initial time $\tau_0=0.6$ fm/$c$. These values are about 2 orders of magnitude higher than the $\hat q$ in the cold nuclear matter extracted from deeply inelastic scattering experiments \cite{Wang:2002ri,Chang:2014fba}. The temperature and energy dependence of $\hat q$ are given by each model. Differences between different models represent theoretical uncertainties. It is interesting to point out that the above value is about the same as Baier's first numerical estimate in Fig.~\ref{fig:qhat1} at a temperature $T=400$ MeV and the energy density $\epsilon=38$ GeV/fm$^3$.

Many other efforts on the extraction of $\hat q$ \cite{Andres:2016iys,Feal:2019xfl,Xie:2020zdb,Ke:2020clc} have been made since the first study by the JET Collaboration. One should, however, be careful when comparing these results\cite{Apolinario:2022vzg}, in particular about the flavor dependence (quark versus gluons), the incorporation of realistic hydrodynamic evolution of the medium and the assumed temperature dependence.  In order to have a model-independent extraction of the temperature and energy dependence of $\hat q$, one needs to carry out more sophisticated analyses such as Bayesian inference and uncertainty estimation. Such a Bayesian analysis was carried out by the JETSCAPE Collaboration based on all data on single inclusive hadron spectra at both RHIC and LHC energies and a parametrization of the temperature and energy dependence \cite{JETSCAPE:2021ehl}. The extracted values are consistent with that from the JET Collaboration within the combined uncertainties. In this Bayesian
analysis, the prior values of $\hat q$ in different regions of temperature and jet energy are not completely independent when they are sampled according to a parameterized temperature and jet-energy dependence. Such long-range correlation can lead to additional uncertainties of the $\hat q$ in regions not well constrained by the existing data. If an information field approach is adapted to model the prior distributions that are free of the long-range correlation, the extracted $\hat q$ shows a stronger temperature dependence \cite{Xie:2022ght,Xie:2022fak} as shown in Fig.~\ref{fig:qhat} (\textit{Right}). Data on $\gamma$-hadron and dihadron correlations are also shown to have a strong impact on the extracted $\hat q$ though it is still limited by the experimental uncertainties and finite ranges of kinematics of the available data.  

As a footnote to the Bayesian inferences of $\hat q$, we want to emphasize that model uncertainties have not been introduced in the existing efforts. In the compilation collected by Apolinario {\it et al}\cite{Apolinario:2022vzg}, one can see a large range of model uncertainties even after taking into account the flavor dependence and discrepancy in the hydrodynamic description of the QGP evolution.  In addition, event-by-event fluctuations will also have some impact on jet quenching.  One should marginalize these uncertainties in future efforts.

\subsubsection{Centrality dependence of quenched hadron spectra}
While the study by the US DOE JET Collaboration \cite{JET:2013cls} remained focused on the $0 - 5\%$ most central collisions, several subsequent studies have considered in detail the centrality dependence of quenching effects. In Ref.~\cite{Andres:2016iys}, the authors compared the centrality dependence at RHIC and at the LHC, and they noted a possible tension between both dependencies if the quenching parameter is assumed to scale with $\epsilon^{3/4} \sim T^3$. In Ref.~\cite{Huss:2020whe}, the authors compared thirteen different model implementations of parton energy loss in the literature by tuning to 
the nuclear modification factor at fixed $p_T = 50$ GeV for minimum bias Pb+Pb collisions at the LHC and then comparing without further model adjustments to data on the $p_T$-dependence, the centrality dependence and to data for Xe+Xe collisions at the LHC. The main finding was that these dependencies were well reproduced for centrality bins in the range up to $50\% $ for Pb+Pb and Xe+Xe, but that all models failed to reproduce the nuclear modification factor measured in peripheral collisions ($> 70\%$ centrality). 

There is some understanding that this failure to reproduce quenching data in small collision systems is at least partly due to the fact that the standard nuclear modification factor $R^{\rm h}_{\rm AA}$ may not be adequate for characterizing jet quenching in very small collision systems~\cite{Loizides:2017sqq}. The likely culprit is the nuclear overlap function $\langle T_{\rm AA}\rangle$ which is a geometric measure of the centrality-dependent size of the collision systems but which is inferred experimentally from the event multiplicity. In sufficiently large collision systems, impact parameter (i.e. nuclear overlap) correlates well with event multiplicity. In smaller systems, however,  event multiplicity also depends on the high-$p_T$ process on which one triggers. Kinematic biases in the centrality selection have been identified also in other small collision systems~\cite{Armesto:2015kwa}. Monte Carlo studies show~\cite{Loizides:2017sqq} that this effect indeed invalidates a simple interpretation of  $R_{\rm AA}^{\rm h}(p_T,y)$ for peripheral collisions of $> 70\% $ centrality, and they can be used to correct~\cite{Loizides:2017sqq} the normalization factor $\langle T_{\rm AA}\rangle$ in peripheral collisions.

A better clarification of this issue is wanted: Both jet quenching and collective flow invoke final state interactions with the QGP. Collective flow has been observed in peripheral heavy-ion collisions and in systems as small as p+Pb at the LHC. It thus follows on general ground that energetic partons must undergo interactions with the QGP in small systems, too. Therefore, either the resulting quenching in small collision systems is verified experimentally, or it is understood why the expected quenching is smaller than the current experimental sensitivity. One way to increase the experimental sensitivity is to consider inclusive minimum-bias nuclear modification factors that can be constructed as ratios of the cross sections measured in A+A and p+p normalized by $A^2$. These minimum bias nuclear modification factors are free of uncertainties related to the centrality-dependence of $\langle T_{\rm AA}\rangle$. It has been shown~\cite{Huss:2020dwe} that for small systems such as O+O collisions at the LHC, the no-quenching baseline can be determined theoretically to better than $5 \%$ in NLO calculations, while extrapolations of all existing quenching models predict effects larger than $5 \%$. The short O+O run at LHC in the summer of 2025  is expected to extend the study of jet quenching phenomenology to smaller collision systems. We also note that beside p+p, O+O is the only collision system for which both RHIC (data are on tape from the RHIC 2021 run) and the LHC took data.

\section{Jet quenching and medium modification of jets}

The high colliding energies at LHC not only have extended the kinematic range of single inclusive hadron spectra, but also make it possible to study reconstructed full jets according to different jet finding algorithms. Jet production observables are calorimetric measurements, they are insensitive to non-perturbative hadronization processes and they can be systematically calculated in perturbative QCD for p+p collisions. Therefore, they can be used as standard candles for medium modification in heavy-ion collisions.

With the RHIC discovery of jet quenching on the level of leading hadrons and jet-like hadron correlations, theorists had understood well before the start of LHC operations that the entire parton shower associated to jet measurements may be subject to significant medium-modifications if jets are embedded in nucleus-nucleus collisions at the LHC. 
Prior to the start of LHC data taking, theorists anticipated medium-modifications to essentially all characteristics of jets in nucleus-nucleus collisions, including a broad range of jet substructure measurements~\cite{Zapp:2008gi}, jet multiplicity distributions~\cite{Borghini:2005em,Borghini:2005mp} and the energy distribution inside jets~\cite{Vitev:2008rz,Vitev:2009rd}.  Some effects of early works are still conceivable from today's perspective, but have not found an experimental confirmation yet.  They include e.g.
characteristic asymmetries of jet fragmentation in response to collective flow fields~\cite{Armesto:2004pt}, modification of hadrochemical composition of jets due to jet-medium interaction~\cite{Sapeta:2007ad}, or effects of jet interaction with turbulent color fields~\cite{Majumder:2006wi}. In 2010, when the news finally broke of a very strong dijet asymmetry discovered by ATLAS~\cite{ATLAS:2010isq} and CMS~\cite{CMS:2012ytf}, see Fig.~\ref{fig:dijet}, the theory community was not only excited and surprised about the large size of the effect, but was also prepared to follow up immediately with a significant number of analyses~\cite{Casalderrey-Solana:2010bet,Qin:2010mn,Cacciari:2011tm,Young:2011qx,Lokhtin:2011qq,Renk:2011qf}. The LHC era of studying medium-modified parton showers had begun.

\subsection{Jet finding algorithms and background subtraction}
At the time of the first quenched jet measurements at the LHC, particle physics saw a revolution in the very algorithmic definition of what constitutes jets, see Ref.~\cite{Larkoski:2017jix} for a review. What started as an algorithmic reassessment of the previously used $k_T$-jet finding~\cite{Cacciari:2005hq} and anti-$k_T$ jet finding~\cite{Cacciari:2008gp} algorithms, had grown by 2010 into a versatile FastJet toolbox~\cite{Cacciari:2011ma} that algorithmically implemented the modern thinking that jets are defined via reconstruction algorithms and that reconstruction algorithms can be adapted depending on the physics under investigation. These algorithmic jet definitions are infrared and collinear safe, thus ensuring that jet production cross sections in elementary p+p collisions are systematically calculable in perturbative QCD and that effects of non-perturbative hadronization processes are minimized~\cite{Sterman:1977wj}. 

However, jet finding algorithms define more than perturbatively calculable jet observables in heavy-ion collisions. Of relevance for heavy-ion collisions is, for instance, the concept of jet catchment area~\cite{Cacciari:2008gn}, the region  in the $\Delta\eta \times \Delta \phi$ phase space within which a particular jet is reconstructed. The jet catchment area is not an infrared and collinear safe quantity, but it is an important tool for pile-up subtraction in elementary p+p collisions~\cite{Cacciari:2007fd} and for background subtraction in heavy-ion collisions~\cite{Cacciari:2010te}. This area turns out to be approximately circular for anti-$k_T$ reconstructed jets, but not for $k_T$-reconstructed jets. Also in comparison with Monte Carlo truth, both anti-$k_T$ and Cambridge/Aachen (C/A) algorithms perform better with respect to the transverse momentum offsets and dispersions~\cite{Cacciari:2010te}. This explains the preponderance of  anti-$k_T$ reconstructed and C/A reclustered jet measurements in heavy-ion collisions. 

In central Pb+Pb collisions at the LHC, any small jet cone of radius   $r=\sqrt{\Delta\eta^2+\Delta\phi^2} = 0.4$ covers a predominantly soft, high-multiplicity background activity of $O(100 \;{\rm GeV})$. In central heavy-ion collisions, jets must be reconstructed on top of this background that fluctuates event by event. It was fortunate that the first observed dijet asymmetries at the LHC were so large that they were clearly visible in single event displays -- this convinced ATLAS after multiple cross checks to publish their discovery immediately\cite{ATLAS:2010isq}.  CMS had made essentially the same observation around the same time but it did proceed to publication only after being able to check after a few months that the energy lost from the dijet can be recovered event-by event within the same hemisphere of the event in soft large-angle radiation~\cite{CMS:2011iwn}. The history of quenched jet measurements is, since the very first measurements at the LHC, a history of understanding how to identify quenching signatures on top of a fluctuating high-multiplicity background.  

FastJet is nowadays the basis of a large number of refined jet substructure measurements, including jet fragmentation functions, $r$-differential jet shapes, measurements of girth and broadening, groomed or ungroomed jet mass measurements and soft-drop~\cite{Larkoski:2014wba} measurements, see Ref.\cite{Salam:2019ayd} for technical definitions and links to the recent literature. These techniques make it possible, at least in principle, to dissect the medium-modification of quenched jets up to the level of elementary medium-modified splittings~\cite{CMS:2017qlm,Brewer:2025wol}, or to reconstruct likely branching histories and their space-time embedding in heavy-ion collisions~\cite{Apolinario:2024hsm}.

\subsection{Monte Carlos simulations of jet transport}
\label{MCtools}

Because of the multi-parton nature of reconstructed jets, analytic or semi-analytic calculations of jet observables become difficult if not impossible. One has to resort to Monte Carlo methods, in particular for jet evolution inside QGP. There have been several Monte Carlo models that attempt to simulate medium-induced multiple gluon emissions \cite{Armesto:2007dt,Renk:2008pp,Armesto:2009fj} and describe jet evolution in QGP medium.  Parton cascade models by Geiger \cite{Geiger:1991nj}, Zhang \cite{Zhang:1997ej} (which was incorporated into the AMPT model \cite{Lin:2004en}), Xu and Greiner \cite{Xu:2004mz} were also developed to address the questions of parton interaction and thermalization in an evolving QGP. Sophisticated Monte Carlo models were only developed later on to take into account both the medium-induced multiple gluon emissions with proper treatment of the LPM interference and elastic scatterings to specifically study jet transport in QGP.

\begin{enumerate}
    \item 
 {\it JEWEL}: Jet Evolution with Energy Loss (JEWEL) \cite{Zapp:2008af,Zapp:2008gi,Zapp:2012ak} is a Monte Carlo model developed to specifically implement the LPM interference in induced gluon emissions in a probabilistic (and local) way depending on the scale of the initial scattering, the scale of medium interaction and the formation time of the gluon radiation. The interaction with medium during the jet shower development is modeled by extending the collinear factorized $ 2 \rightarrow 2 $ pQCD matrix elements to the infrared limit. The parton shower is based on PYTHIA Monte Carlo and the medium was initially modeled according to the Bjorken scaling model \cite{Bjorken:1982qr}, but later versions can adopt any medium with dynamic evolution. JEWEL keeps the information of the recoil partons from each parton medium interaction. In the default version of the model, recoil partons are not rescattered further on the medium. In the version ``recoil on", the code runs inevitably much slower since a more complicated dynamics is followed. JEWEL has a smooth transition from vacuum like jet evolution into a regime where medium modifications occur. 
 
\item {\it MARTINI}: The Modular Algorithm for Relativistic Treatment of heavy IoN Interactions (MARTINI) \cite{Schenke:2009gb} is a comprehensive Monte Carlo event generator developed by the McGill group to simulate jet evolution in the QGP medium. The multiple parton scattering and medium-induced gluon radiation are based on the McGill-AMY approach \cite{Arnold:2002ja,Qin:2007rn} with coupled 
Fokker-Planck type rate equations that incorporate the LPM interference for gluon emissions in the medium  as well as elastic scatterings. Splitting rates depend on the local temperature of the medium. The main components of MARTINI include 2+1D or 3+1D hydrodynamics, PYTHIA~8.1 and the  McGill-AMY parton evolution scheme \cite{Qin:2009bk}. The McGill-AMY formalism includes the LPM effect in the transition rates which are exact only in the infinite volume limit which is different from other approaches that will have path-length dependence in the multiple gluon emissions. It also does not include the interference between medium and vacuum showers as in other Monte Carlo simulations. 

\item {\it LBT} and {\it CoLBT}: The Linear Boltzmann Transport (LBT) model \cite{Li:2010ts,He:2015pra,Luo:2023nsi} is developed to study jet interaction and propagation in QGP with a particular emphasis on thermal recoil partons and their further interaction and propagation through the medium. It is based on Boltzmann equations for the evolution of both jet shower partons and the recoil partons. The collision kernels in the Boltzmann equations contain the complete set of leading order (LO) elastic $2\to 2$ scattering processes with the collision matrices regulated by the Debye screening mass. Inelastic processes  $2\rightarrow 2+n$ with multiple gluon radiation are implemented according to the high-twist formula \cite{Guo:2000nz,Wang:2001ifa} for induced gluon emission that contains LPM interference according to the formation time and the distance between the last scattering with gluon emission. During each scattering, the initial thermal parton recorded as ``negative'' partons are also allowed to propagate in the medium according to the Boltzmann equation. The energy and momentum of these ``negative" partons should be subtracted from all final observables to account for the back-reaction in the Boltzmann transport equations. They are part of the jet-induced medium response and manifest as the diffusion wake behind the propagating jet shower partons \cite{Wang:2013cia,Li:2010ts,He:2015pra}. 

In the LBT model, the densities of jet shower, recoil and ``negative'' partons are assumed to be small ($\delta f\ll f$) so that one can neglect interaction among them. One considers only interactions between jet shower and recoil partons with medium partons. The thermal medium evolves independently according to a hydrodynamic model that provides spatial and time profile of the local temperature and fluid velocity. Such a linear approximation breaks down when recoil and ``negative'' parton density becomes comparable to the local thermal parton density.  To extend LBT beyond the linear approximation, a coupled LBT and hydrodynamic (CoLBT-hydro) model \cite{Chen:2017zte} was developed. In CoLBT, soft partons from LBT are fed back to the bulk medium as a source term in the hydrodynamic equations while more energetic partons propagate simultaneously with the hydrodynamic evolution of the medium with the source term updated in real time.  A hybrid fragmentation-recombination model \cite{Zhao:2021vmu} for hadronization is also implemented in CoLBT.

\item {\it Hybrid Model}\cite{Casalderrey-Solana:2014bpa}: 
In contrast to the other models on this list, the Hybrid model aims at implementing a picture of jet quenching that is inspired by holographic calculations in a strongly coupled QGP rather than being inspired by the pQCD framework. The initial jet production and vacuum showering in this hybrid model follows that of PYTHIA,  the interactions between the shower partons and the medium is modeled after a holographic calculation \cite{Chesler:2014jva} of the
energy loss of energetic partons in a strongly coupled plasma that is parametrized by a stopping distance. The final partons with reduced energy hadronize according to the Lund string model in PYTHIA. The lost energy from jet shower partons is then fed back into the medium as completely thermalized and the contribution to the final hadron spectra is calculated from the freeze-out surface of the fire-ball, analogous to the jet-induced medium response (recoils and ``negative'' partons) from the jet transport models.

\item {\it JETSCAPE} \cite{Putschke:2019yrg} is not a single Monte Carlo model but rather a framework for integrating different components of initial jet production, jet showering in medium, hydrodynamic evolution of QGP medium and hadronization into a Monte Carlo package for simulating jet transport in heavy-ion collisions. Nevertheless, the most used variant of the package for phenomenological studies of jet quenching by the JETSCAPE Collaboration is based on 
Modular-All-Twist-Transverse-and-Elastic-
scattering-induced-Radiation (MATTER) \cite{Majumder:2013re} for the initial medium modified jet showering at the initial large scale, LBT for later parton transport through QGP medium described by a hydrodynamical model for evolution and a modified string fragmentation for hadronization.

\item {\it JetMed}~\cite{Caucal:2018ofz,Caucal:2020xad,Caucal:2020uic} implements a medium-modified parton shower based on a factorized picture~\cite{Caucal:2018dla} of vacuum-like and medium-induced parton splittings.  Vacuum emissions are generated by an angular-ordered shower  starting from $\theta_{\rm max} = 1$, with a minimal transverse momentum cut-off $k_{\perp, {\rm min}} = 0.25$ GeV. In the region $k_\perp^3 \theta < 2 \hat{q}$, medium-induced emissions are dominant and vacuum emissions are hence vetoed. Emissions with $\theta < \theta_c  = 2 /\sqrt{\hat{q} L^2}$ lose energy coherently with their emitter and are treated as if they were happening outside the medium. Medium-induced emissions are then generated for each parton produced in vacuum emissions. To this end, collinear splittings are ordered in an emission time $0 < t < L$ and computed with multiple-soft splitting kernels. Partons produced both from the vacuum-like and medium-induced splittings acquire transverse momentum according to Brownian motion while traversing the medium. For each parton that leaves the medium, vacuum-like emissions outside the medium are restarted with $\theta_{\rm max}$. That the first emission outside the medium does not obey angular ordering is a consequence of the loss of coherence in the medium. Emissions with $k_{\perp} \theta > 2/L$ or $\theta > \theta_c$ are vetoed. The generated partons are then passed to FastJet.

\end{enumerate}

These Monte Carlo models have been used to study many aspects of jet quenching in heavy-ion collisions, especially cone-size dependence of jet suppression, substructures, jet splitting of groomed jets, angular or energy-energy correlators and jet-induced medium response. These studies are essential to understand the underlying jet-medium interaction and properties of the QGP medium itself.

\subsection{The qualitative tale of jet suppression}
In this subsection, we aim at highlighting in a simple qualitative narrative the main physics picture that has emerged from measurements of jet suppression at the LHC. Many aspects of this picture have been corroborated in jet quenching Monte Carlo studies. 
    
As jet partons undergo both vacuum and medium-induced splittings during their propagation in the QGP medium, some large angle jet shower partons and radiated gluons will fall out of the jet cone, leading to a reduction of the final reconstructed jet energy within the jet cone. One of the consequences of this reduced energy inside the jet cone due to jet medium interaction is the increased dijet asymmetry, the energy difference between the leading and sub-leading jet in dijet events. Because of trigger-bias as in the high-$p_T$ dihadron production, the jet-pair is usually produced close to the surface of the QGP fireball and the leading (triggered) jet escapes without much energy loss. The sub-leading jet, on the other hand, has to go through the whole volume of the hot matter and loses a significant amount of energy, leading to an increased dijet asymmetry as compared to in p+p collisions. This was the first jet measurement at LHC by the ATLAS Collaboration \cite{ATLAS:2010isq}. Furthermore, the analysis of jet-track correlation by the CMS Collaboration \cite{CMS:2011iwn} showed that the increased dijet asymmetry or energy loss of the sub-leading jet is accompanied by energy flow to large angles carried by soft tracks.  

 \begin{figure}[t]
    \centering
    \includegraphics[width=0.51\textwidth]{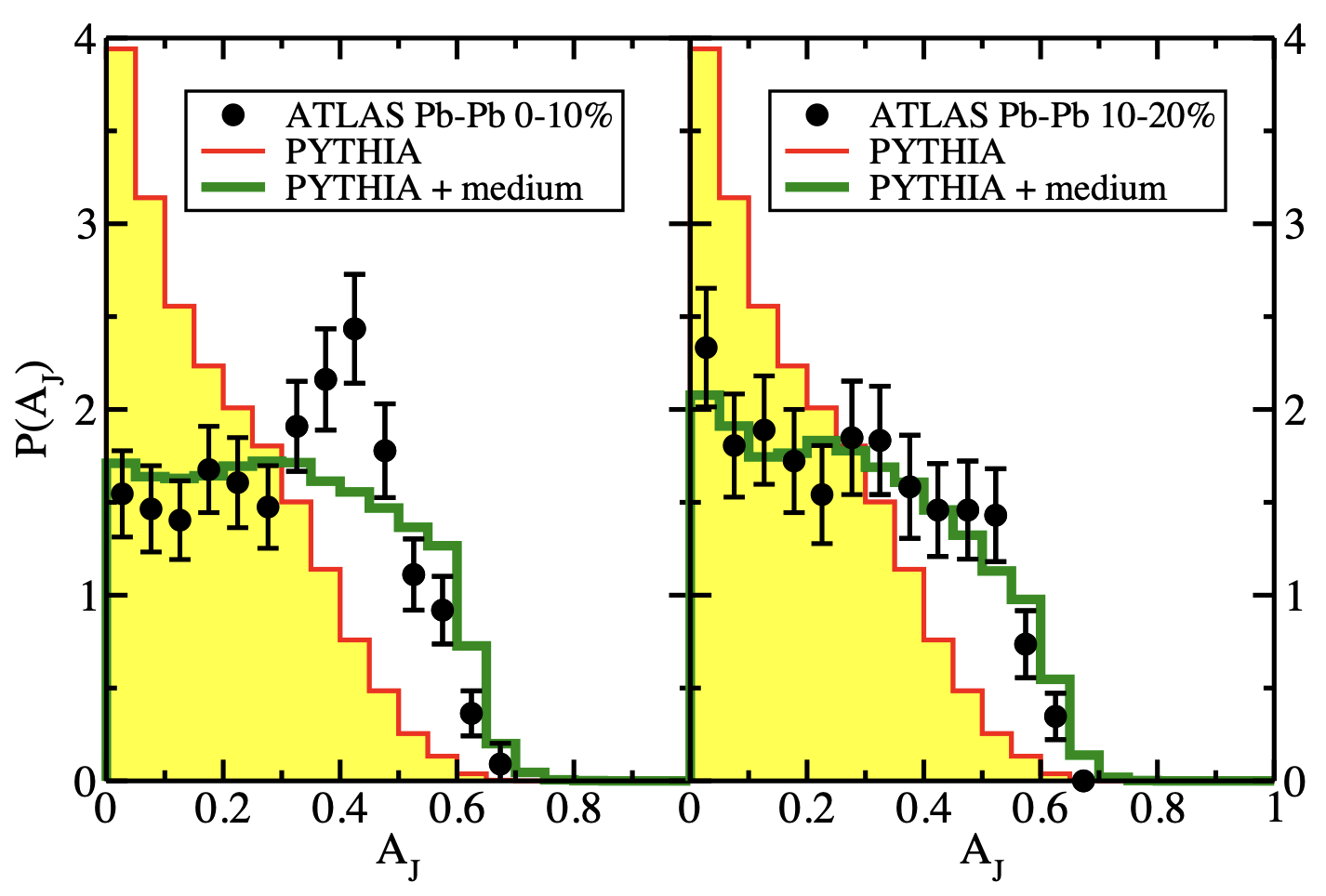}
    \includegraphics[width=0.45\textwidth]{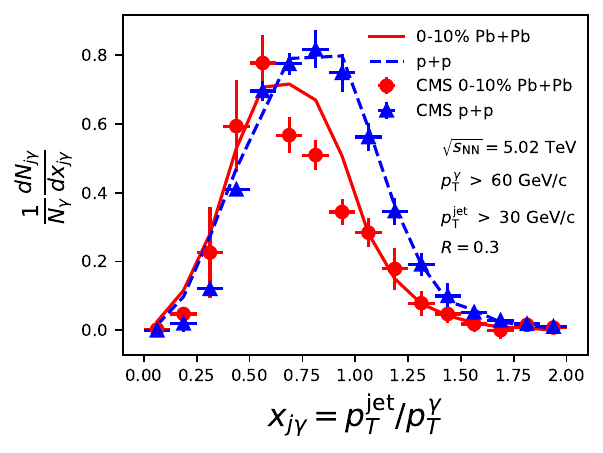}
    \caption{(\textit{Left}) Dijet and (\textit{Right}) $\gamma$-jet asymmetry distributions in Pb+Pb collisions at the LHC energies from ATLAS \cite{ATLAS:2010isq} and CMS \cite{CMS:2012ytf} experiment as compared to theoretical calculations. Figures from Refs.~\cite{Qin:2010mn,Yang:2022nei}.}
    \label{fig:dijet}
    \end{figure}
    
Similarly, in $\gamma/Z^0$-jet events, experiments at LHC also observed the increased $\gamma/Z^0$-jet asymmetry due to jet energy loss \cite{CMS:2012ytf,CMS:2017eqd,ATLAS:2018dgb}.  Suppression of $\gamma$-triggered jets was also recently measured in Au+Au collisions at RHIC \cite{STAR:2023pal,STAR:2023ksv}. These increased asymmetries and  suppression of $\gamma/Z^0$-jets can be described by radiative parton energy loss \cite{Qin:2010mn,Young:2011qx,He:2011pd,Dai:2012am,Wang:2013cia,Luo:2018pto,Zhang:2018urd,Yang:2022nei} as shown in Fig.~\ref{fig:dijet}, only when the dissipation of soft gluons to large angles through elastic scattering or thermalization is considered. Furthermore, dijet and $\gamma/Z^0$-jet asymmetries are also influenced by the fluctuation of jet energy loss \cite{Milhano:2015mng,Escobedo:2016vba}.

\begin{figure}[t]
    \centering
    \includegraphics[width=0.60\textwidth]{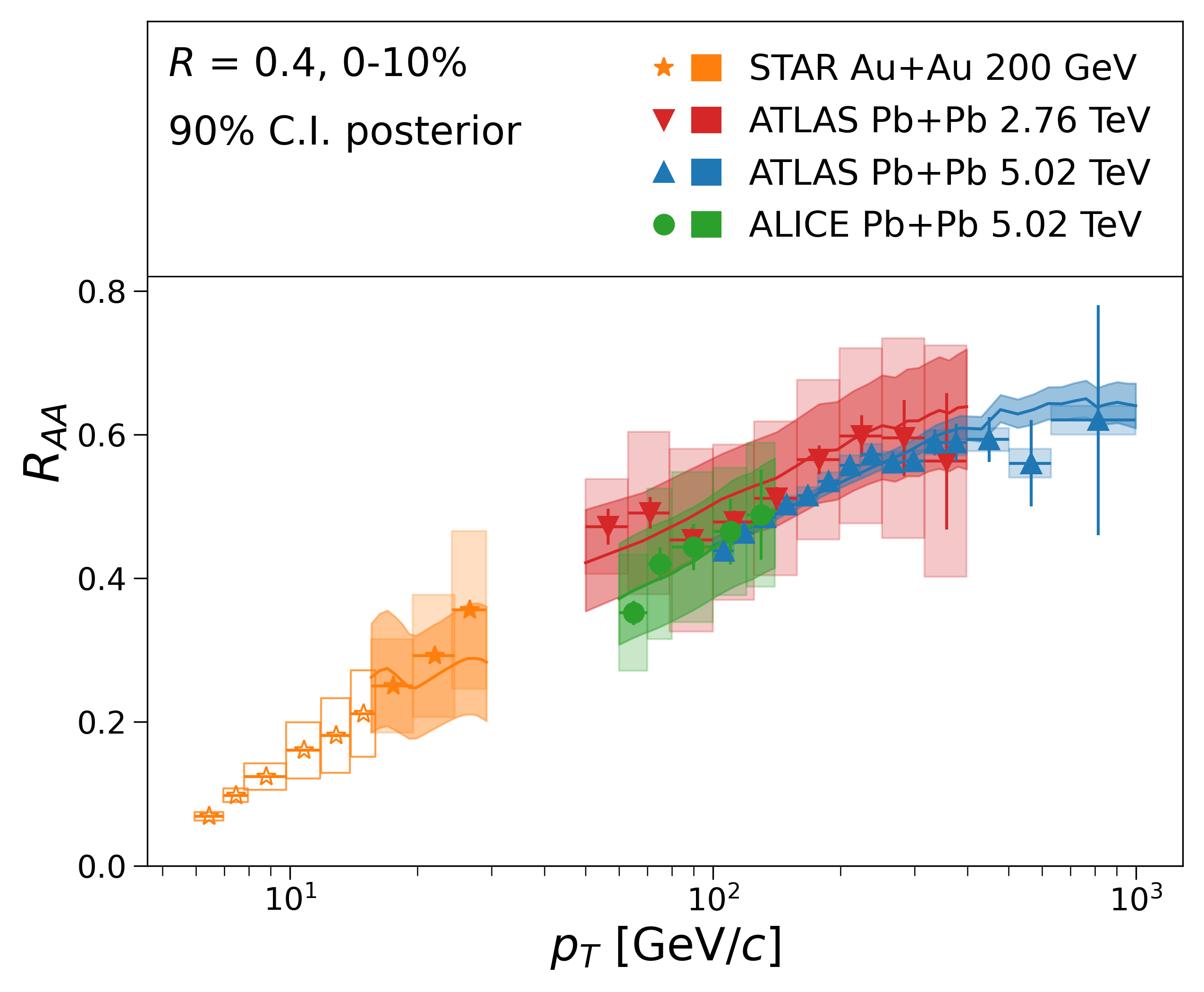}
    \caption{Nuclear modification factors $R_{AA}$ for single inclusive jet spectra (full jets by ATLAS and charged jets by ALICE and STAR) in central heavy-ion collisions at LHC \cite{ATLAS:2014ipv,ATLAS:2018gwx,ALICE:2019qyj} and RHIC \cite{STAR:2020xiv} energies. The lines with error bands are Bayesian fits according to a simple jet energy loss model. Figure from Ref.~\cite{Wu:2023azi}. }
    \label{fig:eloss}
    \end{figure}
    
Jet energy loss due to dissipation of energy to the outside of jet cone during the jet medium interaction also causes the suppression of single inclusive jet spectra in heavy-ion collisions. This suppression has been observed by all experiments~\cite{ATLAS:2012tjt,ALICE:2013dpt,ATLAS:2014ipv,ALICE:2015mjv,ATLAS:2018gwx,CMS:2016uxf,ALICE:2019qyj} (ALICE, ATLAS and CMS) at LHC as well as at RHIC~\cite{STAR:2020xiv} (STAR) lately. Shown in Fig.~\ref{fig:eloss} is the nuclear modification factor $R_{AA}$ for single inclusive jet spectra as a function of jet $p_T$ at different colliding energies. For a given centrality of the nucleus-nucleus collisions, the modification follows a seemingly common $p_T$ dependence within the uncertainties of the data for all colliding energies. There are many interplaying mechanisms \cite{He:2018xjv} behind the $p_T$ dependence and it is not necessarily independent of the colliding energy. It depends on the single inclusive jet spectra of the initial jet production in p+p collisions and the energy dependence of the effective jet energy loss. According to a Bayesian fit to a simple model of jet energy loss \cite{Wu:2023azi} as shown in Fig.~\ref{fig:eloss}, the effective jet energy loss 
\begin{equation}
    \Delta E/\rho \propto L^{0.59} p_T^{0.13}\ln p_T
\end{equation}
scaled by the parton density $\rho \sim dN_{ch}/d\eta/L^2$ depends on both the system size $L\sim \langle N_{\rm part}\rangle^{1/3}$ and the jet energy which is slightly stronger than a logarithmic dependence. Note that the $L^2$-dependence of energy loss by a single parton does not apply to the jet energy loss here since a jet is a collection of partons and some of the energy loss by the leading parton remains inside the jet cone.
Because of the system size or path length dependence of the effective jet energy loss, the single inclusive jet spectra should be anisotropic in azimuthal angle with respect to the event plane~\cite{He:2022evt,Mehtar-Tani:2024jtd} as observed by ATLAS~\cite{ATLAS:2013ssy,ATLAS:2021ktw} and CMS experiment~\cite{CMS:2022nsv}. 

\subsection{Coherence, decoherence and elastic recoil}

Unlike the suppression of high-$p_T$ light hadron spectra which is caused by radiative parton energy loss from leading partons described by quenching weights [Eq.~\eqref{eq:weight}], the suppression of jet spectra and the effective jet energy loss through dissipation of energy outside of the jet cone are caused by a broader range of physical mechanisms, including in particular medium-induced large angle gluon radiation and elastic scattering of jet shower partons. In this case, the angular distribution of medium-induced gluon bremsstrahlung becomes important. Equally relevant is the non-perturbative nature of parton thermalization and dissipation which can be better described by hydrodynamics.

Because of quantum coherence, parton showering in vacuum is angularly ordered \cite{Mueller:1981ex,Bassetto:1982ma} which limits phase space available for parton splitting and leads to a strong suppression of soft-gluon emissions. For medium induced gluon emissions, the LPM interference suppresses gluon bremsstrahlung whose formation time is longer than typical mean-free path and the medium size. Since the formation time depends on the relative angle between the leading parton, radiated gluon and the medium transverse momentum transfer, LPM interference suppresses small angle gluon bremsstrahlung. Furthermore, the angular ordering of the off-springs of two emitters can also be influenced by color rotation associated with multiple interaction with the medium \cite{Mehtar-Tani:2011hma,Mehtar-Tani:2012mfa}, leading to a gradual decoherence of the antenna-like radiation. Before this occurs, partons within an angular scale radiate coherently as a single emitter \cite{Casalderrey-Solana:2012evi}. When multiple branchings are taken into account for soft partons, medium-induced branchings are shown to be quasi-democratic, transporting energy to the large angle in the medium like a wave turbulence with a scaling spectrum \cite{Blaizot:2013hx}. Before these large angle emissions (formation time) and after jet partons exit the medium, the jet is argued to branch like in the vacuum \cite{Caucal:2018dla}. 

Furthermore, both leading and emitted partons in a jet will go through multiple elastic scatterings which will change the final parton angular distributions and drive the soft jet partons toward thermal equilibrium \cite{Mehtar-Tani:2022zwf}. Furthermore, the thermal recoil partons from these elastic scatterings that are correlated with the jet partons can become a part of the jet, contributing to the total jet energy within a jet cone \cite{He:2015pra,He:2018xjv,Luo:2023nsi}.

These interplays between color coherence and decoherence, momentum broadening, parton thermalization and contributions of the elastic thermal recoil partons should all influence the angular distribution of the emitted partons with respect to the originating jet parton and lead to modification of the substructures of jets such as the jet shape, fragmentation functions, cone size dependence of the jet suppression and energy-energy correlators. Numerical Monte Carlo simulations of jet transport in a realistic evolving QGP medium become necessary to study the interplay of these effects.

\begin{figure}[t]
    \centering
    \includegraphics[width=0.48\textwidth]{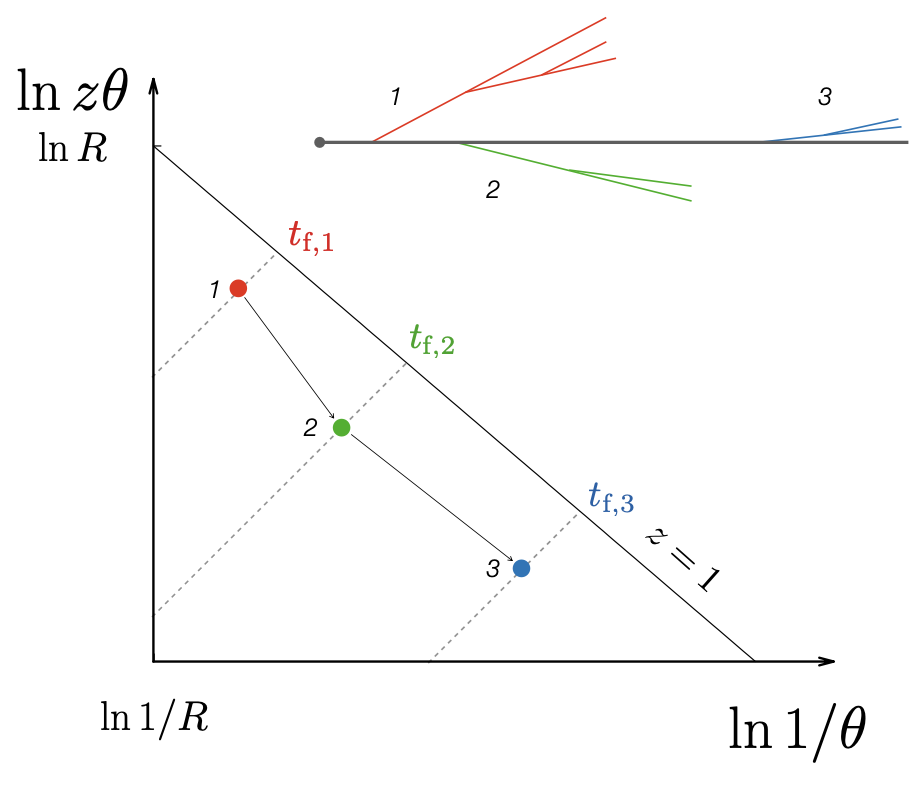}
    \includegraphics[width=0.48\textwidth]{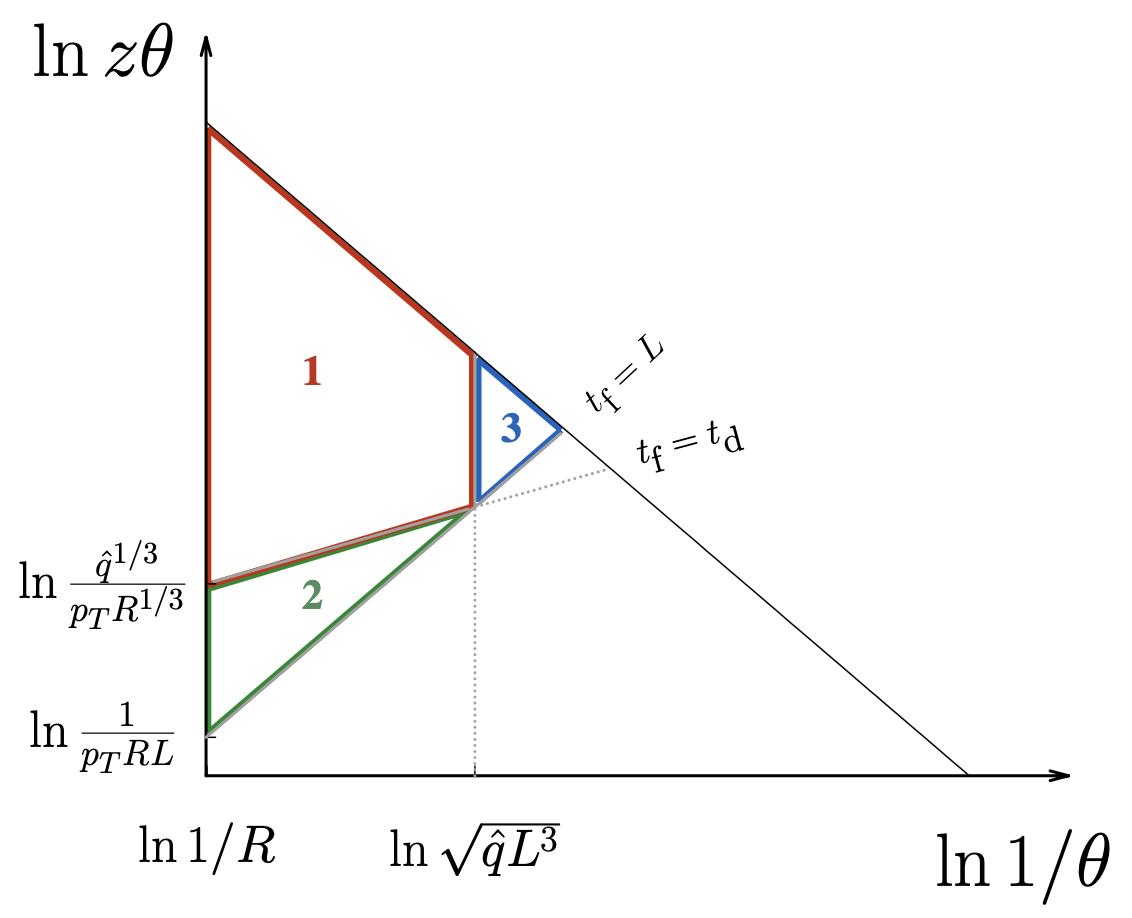}
    \caption{ (\textit{Left}) A schematic picture of the kinematical Lund plane spanned by the logs of the transverse opening angle $\theta$ and the longitudinal momentum fraction $z$. Embedded is a representative jet clustering history with the formation time of primary emissions.  
    (\textit{Right}) Characteristic parametric in-medium scales that separate the Lund plane into regions in which vacuum emissions dominate (region 1), medium-induced splittings dominate (region 2)  emissions that occur within the medium but are never resolved by the medium (region 3) and emissions that occur outside the medium (rest). Figure from \cite{Andrews:2018jcm}.  }
    \label{fig:lund}
    \end{figure}

There have been community-wide efforts to understand how this rich physics implemented in different Monte Carlo simulations fills longitudinal and transverse phase space~\cite{Andrews:2018jcm}. Of particular use for this purpose is to study the distribution of parton splittings in the Lund plane~\cite{Andersson:1988gp}, see Fig.~\ref{fig:lund}. In practice, for any sample of real or simulated jets, one can build a history of splittings by reclustering a jet with a given clustering algorithm. For each $1\to 2$ branching, one can then extract the longitudinal momentum sharing $z$ and the transverse opening angle $\theta$, and one enters the splitting in two-dimensional histogram spanned by the Lund plane $\ln z\theta \times \ln 1/\theta$. In general, in this Lund plane, hard collinear splittings are placed close to the diagonal $z=1$ line and soft large-angle radiation enters close to the $\ln z\theta$-axis.  As seen in Fig.~\ref{fig:lund}, the main parametric ideas about the space-time embedding of medium-modified parton showers in a finite size QGP, about the capability of splittings to resolve the medium and about the interplay between vacuum and in-medium emission are expected to dominate characteristically different regions of the Lund plane. One can then ask to what extent the MC simulations of jet transport discussed in Sec.~\ref{MCtools} differ in the way they fill the Lund plane, or to what extent the physics of medium-modified parton splittings remains experimentally accessible above the soft high-multiplicity background that affects the reclustering history~\cite{Andrews:2018jcm}. Here, we restrict ourselves to highlighting the basic use of this Lund plane representation of parton showers without discussing some of the many existing Monte Carlo simulations in which, for instance,  recoil effects leave clearly visible distinct traces~\cite{Andrews:2018jcm,Chien:2018dfn}.

\subsection{Effects of medium response}

In addition to the medium-induced multiple gluon emissions, jet substructures in heavy-ion collisions are also influenced by recoil partons and the ``negative'' partons due to the back-reaction, which are collectively denoted as medium response. 

\begin{figure}[t]
    \centering
    \includegraphics[width=0.50\textwidth]{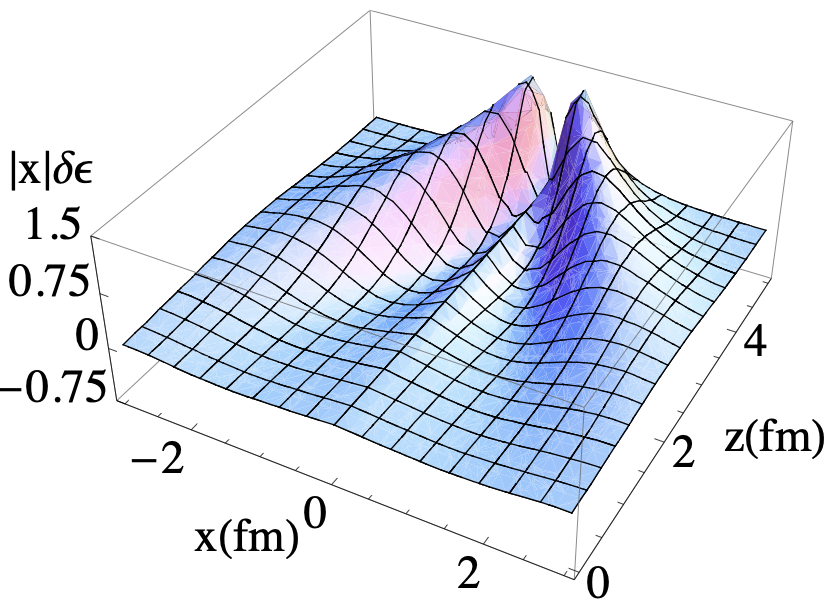}
    \caption{ The energy density of the linear fluid dynamical response to the energy deposited by a quark-initiated shower. Figure from Ref.~\cite{Qin:2009uh}.}
    \label{fig:machcone}
    \end{figure}

Jet-induced medium responses can arise in the form of Mach-cone excitations by massless partons that travel at the speed of light in the QGP medium whose velocity of sound $c_s<1$ is smaller than that of the propagating jets \cite{Stoecker:2004qu,Casalderrey-Solana:2004fdk,Ruppert:2005uz}.  Such Mach-cone excitations in general consist of the wake front at an angle $\theta_M=\arccos c_s$ relative to the propagating parton and a diffusion wake in the opposition direction as shown by both hydrodynamic simulations and linear response calculations that take into account of the energy deposition from both elastic scattering and induced gluon emission \cite{Qin:2009uh} as shown in Fig.~\ref{fig:machcone}. In the numerical solution of single parton propagation in a strongly coupled plasma as described by a gravity dual in AdS/CFT, similar Mach cone, both the wake front and diffusion wake, can be generated as shown in Fig.~\ref{fig:ads} (\textit{Right}). The Mach cone excitation is therefore a very general and generic feature of energy propagation excited by a supersonic projectile in any type of medium.

\begin{figure}[t]
    \centering
    \includegraphics[width=0.40\textwidth]{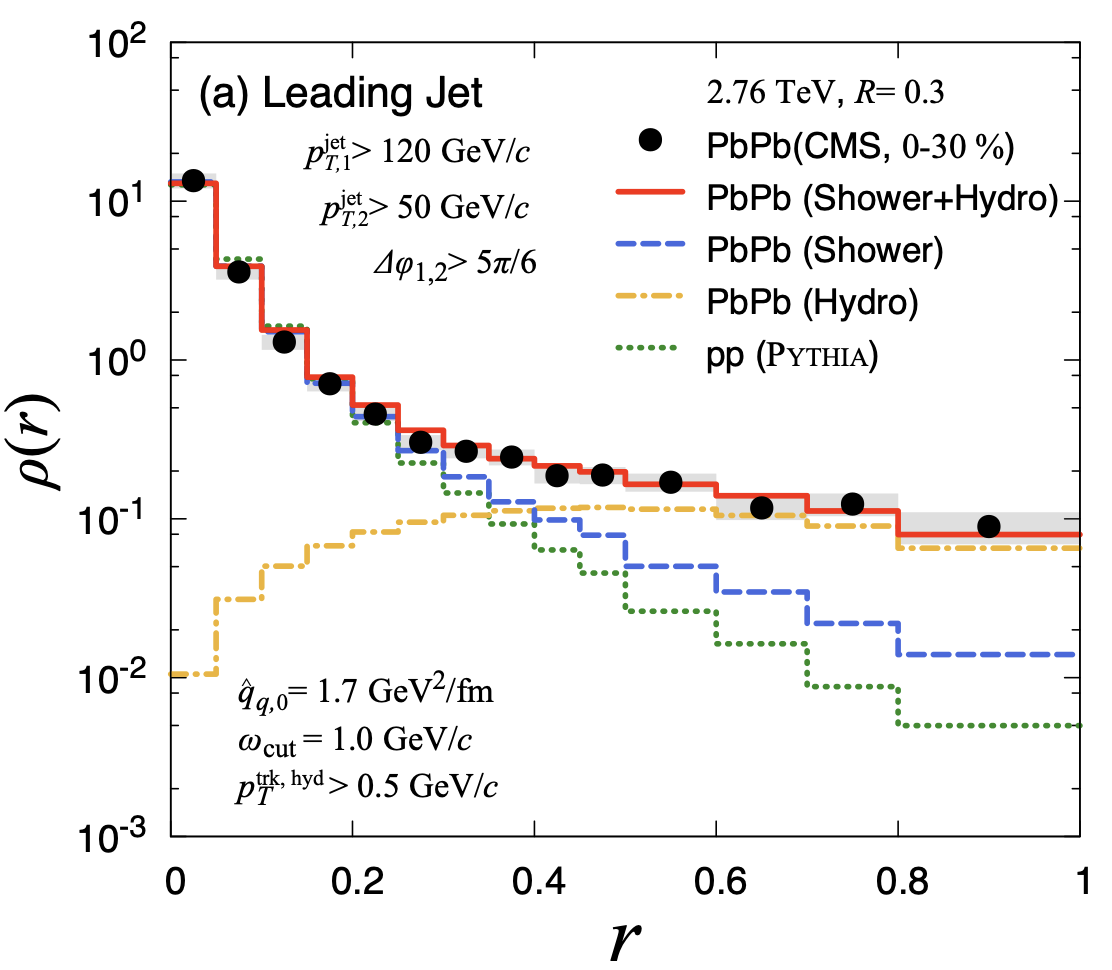}
    \includegraphics[width=0.55\textwidth]{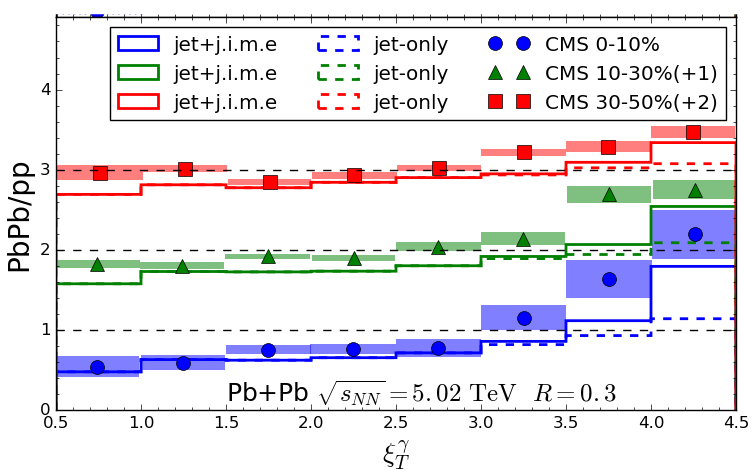}
    \caption{ (\textit{Left}) The jet shape function relative to the direction of a leading jet in dijet events in 0-30\% central Pb+Pb collisions at $\sqrt{s}=2.76$ TeV from the jet-fluid model calculation and CMS experiment~\cite{CMS:2013lhm}. (\textit{Right}) The ratio between charged fragmentation functions of $\gamma$-jets in Pb+Pb and p+p collisions at $\sqrt{s}=5.02$ TeV as a function of $\xi_T^\gamma=\ln{(p_T^\gamma/p_T)}$ from CoLBT calculations and CMS experiment~\cite{CMS:2018mqn}. Figures from Refs.~\cite{Tachibana:2017syd,Chen:2020tbl}. }
    \label{fig:rho-ff}
    \end{figure}

The early interests in jet-induced Mach-cones were mainly inspired by the experimental data on dihadron azimuthal angle correlations that showed an intriguing double-peak structure \cite{PHENIX:2005zfm} which later turned out to be the consequence of the triangular flow \cite{Alver:2010gr} due to the geometrical fluctuations of the initial dense matter in heavy-ion collisions. 
While unambiguous signals of the Mach-cone excitation are still elusive in both experimental measurements and simulations with realistic hydrodynamic evolution of the medium \cite{Betz:2010qh}, its effects, which cannot be distinguished from that of medium-induced gluon bremsstrahlung,  on jet substructures have been investigated extensively, see a recent review \cite{Cao:2020wlm} and the references therein. Hadrons from the Mach-cone excitation are found to be at large angles relative to the jet direction. They lead to significant enhancement of the jet shape function at large radial distance $r=[(\phi-\phi_{\rm jet})^2+(\eta-\eta_{\rm jet})^2]^{1/2}$ as shown in Fig.~\ref{fig:rho-ff}~(\textit{Left}) from a jet-fluid model calculation \cite{Tachibana:2017syd}. Jet-track correlations in dijet events at LHC~\cite{CMS:2011iwn} also show that the lost energy by the sub-leading jet flows to large angles carried by soft tracks. Since these hadrons from the jet-induced medium response are in general from a thermal source, they are mostly soft hadrons at large angles relative to the jet. They therefore lead to enhancement of the effective fragmentation functions at small longitudinal momentum fraction $z_h=p_T/p_T^{\rm jet}$ or $z_T^\gamma=p_T/p_T^\gamma$ in $\gamma$-jets as shown by the CMS experimental data \cite{CMS:2018mqn} in Fig.~\ref{fig:rho-ff}~(\textit{Right}) when compared to the CoLBT calculations \cite{Chen:2020tbl}. The medium response is also shown to be responsible for enhancement of energy-energy correlators at large angles \cite{Yang:2023dwc,Yang:2023dwc}.

\begin{figure}[t]
    \centering
    \includegraphics[width=0.38\textwidth]{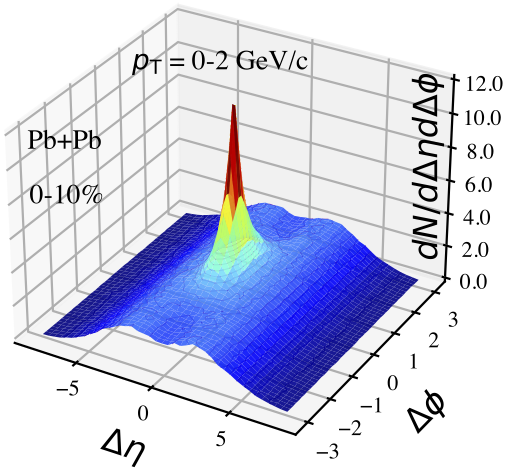}
    \includegraphics[width=0.28\textwidth]{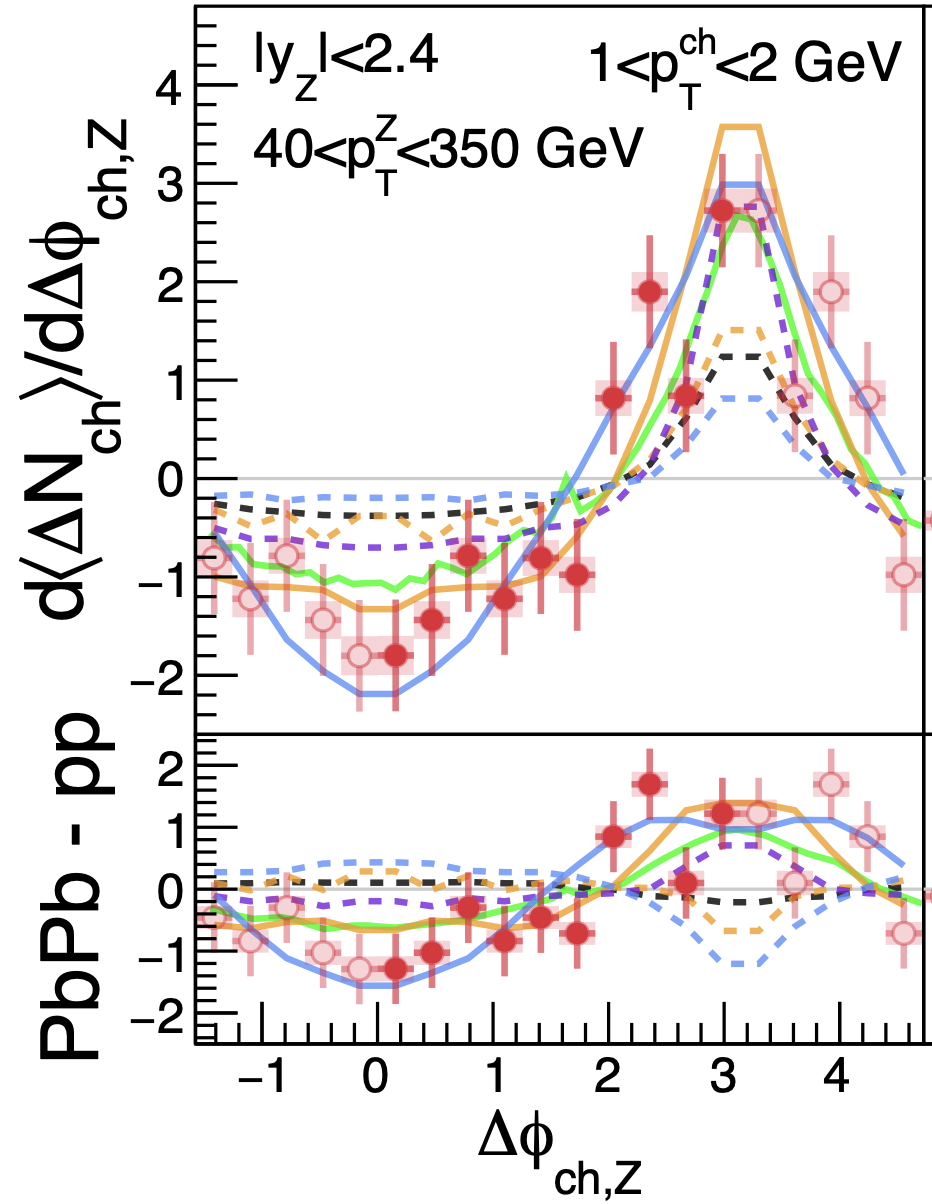}\hspace{-2pt}
     \includegraphics[width=0.28\textwidth]{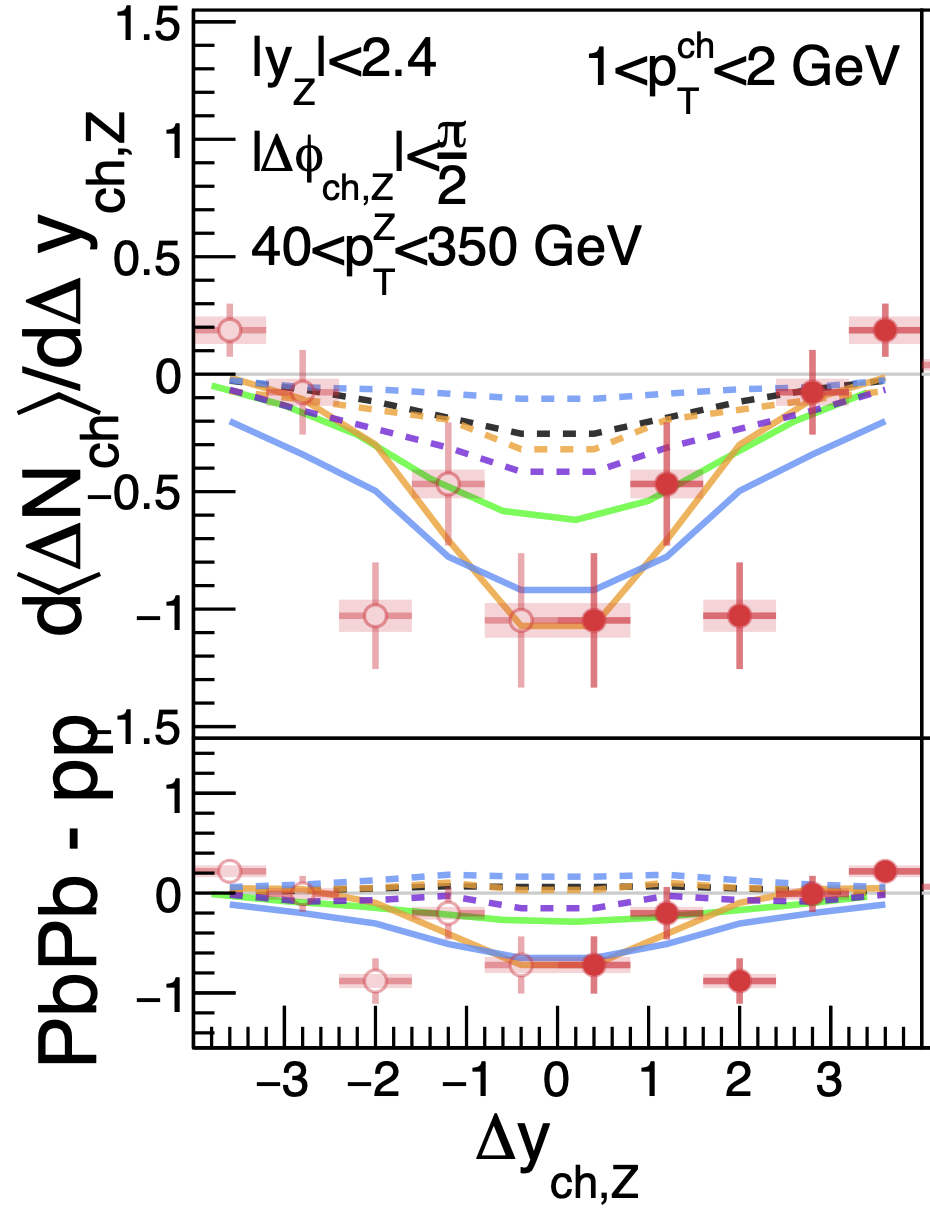}
    \caption{(\textit{Left}) Jet-hadron ($p_T<2$ GeV/$c$) correlation in rapidity $\eta$ and azimuthal angle $\phi$ in Pb+Pb $\gamma$-jet events from CoLBT simulations. (\textit{Right}) The $Z$-hadron correlation  as a function of azimuthal angle and rapidity in (upper) Pb+Pb collisions and (lower) the difference between Pb+Pb and p+p collisions from CMS experiments \cite{CMS:2025dua} as compared to model calculations. Figures adopted from Refs~\cite{Yang:2022nei,CMS:2025dua}.}
    \label{fig:dfw}
    \end{figure}

The enhancement of soft hadrons inside jets at large angles, however, is not unique since the hadronization of the emitted gluons induced by jet-medium interaction also contribute to the same kind of medium modifications of the jet shape and fragmentation functions \cite{Blaizot:2013hx}. It therefore cannot be considered as an unambiguous signal of the jet-induced Mach-cone type medium response. On the other hand, the diffusion wake accompanying the Mach-cone has been shown to lead to a depletion of soft hadrons in the opposite direction of the propagating jet \cite{Chen:2017zte,Pablos:2019ngg,Yang:2021qtl}. This signal will be unique since there is no other competing mechanisms that can deplete hadron yield in the back direction of the propagating jet. Since jets, the induced Mach-cone and diffusion wake are all 3D objects in space, the 2-dimensional jet-hadron correlation in both azimuthal angle $\phi$ and rapidity $\eta$ can be used to image the depletion of soft hadrons due to the diffusion wake. As shown in Fig.~\ref{fig:dfw}~(\textit{Left}) from CoLBT simulations \cite{Yang:2022nei}, the jet-hadron correlation as a function of rapidity and azimuthal angle in $\gamma$-jets in Pb+Pb collisions has a valley structure caused by the diffusion wake on top of a ridge from the initial multiple parton interaction in the $\gamma$ direction (the opposite direction of the jet). Such a valley-on-ridge structure is absent in p+p collisions and can be considered as an unambiguous signal of the diffusion wake. 

In the most recent analysis of $Z-$hadron correlation $dN_{\rm ch}/d\Delta\eta d\Delta\phi$ in $Z$-jet events in Pb+Pb collisions by the CMS Collaboration \cite{CMS:2025dua} the predicted valley structure is indeed observed as shown in Fig.~\ref{fig:dfw} (middle and right). Recently, Yang and Wang \cite{Yang:2025dqu,Yang:2025xni} propose to measure the difference between jet-hadron correlations in events with different jet rapidities while the rapidity of the trigger remains the same. This difference, defined as the rapidity asymmetry, is background free and allows more precision measurements of the diffusion wake which can provide new constraints on the QGP properties such as the transport coefficients and EoS.

\section{Concluding discussion and outlook}

\subsection{A summary of experimental results}

Since its inception about 40 years ago, jet quenching has been developed from a theoretical concept into a fledging area of research in the early years and now a powerful and advanced tool for the study of properties of the QGP in high-energy heavy-ion collisions. We know now that both high-$p_T$ hadron and jet spectra are strongly suppressed in heavy-ion collisions at both RHIC and LHC energies due to jet-medium interactions. We know that the jet transport parameter $\hat q$ in the initial stage of the QGP formed in heavy-ion collisions is about two orders of magnitude higher than in cold nuclei. Jet substructures are modified due to interactions between the medium and jet-shower partons and thermal recoil partons, transferring jet energy to large angles. Because of the modification of jet substructures, the suppression of single and dijet spectra depends on the jet cone-size. For the first time recently we see direct evidence of the jet-induced diffusion wake in jet-hadron correlations. This review has focused on theoretical developments and an extensive review of all experimental data on jet quenching lies outside its scope. Instead, we will list here in chronological order some (not exhaustive) major experimental results related to what we have discussed in this review:

 \begin{itemize}
     \item[{\bf 1998}] {\hspace{0.7cm}}
     $d\sigma_{\pi^0}/d\eta d^2p_T$ $\pi^0$, S+S and Pb+Pb at $\sqrt{s}=20, 17.8$ GeV, no suppression observed, WA80\cite{WA80:1998xbn}, WA98\cite{WA98:1998psk}. 
     
     \item[{\bf 2001}] {\hspace{0.7cm}}
     $R_{AA}^h$ for single inclusive charged hadrons and $\pi^0$, Au+Au at $\sqrt{s}=130$ GeV, strong suppression, PHENIX \cite{PHENIX:2001hpc} and STAR (2002)\cite{STAR:2002ggv}.
         
    \item[{\bf 2002}] {\hspace{0.7cm}} $p({\bar p})/\pi^\pm, K^\pm/\pi^\pm$, Au+Au at $\sqrt{s}=130$ GeV, enhancement of $p/\pi^\pm, K^\pm/\pi^\pm$ at high $p_T$, PHENIX\cite{PHENIX:2001vgc}.
    
     \item[{\bf 2003}] {\hspace{0.7cm}} $R_{AA}^h$ for single inclusive charged hadrons and $\pi^0$, Au+Au at $\sqrt{s}=200$ GeV, strong suppression, PHENIX \cite{PHENIX:2003qdj} and STAR\cite{STAR:2003fka}.
     
     \item[{\bf 2003}] {\hspace{0.7cm}} $I_{AA}^{hh}$ modification factor for dihadron correlation, Au+Au at $\sqrt{s}=200$ GeV, disappearance of back-to-back dihadrons, STAR \cite{STAR:2002svs}.
     
     \item[{\bf 2003}] {\hspace{0.7cm}} $R_{dAu}^h$ for single inclusive hadrons, d+Au at $\sqrt{s}=200$ GeV, no suppression, PHENIX \cite{PHENIX:2003qdw}, STAR\cite{STAR:2003pjh}, PHBOS~\cite{PHOBOS:2003uzz}, BRAHMS (2004)~\cite{BRAHMS:2004xry}.
     
     \item[{\bf 2003}] {\hspace{0.7cm}}  $v_2$ of single hadrons at high $p_T$, Au+Au at $\sqrt{s}=200$ GeV, large $v_2$ at high $p_T$, STAR \cite{STAR:2002pmf}.

     \item[{\bf 2005}] {\hspace{0.7cm}} $R_{AA}^\gamma$ modification factor for high-$p_T$ direct photons, Au+Au at $\sqrt{s}=200$ GeV, direct photons are not suppressed in contrast to hadrons, PHENIX\cite{PHENIX:2005yls}.
     
    \item[{\bf 2006}] {\hspace{0.7cm}} $R_{AA}^e$ large $p_T$ leptons from charm semi-leptonic decay, Au+Au at $\sqrt{s}=200$ GeV, suppression of high-$p_T$ charm mesons, PHENIX \cite{PHENIX:2005nhb} and STAR \cite{STAR:2006btx}.
    
     \item[{\bf 2009}] {\hspace{0.7cm}} $I_{AA}^{\gamma h}$ modification factor for $\gamma$-hadron spectra, Au+Au at $\sqrt{s}=200$ GeV, hadrons in $\gamma$-jets suppressed, PHENIX \cite{PHENIX:2009cvn}, STAR\cite{STAR:2009ojv},
     STAR\cite{STAR:2016jdz} (2016), PHENIX\cite{PHENIX:2020alr} (2020)
     
    \item[{\bf 2010}] {\hspace{0.7cm}} $A_{jj}$ dijet asymmetry, Pb+Pb at $\sqrt{s}=2.76$ TeV, large momentum asymmetry between leading and subleading jets, ATLAS \cite{ATLAS:2010isq}.

    \item[{\bf 2011}] {\hspace{0.7cm}} $A_{jj}$ and jet-track correlation, Pb+Pb at $\sqrt{s}=2.76$ TeV, increased momentum imbalance between leading and subleading jets and flow of energy to large angle by soft tracks, CMS \cite{CMS:2011iwn}.

    \item[{\bf 2012}] {\hspace{0.7cm}} $R_{pPb}^h$ for single inclusive hadrons, p+Pb at $\sqrt{s}=5.02$  TeV, no suppression at high $p_T$, ALICE \cite{ALICE:2012mj}, CMS\cite{CMS:2015ved} (2015), ATLAS\cite{ATLAS:2022kqu} (2022). 

    \item[{\bf 2012}] {\hspace{0.7cm}} $R_{AA}^D$ modification factors for $D^0,D^+$ and $D^{*+}$, Pb+Pb at $\sqrt{s}=2.76$  TeV, strong suppression, ALICE~\cite{ALICE:2012ab}.

    \item[{\bf 2013}] {\hspace{0.7cm}} $R_{AA}^{\rm jet}$ for single inclusive jets, Pb+Pb at $\sqrt{s}=2.76$ TeV, strong suppression, ATLAS\cite{ATLAS:2012tjt,ATLAS:2014ipv}, 
    ALICE ($R_{CP}^{\rm jet}$ charged jets) \cite{ALICE:2013dpt}, ALICE (charged jets 2015)\cite{ALICE:2015mjv},
    CMS\cite{CMS:2016uxf}(2017).
    
     \item[{\bf 2013}] {\hspace{0.7cm}} $R_{AA}^{h^\pm}$ for single inclusive hadrons, Pb+Pb at $\sqrt{s}=2.76$ TeV, Strong suppression, ALICE \cite{ALICE:2012aqc}.
     
     \item[{\bf 2013}] {\hspace{0.7cm}} $A_{\gamma j}$ $\gamma$-jet asymmetry, Pb+Pb at $\sqrt{s}=2.76$ TeV, increased $\gamma$-jet momentum inbalance, CMS\cite{CMS:2012ytf}.

     \item[{\bf 2013}] {\hspace{0.7cm}} $v_n^{\rm jet}$ azimuthal anisotropy of single inclusive jets, Pb+Pb at $\sqrt{s}=2.76$ TeV, sizable anisotropy observed, ATLAS \cite{ATLAS:2013ssy} (similar single jet anisotropy at $\sqrt{s}=5.02$ TeV \cite{ATLAS:2021ktw}). 

    \item[{\bf 2014}] {\hspace{0.7cm}} $R_{AA}^{D^0}$ modification factor of $D^0$ meson, Au+Au at $\sqrt{s}=200$ GeV, strong suppression of $D^0$ meson spectra, STAR~\cite{STAR:2014wif}, STAR (2019 HFT measurements)~\cite{STAR:2018zdy}.
    
     \item[{\bf 2014}] {\hspace{0.7cm}} $D(z)$ fragmentation function of single inclusive jets, Pb+Pb at $\sqrt{s}=2.76$ TeV, enhancement of hadrons at both low and large $z=p_T^h/p_T^{\rm jet}$, ATLAS \cite{ATLAS:2014dtd,ATLAS:2017nre}, CMS\cite{CMS:2014jjt} \footnote{The first measurement of $D(z)$ in single inclusive jets by CMS \cite{CMS:2012nro} did not observe significant modification.}.

     \item[{\bf 2014}] {\hspace{0.7cm}} $\rho(r)$ jet shape of single inclusive jets, Pb+Pb at $\sqrt{s}=2.76$ TeV, enhancement of $\rho(r)$ at large $r$, CMS\cite{CMS:2013lhm}.

  \item[{\bf 2014}] {\hspace{0.7cm}} $A_{jj}$ dijet asymmetry, p+Pb at $\sqrt{s}=5.02$ TeV, null result consistent with pQCD and nuclear modification factor CMS\cite{CMS:2014qvs},  
  ATLAS \cite{ATLAS:2019jgo} (2019).

     \item[{\bf 2015}] {\hspace{0.7cm}} $R_{pPb}^{\rm jet}$ modification factor for single inclusive jets, p+Pb at $\sqrt{s}=5.02$ TeV, no suppression, ATLAS \cite{ATLAS:2014cpa}, CMS \cite{CMS:2016svx} (2016).

    \item[{\bf 2015}] {\hspace{0.7cm}} $R_{AA}^{h^\pm}$ for single inclusive hadrons in large $p_T$ range, Pb+Pb at $\sqrt{s}=2.76$ TeV, suppression decreases at large $p_T$, ATLAS \cite{ATLAS:2015qmb}, CMS \cite{CMS:2016xef} (2017).

    \item[{\bf 2017}] {\hspace{0.7cm}} $v_2(D^0)$ elliptic anisotropy for high-$p_T$ $D^0$ mesons, Au+Au at $\sqrt{s}=200$ GeV, large $v_2$ observed, STAR~\cite{STAR:2017kkh}.
    
    \item[{\bf 2017}] {\hspace{0.7cm}} $A_{Zj}$ $Z$-jet asymmetry, Pb+Pb at $\sqrt{s}=5.02$ TeV, increased $Z$-jet momentum imbalance,  CMS \cite{CMS:2017eqd}.

    \item[{\bf 2017}] {\hspace{0.7cm}} $R_{AA}^B$ modification factor for $B$ mesons, Pb+Pb at $\sqrt{s}=$5.02 TeV, strong suppression, CMS\cite{CMS:2017uoy}, CMS (non-prompt $J/\psi$, non-prompt $D^0$ 2018)\cite{CMS:2017uuv,CMS:2018bwt}, ATLAS ($\mu$-decay 2022)\cite{ATLAS:2021xtw}, ALICE(non-prompt $D^0$ 2022)\cite{ALICE:2022tji}.

    \item[{\bf 2017}] {\hspace{0.7cm}} $z_g$-dependence of medium-modified splitting function via SoftDrop groomed jets, CMS\cite{CMS:2017qlm}.

    \item[{\bf 2018}]{\hspace{0.7cm}} $D(z)$ $\gamma$-jet fragmentation function, Pb+Pb at $\sqrt{s}=5.02$ TeV, suppression of leading hadrons and enhancement of soft hadrons, CMS\cite{CMS:2018mqn}, ATLAS\cite{ATLAS:2019dsv}(2019).

    \item[{\bf 2018}] {\hspace{0.7cm}} $R_{AA}^D$ modification factor for $D$ mesons, Pb+Pb at $\sqrt{s}=5.02$ TeV, strong suppression of $D$ mesons, CMS\cite{CMS:2017qjw}, ALICE\cite{ALICE:2018lyv}.

    \item[{\bf 2019}] {\hspace{0.7cm}} $R_{AA}^{\rm jet}$ for single inclusive jets, Pb+Pb at $\sqrt{s}=5.02$ TeV, suppression similar as at $\sqrt{s}=2.76$ TeV,
ATLAS\cite{ATLAS:2018gwx}.

    \item[{\bf 2019}] {\hspace{0.7cm}} $\rho(r)$ jet shape for $\gamma$-jet, Pb+Pb at $\sqrt{s}=5.02$ TeV, jet shape enhancement at large $r$, CMS\cite{CMS:2018jco}.
    
    \item[{\bf 2020}] {\hspace{0.7cm}} $R_{AA}^{\rm jet}$ for single inclusive jets, Au+Au at $\sqrt{s}=200$ GeV, strong suppression as single hadrons, STAR \cite{STAR:2020xiv}.

    \item[{\bf 2021}] {\hspace{0.7cm}} $I^{Z^0h}_{AA}(z)$ modification factor for $Z$-hadron correlation, Pb+Pb at $\sqrt{s}=5.02$ TeV, suppression of leading hadron at large $z=p_T^h/p_T^Z$ and enhancement of soft hadron at small z, ATLAS\cite{ATLAS:2020wmg}, CMS(2022)\cite{CMS:2021otx}.

    \item[{\bf 2022}] {\hspace{0.7cm}} $R(\theta)$ ratio of splitting functions between $D^0$-tagged jets and single inclusive jets, p+p at $\sqrt{s}=13$ TeV, observation of the deadcone effect, ALICE \cite{ALICE:2021aqk}.
    
    \item[{\bf 2022}] {\hspace{0.7cm}} $Z$-hadron azimuthal angle correlation, Pb+Pb at $\sqrt{s}=$5.02 TeV, enhancement of the pedestal effect due to multiple parton interaction (MPI), CMS\cite{CMS:2021otx}.

    \item[{\bf 2023}] {\hspace{0.7cm}}$R_{AA}^{\rm jet}$ for $\gamma$-tagged jets, Pb+Pb at $\sqrt{s}=5.02$ TeV, $\gamma$-tagged jets are less suppressed than single inclusive jets, ATLAS \cite{ATLAS:2023iad}.

    \item[{\bf 2024}] {\hspace{0.7cm}} Yield and acoplanarity of recoil jets from high-$p_T$ hadron trigger, Pb+Pb at $\sqrt{s}=5.02$ GeV, enhanced yield and acoplanarity of low $p_T$ recoil jets, ALICE\cite{ALICE:2023qve}.

    \item[{\bf 2025}] {\hspace{0.7cm}} $I_{AA}^{\gamma/\pi^0 {\rm jet}}$  modification factors for $\gamma/\pi^0$-jets, Au+Au at $\sqrt{s}=200$ GeV, suppression of $\gamma/\pi^0$-triggered jets, STAR \cite{STAR:2023pal,STAR:2023ksv}.

    \item[{\bf 2025}]{\hspace{0.7cm}}  $dN_{ch}/d\Delta\phi d\Delta\eta$ $Z$-hadron correlation in azimuthal and rapidity, Pb+Pb at $\sqrt{s}=$5.02 TeV, first direct evidence of jet-induced diffusion wake, CMS \cite{CMS:2025dua}.
    
    \item[{\bf 2025}] {\hspace{0.7cm}} (EEC) energy-energy correlator, Pb+Pb at $\sqrt{s}=$5.02 TeV, enhancement of EEC at large angles, CMS\cite{CMS:2025ydi}.
 \end{itemize}

\subsection{Outlook}
\subsubsection{Looking ahead to more data}
This review is written more than 40 years into studying jet quenching while more data are likely to come in the next decades. At the time of completing this work, the sPHENIX collaboration~\cite{Belmont:2023fau} at RHIC is in the middle of collecting the decisive data set of its high-luminosity Au+Au program, aiming at significantly extending the high-$p_T$ range of single inclusive hadron spectra, jet-like hadron correlations and calorimetrically reconstructed jets at RHIC. At the LHC, data from a few-days O+O run taken one month ago promise to advance our understanding of the onset of jet quenching in small collision systems~\cite{Grosse-Oetringhaus:2024bwr}. For the long-term future of the CERN HL-LHC heavy-ion program, the scientific community has just submitted its plans~\cite{EPPSU2025} of fully exploiting LHC as a heavy-ion collider throughout run 4 and run 5 up to the end of the LHC experimental program currently foreseen in the early 2040s. All LHC experiments foresee significant upgrades in tracking resolution and rate capabilities that will significantly extend our knowledge of the high-$p_T$ sector of heavy-ion collisions. 

Of particular interest is the study of heavy flavored hard probes that are greatly assisted by the increased tracking resolution, {\it e.g.} in ALICE 3~\cite{ALICE:2022wwr}. Since heavy flavor is conserved on QGP timescales, the propagation and fragmentation of heavy quarks offer a unique opportunity for understanding how quenching of a specific colored partonic test charge at high-$p_T$ is related to hydrodynamization, thermalization and recombination phenomena at intermediate and low $p_T$. 

The increase in integrated luminosity expected at the HL-LHC promises many other advances. For instance, we have by now ample experimental evidence that the electro-weak boson triggered jet measurements can provide truly golden opportunities for quantifying jet quenching effects. And yet, many of the more differential jet substructure measurements cannot be studied in these best-gauged observables so far since the integrated luminosity is too low to go differential. For this reason, the LHC heavy-ion physics community has used repeatedly the rate of $Z$-boson triggered jets as a benchmark to justify the integrated Pb+Pb luminosity requested throughout HL-LHC. 

Higher luminosity is of obvious interest for a large number of other jet quenching observables that become either newly available or that can be studied more differentially. For instance, with accumulating a sufficiently large number of $O(500)$ GeV jets and reclustering their branching histories, it becomes possible to study medium-modified jet evolution in the Lund diagram well beyond the level of the hardest parton splitting. Will this reveal new physics or will it ``only" corroborate the understanding summarized in this review? It is only by doing that we will know. The HL-LHC is likely to extend the above time-line with two more decades of exciting experimental insights. And in the second part of this century, nuclear beams at the future circular collider may extend these studies to a qualitatively novel regime~\cite{Chang:2015hqa,Dainese:2016gch}.

\subsubsection{Looking ahead to more theory and data comparison}

The main goal of jet quenching studies is to use jet quenching as a tool for quantifying properties of the QGP
formed in high-energy heavy-ion collisions. We have already seen the values of the jet transport parameter $\hat q$ extracted from comparisons between experimental data and theoretical calculations of hadron and jet spectra based on a wide range of models of jet-medium interaction. To improve the accuracy of these extractions based on direct fits or Bayesian inferences, one needs to improve the current theoretical framework beyond the leading order pQCD and leading eikonal approximation to take into account of the gradient corrections and flow velocity dependence. The wealth of jet substructure and jet multiplicity measurements needs to be included in Bayesian inference studies. One also needs to extend the kinematic ranges and statistics of a wide variety of jet observables, especially direct $\gamma/Z$ triggered jets with narrower distribution of the jet initial energy. Measurements of jet observables in geometrically engineered events with given azimuthal anisotropic flows $v_n$ can also provide additional constraints on the system size or path-length dependence. Tomographic tools coupled with machine learning have been developed to image density and flow velocity distributions in the expanding QGP matter \cite{He:2020iow,Yang:2022yfr,Xiao:2024ffk}. These include coincidental measurements of soft and hard hadron spectra correlated with jet production. These measurements not only provide more differential studies of the hard jets but also the medium response induced by jet-medium interaction. New observables should still be needed for direct access to the jet-induced Mach-cone wave front. Sensitivities of these observables on Mach-cone and diffusion wake to the equation of state and bulk transport properties should also be studied. Experiments on jet propagation inside cold nuclei at future electron-ion colliders (EIC) can also provide comparative study on the difference between hot and cold nuclear medium.

In addition, a significant range of further central questions may become accessible in the coming decades in an improved interplay between theory and experiment. These questions include:

\begin{itemize}
\item
Can we unambiguously identify Rutherford-like scattering events in jet-medium interactions? Can we quantify in this way from which momentum transfer scale upwards particle-like QGP constituents reveal themselves in jet-medium interactions? We note in this context the qualitatively different pictures of how the QGP interacts with jets in the weakly coupled (see Section~\ref{weakjet}) and in the strongly coupled (see Section~\ref{strongjet}) regimes.
\item
Are there, beyond the improved extraction of $\hat q$, other in-medium transport properties that can be quantified in improved Bayesian analyses? Opportunities in this direction range, in principle, from independent determination of the velocity of sound via analyses of Mach-cone like structures to extending the study of Langevin transport parameters (and their necessary extensions) to the phenomenology of relativistically moving heavy quarks.
\item
How can we better relate the jet quenching phenomena to an all encompassing dynamics of out-of-equilibrium evolution within which jet quenching is only the high-$p_T$ limit of a dynamics that drives - at late times - highly energetic probes into equilibrium?
\item
Rather than using jet quenching to test the QGP, what are the perspectives of using the QGP to test as far untested aspects of jet evolution? In the perturbative regime, this includes e.g. tests of the concept of QCD formation time that is inaccessible in vacuum. In the non-perturbative regime, jet quenching in medium has the potential of providing insights into the dynamics of hadronization that are relevant beyond heavy-ion physics.
\end{itemize}

\begin{acknowledgement}
We thank A. Dainese, J. W. Harris, P. Jacobs, B. Muller, T. Nayak, D. Perepelitsa, K. Rajagopal, G. M. Roland, C. Salgado, B. Scheihing-Hitschfeld, M. van Leeuwen, J. Velkovska and K. Zapp for constructive input and criticism at various stages of this work. XNW is supported in part by the National Science Foundation of China under Grant No. 1193507, by the Guangdong Major Project of Basic and Applied Ba-
sic Research under Grant No. 2020B030103008 and by the Alexander von Humboldt Foundation through the Humboldt Research Award during the research visit to the Institut f\"{u}r Theoretische Physik, Johann Wolfgang Goethe–Universit\"{a}t.
\end{acknowledgement}

\bibliographystyle{atlasnote}
\bibliography{biblio}

\end{document}